\documentclass{lmcs}
\pdfoutput=1
\usepackage[utf8]{inputenc}

\usepackage{lastpage}
\lmcsdoi{21}{1}{12}
\lmcsheading{}{\pageref{LastPage}}{}{}%
{Dec.~13,~2024}{Jan.~31,~2025}{}


\usepackage{hyperref}

\usepackage{wrapfig}

\usepackage{amsmath}
\usepackage{mathrsfs}
\usepackage{placeins}
\usepackage[all]{xy}
\usepackage{tikz}
\usetikzlibrary{arrows,cd}
\usepackage{tikzit}

\tikzstyle{black}=[circle, draw=black, fill=black, inner sep=0pt, minimum size=4pt]
\tikzstyle{basic box}=[draw, fill=white, rectangle, minimum height=1.2em, minimum width=1em]
\tikzstyle{gate}=[shape=rectangle, text height=1.5ex, text depth=0.25ex, yshift=0.5mm, fill=white, draw=black, minimum height=5mm, yshift=-0.5mm, minimum width=5mm, font={\small}, tikzit category=circuit]
\tikzstyle{big gate}=[shape=rectangle, text height=1.5ex, text depth=0.25ex, yshift=0.5mm, fill=white, draw=black, minimum height=10mm, yshift=-0.5mm, minimum width=5mm, font={\small}, tikzit category=circuit]
\tikzstyle{Z dot}=[inner sep=0mm, minimum size=2mm, shape=circle, draw=black, fill={rgb,255: red,221; green,255; blue,221}, tikzit category=zx]
\tikzstyle{Z phase dot}=[minimum size=5mm, font={\footnotesize\boldmath}, shape=rectangle, rounded corners=2mm, inner sep=0.2mm, outer sep=-2mm, scale=0.8, tikzit shape=circle, draw=black, fill={rgb,255: red,221; green,255; blue,221}, tikzit draw=blue, tikzit category=zx]
\tikzstyle{X dot}=[Z dot, shape=circle, draw=black, fill={rgb,255: red,255; green,136; blue,136}, tikzit category=zx]
\tikzstyle{X phase dot}=[Z phase dot, tikzit shape=circle, tikzit draw=blue, fill={rgb,255: red,255; green,136; blue,136}, font={\footnotesize\boldmath}, tikzit category=zx]
\tikzstyle{hadamard}=[fill=yellow, draw=black, shape=rectangle, inner sep=0.6mm, minimum height=1.5mm, minimum width=1.5mm, tikzit category=zx]
\tikzstyle{paulibox}=[fill={rgb,255: red,221; green,221; blue,255}, draw=black, shape=rectangle, inner sep=0.6mm, minimum height=5mm, minimum width=5mm, font={\footnotesize}, text height=1.5ex, text depth=0.25ex, tikzit category=zx]
\tikzstyle{vertex}=[inner sep=0mm, minimum size=1mm, shape=circle, draw=black, fill=black, tikzit category=misc]
\tikzstyle{vertex set}=[inner sep=0mm, minimum size=1mm, shape=circle, draw=black, fill=white, font={\footnotesize\boldmath}, tikzit category=misc]
\tikzstyle{small black dot}=[fill=black, draw=black, shape=circle, inner sep=0pt, minimum width=1.2mm, tikzit category=circuit, font=\footnotesize, text=white]
\tikzstyle{cnot ctrl}=[fill=black, draw=black, shape=circle, inner sep=0pt, minimum width=1.2mm, tikzit category=circuit]
\tikzstyle{cnot targ}=[fill=white, draw=white, shape=circle, tikzit category=circuit, label={center:$\oplus$}, inner sep=0pt, minimum width=2.1mm, tikzit fill={rgb,255: red,102; green,204; blue,255}, tikzit draw=black]
\tikzstyle{ket}=[fill=white, draw=black, shape=regular polygon, regular polygon sides=3, regular polygon rotate=-30, scale=0.7, inner sep=1pt, tikzit category=circuit, tikzit shape=rectangle, tikzit fill=green]
\tikzstyle{bra}=[fill=white, draw=black, shape=regular polygon, regular polygon sides=3, regular polygon rotate=30, scale=0.7, inner sep=1pt, tikzit category=circuit, tikzit shape=rectangle, tikzit fill=red]
\tikzstyle{scalar}=[shape=rectangle, text height=1.5ex, text depth=0.25ex, yshift=0.5mm, fill=white, draw=black, minimum height=5mm, yshift=-0.5mm, minimum width=5mm, font={\small}]
\tikzstyle{clabel}=[fill=white, draw=none, shape=rectangle, tikzit fill={rgb,255: red,56; green,255; blue,242}, font={\footnotesize}, inner sep=1pt, tikzit category=labels]
\tikzstyle{empty diagram}=[draw={gray!40!white}, dashed, shape=rectangle, minimum width=1cm, minimum height=1cm, tikzit category=misc]
\tikzstyle{amap}=[fill=white, draw=black, shape=NEbox, tikzit category=asymmetric, tikzit fill=yellow, tikzit shape=rectangle]
\tikzstyle{amap conj}=[fill=white, draw=black, shape=NWbox, tikzit category=asymmetric, tikzit fill=green, tikzit shape=rectangle]
\tikzstyle{amap adj}=[fill=white, draw=black, shape=SEbox, tikzit category=asymmetric, tikzit fill=red, tikzit shape=rectangle]
\tikzstyle{amap trans}=[fill=white, draw=black, shape=SWbox, tikzit category=asymmetric, tikzit fill=orange, tikzit shape=rectangle]
\tikzstyle{astate}=[fill=white, draw=black, shape=NEtriangle, tikzit category=asymmetric, tikzit shape=circle, tikzit fill=yellow]
\tikzstyle{astate conj}=[fill=white, draw=black, shape=NWtriangle, tikzit category=asymmetric, tikzit shape=circle, tikzit fill=green]
\tikzstyle{astate adj}=[fill=white, draw=black, shape=SEtriangle, tikzit category=asymmetric, tikzit shape=circle, tikzit fill=red]
\tikzstyle{astate trans}=[fill=white, draw=black, shape=SWtriangle, tikzit category=asymmetric, tikzit shape=circle, tikzit fill=orange]

\tikzstyle{hadamard edge}=[-, dashed, dash pattern=on 2pt off 0.5pt, thick, draw={rgb,255: red,68; green,136; blue,255}]
\tikzstyle{box edge}=[-, dashed, dash pattern=on 2pt off 0.5pt, thick, draw={rgb,255: red,203; green,192; blue,225}]
\tikzstyle{brace edge}=[-, tikzit draw=blue, decorate, decoration={brace,amplitude=1mm,raise=-1mm}]
\tikzstyle{diredge}=[->]
\tikzstyle{double edge}=[-, double, shorten <=-1mm, shorten >=-1mm, double distance=2pt]
\tikzstyle{gray edge}=[-, {gray!60!white}]
\tikzstyle{pointer edge}=[->, very thick, gray]
\tikzstyle{boldedge}=[-, line width=1.6pt, shorten <=-0.17mm, shorten >=-0.17mm]
\tikzstyle{bidir edge}=[<->, very thick, draw={rgb,255: red,191; green,191; blue,191}]

\usepackage{cleveref}
\usepackage{stmaryrd}

\makeatletter
\DeclareFontFamily{OMX}{MnSymbolE}{}
\DeclareSymbolFont{MnLargeSymbols}{OMX}{MnSymbolE}{m}{n}
\SetSymbolFont{MnLargeSymbols}{bold}{OMX}{MnSymbolE}{b}{n}
\DeclareFontShape{OMX}{MnSymbolE}{m}{n}{
    <-6>  MnSymbolE5
   <6-7>  MnSymbolE6
   <7-8>  MnSymbolE7
   <8-9>  MnSymbolE8
   <9-10> MnSymbolE9
  <10-12> MnSymbolE10
  <12->   MnSymbolE12
}{}
\DeclareFontShape{OMX}{MnSymbolE}{b}{n}{
    <-6>  MnSymbolE-Bold5
   <6-7>  MnSymbolE-Bold6
   <7-8>  MnSymbolE-Bold7
   <8-9>  MnSymbolE-Bold8
   <9-10> MnSymbolE-Bold9
  <10-12> MnSymbolE-Bold10
  <12->   MnSymbolE-Bold12
}{}

\let\llangle\@undefined
\let\rrangle\@undefined
\DeclareMathDelimiter{\llangle}{\mathopen}%
                     {MnLargeSymbols}{'164}{MnLargeSymbols}{'164}
\DeclareMathDelimiter{\rrangle}{\mathclose}%
                     {MnLargeSymbols}{'171}{MnLargeSymbols}{'171}
\makeatother

\newcommand{\emptyword}{\epsilon}

\newcommand\hyp{\mathbf{Hyp}_{\Sigma,\mathcal{C}}}
\newcommand\disc{D}

\newcommand{\JSGraph}{\mathbf{JSGraph}}
\newcommand\csp{\mathrm{Csp}_{\disc}}
\newcommand{\cspobj}[5]{#1 \xrightarrow{#2} #3 \xleftarrow{#4} #5}
\newcommand\cspf{\mathrm{Csp}(\mathbb{F})}
\newcommand\rmacspf{\mathrm{RMCsp}(\mathbb{F})}
\newcommand\rmacsphyp{\mathrm{RMACsp}_{\disc}\left(\hyp\right)}
\newcommand\rmacspemptyhyp{\mathrm{RMACsp}_{\disc}\left(\mathbf{Hyp}_{\emptyset,\mathcal{C}}\right)}
\newcommand\csphyp{\csp\left(\hyp\right)}

\newcommand\iso[1]{\left\llangle #1 \right\rrangle}
\newcommand\cprd{\mathbf{S}_{\Sigma,\mathcal{C}} + \mathbf{CMon}_\mathcal{C}}
\newcommand\sfun[1][\cdot]{\left\llbracket{#1}\right\rrbracket}
\newcommand\cfun[1][\cdot]{\left| #1 \right|}
\tikzset{every picture/.style={line width=0.75pt}} 

\newcommand\bad{left-amonogamous }

\newcommand\rew{\hookrightarrow}
\newcommand{\R}{\mathscr{R}}

\newcommand\frob[1]{\left\llangle #1\right\rrangle}

\newcommand{\from}{:\,}

\usepackage{graphicx}

\newcommand{\perm}[1]{\mathbf{Perm}_{\scriptscriptstyle #1}}

\newcommand{\PROP}[1]{\mathbf{Prop}_{\scriptscriptstyle #1}} 
\newcommand{\SMCAT}{\ensuremath{\mathsf{SmCat}}} 

\newcommand{\syntax}[1]{\mathbf{S}_{\scriptscriptstyle #1}}

\newcommand{\col}{\mathcal{C}}

\newcommand{\tns}{\oplus}

\newcommand{\diagbox}[3]{
\begin{tikzpicture}
	\begin{pgfonlayer}{nodelayer}
		\node [style=basic box] (0) at (0, 0) {$#1$};
		\node [style=none] (1) at (0.75, 0) {};
		\node [style=none] (2) at (-0.75, 0) {};
		\node [style=none] (3) at (0.65, 0.15) {\scriptsize $#3$};
		\node [style=none] (4) at (-0.65, 0.15) {\scriptsize $#2$};
	\end{pgfonlayer}
	\begin{pgfonlayer}{edgelayer}
		\draw (2.center) to (0);
		\draw (0) to (1.center);
	\end{pgfonlayer}
\end{tikzpicture}
}

\newcommand{\idx}[1]{
  \tikz \draw (0, 0) -- (0.5, 0) node [midway, above] {\scriptsize $#1$};
}

\newcommand{\cat}{\mathbf{C}}

\newcommand{\drcorner}{{\ar@{}[dr]|(.8){\text{\large $\ulcorner$}}}}

\begin{document}

\title[Rewriting for SMCs with Commutative (Co)Monoid Structure]{Rewriting for Symmetric Monoidal Categories with Commutative (Co)Monoid Structure} 
\thanks{We thank Tobias Fritz for helpful discussions, and the anonymous reviewers of CALCO and LMCS for their suggestions. FZ acknowledges support from \textsc{epsrc} grant EP/V002376/1, \textsc{miur} PRIN P2022HXNSC, and \textsc{aria} Safeguarded AI TA1.1 grant n.8777242. Part of this work was conducted while FZ was affiliated with University of Bologna, Italy.}

\author[A.~Milosavljevi{\'c}]{Aleksandar Milosavljevi{\'c}}
\author[R.~Piedeleu]{Robin Piedeleu\lmcsorcid{0000-0002-3945-2704}}
\author[F.~Zanasi]{Fabio Zanasi\lmcsorcid{0000-0001-6457-1345}}

\address{University College London, United Kingdom}	
\email{aleksmil000@gmail.com, r.piedeleu@ucl.ac.uk, f.zanasi@ucl.ac.uk}  







\begin{abstract}
String diagrams are pictorial representations for morphisms of symmetric monoidal categories. They constitute an intuitive and expressive graphical syntax, which has found application in a very diverse range of fields including concurrency theory, quantum computing, control theory, machine learning, linguistics, and digital circuits. Rewriting theory for string diagrams relies on a combinatorial interpretation as double-pushout rewriting of certain hypergraphs. As previously studied, there is a `tension' in this interpretation: in order to make it sound and complete, we either need to add structure on string diagrams (in particular, Frobenius algebra structure) or pose restrictions on double-pushout rewriting (resulting in `convex' rewriting). From the string diagram viewpoint, imposing a full Frobenius structure may not always be natural or desirable in applications, which motivates our study of a weaker requirement: commutative monoid structure. In this work we characterise string diagram rewriting modulo commutative monoid equations, via a sound and complete interpretation in a suitable notion of double-pushout rewriting of hypergraphs.
\end{abstract}

\maketitle

\section{Introduction}

String diagrams are a graphical language for morphisms of categories. Their use has been popularised in the context of monoidal categories, by the seminal works of Kelly, Laplaza, Joyal, and Street~\cite{Kelly1980CoherenceFC,JOYAL199155}. In more recent years, string diagrams have found applications in diverse fields, including quantum computation~\cite{kissinger-quantum}, digital~\cite{Ghica17} and electrical circuits~\cite{BoisseauSobocinski21,BonchiPSZ19}, machine learning~\cite{CruttwellGGWZ22}, concurrency theory~\cite{BonchiHPSZ19}, control theory~\cite{BaezErbele-CategoriesInControl,Bonchi0Z21}, and linguistics~\cite{SadrzadehCC13} among others. Compared to traditional syntax, string diagrams allow one to neatly visualise resource-exchange and message-passing between different parts of a system, which is pivotal in studying the subtle interactions that arise in concurrent processes and quantum computation, for example. Moreover, we can reason with string diagrams both combinatorially and as syntactic, inductively defined objects, which enables forms of compositional analysis typical of programming language semantics. We refer to \cite{piedeleu2023introduction} for a recent survey of string diagrams in computer science, and \cite{Selinger2011,Hinze_Marsden_2023} for a survey of diagrammatic languages to account for various kinds of categorical structures. 

A cornerstone of string diagrammatic approaches is the possibility of performing \emph{diagrammatic reasoning}: transforming a string diagram according to a certain rewrite rule, which replaces a sub-diagram with a different one. A set of such rules, which typically preserve the semantics of the model, may represent for instance a compilation procedure~\cite{MuroyaG19}, the realisation of a specification~\cite{Bonchi0Z21}, a refinement of system behaviour~\cite{BonchiHPS17}.

Compared to traditional term rewriting, a mathematical theory of string diagram rewriting poses new challenges. Formally, string diagrams are graphical representations of morphisms in a category, typically assumed in applications to be a symmetric monoidal category (SMC). In order to perform a rewrite step, we need to match the left-hand side of a rewrite rule to a sub-diagram of a given string diagram. For instance, consider the rewrite rule as on the left below, and the string diagram on the right.
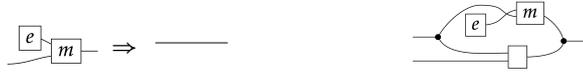
\begin{equation*}
\begin{tikzpicture}[scale=0.7]
	\begin{pgfonlayer}{nodelayer}
		\node [style=none] (104) at (-6.5, -0.75) {};
		\node [style=none] (105) at (-1.5, 0) {};
		\node [style=basic box] (106) at (-3.25, 0) {$m$};
		\node [style=none] (107) at (-3, 0) {};
		\node [style=none] (108) at (-3.5, 0.25) {};
		\node [style=none] (109) at (-3.5, -0.25) {};
		\node [style=none] (110) at (-5, 0.75) {};
		\node [style=basic box] (111) at (-5.25, 0.75) {$e$};
		\node [style=none] (120) at (-0.25, 0) {$\Rightarrow$};
		\node [style=none] (122) at (1, 0) {};
		\node [style=none] (123) at (4, 0) {};
	\end{pgfonlayer}
	\begin{pgfonlayer}{edgelayer}
		\draw [in=-180, out=0] (107.center) to (105.center);
		\draw [in=180, out=0] (104.center) to (109.center);
		\draw [in=180, out=0] (110.center) to (108.center);
		\draw (122.center) to (123.center);
	\end{pgfonlayer}
\end{tikzpicture}\qquad\qquad\qquad\qquad\qquad \begin{tikzpicture}[scale=0.6]
	\begin{pgfonlayer}{nodelayer}
		\node [style=black] (104) at (-4.25, 0) {};
		\node [style=black] (105) at (3.5, 0) {};
		\node [style=basic box] (106) at (1.25, 1.25) {$m$};
		\node [style=none] (107) at (1.5, 1.25) {};
		\node [style=none] (108) at (1, 1.5) {};
		\node [style=none] (109) at (-1.75, 1.75) {};
		\node [style=none] (110) at (-1.75, 0.5) {};
		\node [style=basic box] (111) at (-2, 0.5) {$e$};
		\node [style=none] (124) at (1, -1.25) {};
		\node [style=none] (125) at (5, 0) {};
		\node [style=basic box] (126) at (0.5, -1.25) {};
		\node [style=none] (127) at (0, -1) {};
		\node [style=none] (128) at (0, -1.5) {};
		\node [style=none] (129) at (-5.75, -1.5) {};
		\node [style=none] (130) at (-5.75, 0) {};
		\node [style=none] (147) at (1, 1) {};
	\end{pgfonlayer}
	\begin{pgfonlayer}{edgelayer}
		\draw [in=105, out=0, looseness=0.75] (107.center) to (105);
		\draw [in=180, out=60] (104) to (109.center);
		\draw [in=180, out=0, looseness=1.50] (110.center) to (108.center);
		\draw [in=-105, out=0, looseness=0.75] (124.center) to (105);
		\draw [in=-180, out=-60, looseness=0.75] (104) to (127.center);
		\draw (129.center) to (128.center);
		\draw (130.center) to (104);
		\draw (105) to (125.center);
		\draw [in=0, out=180, looseness=1.75] (147.center) to (109.center);
	\end{pgfonlayer}
\end{tikzpicture}
\end{equation*}
Morally, there is a match for the rule in the string diagram. The issue is that, strictly speaking, such a match does not happen on the nose: we need first to apply the laws of SMCs in order to transform the string diagram into an equivalent one, with the wires into $m$ uncrossed. At this point we have clearly isolated the sub-diagram and are able to perform the rewrite step.
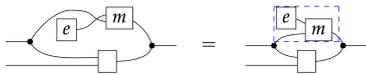
\begin{equation*}
\begin{tikzpicture}[scale=0.6]
	\begin{pgfonlayer}{nodelayer}
		\node [style=black] (104) at (-4.25, 0) {};
		\node [style=black] (105) at (3.5, 0) {};
		\node [style=basic box] (106) at (1.25, 1.25) {$m$};
		\node [style=none] (107) at (1.5, 1.25) {};
		\node [style=none] (108) at (1, 1.5) {};
		\node [style=none] (109) at (-1.75, 1.75) {};
		\node [style=none] (110) at (-1.75, 0.5) {};
		\node [style=basic box] (111) at (-2, 0.5) {$e$};
		\node [style=none] (124) at (1, -1.25) {};
		\node [style=none] (125) at (5, 0) {};
		\node [style=basic box] (126) at (0.5, -1.25) {};
		\node [style=none] (127) at (0, -1) {};
		\node [style=none] (128) at (0, -1.5) {};
		\node [style=none] (129) at (-5.75, -1.5) {};
		\node [style=none] (130) at (-5.75, 0) {};
		\node [style=none] (147) at (1, 1) {};
	\end{pgfonlayer}
	\begin{pgfonlayer}{edgelayer}
		\draw [in=105, out=0, looseness=0.75] (107.center) to (105);
		\draw [in=180, out=60] (104) to (109.center);
		\draw [in=180, out=0, looseness=1.50] (110.center) to (108.center);
		\draw [in=-105, out=0, looseness=0.75] (124.center) to (105);
		\draw [in=-180, out=-60, looseness=0.75] (104) to (127.center);
		\draw (129.center) to (128.center);
		\draw (130.center) to (104);
		\draw (105) to (125.center);
		\draw [in=0, out=180, looseness=1.75] (147.center) to (109.center);
	\end{pgfonlayer}
\end{tikzpicture} \quad = \quad \begin{tikzpicture}[scale=0.6]
	\begin{pgfonlayer}{nodelayer}
		\node [style=black] (131) at (-8.25, 0) {};
		\node [style=black] (132) at (-3.5, 0) {};
		\node [style=basic box] (133) at (-5.5, 1) {$m$};
		\node [style=none] (134) at (-5.25, 1) {};
		\node [style=none] (135) at (-5.75, 1.25) {};
		\node [style=none] (136) at (-5.75, 0.75) {};
		\node [style=none] (137) at (-7.25, 1.75) {};
		\node [style=basic box] (138) at (-7.5, 1.75) {$e$};
		\node [style=none] (139) at (-5.75, -1) {};
		\node [style=none] (140) at (-2, 0) {};
		\node [style=basic box] (141) at (-6.25, -1) {};
		\node [style=none] (142) at (-6.75, -0.75) {};
		\node [style=none] (143) at (-6.75, -1.25) {};
		\node [style=none] (144) at (-10, -1.25) {};
		\node [style=none] (145) at (-10, 0) {};
		\node [style=none] (148) at (-8.75, 2.5) {};
		\node [style=none] (149) at (-8.75, 0.25) {};
		\node [style=none] (150) at (-4.25, 0.25) {};
		\node [style=none] (151) at (-4.25, 2.5) {};
		\node [style=none] (152) at (0, 0) {$\Rightarrow$};
		\node [style=black] (153) at (3.75, 0) {};
		\node [style=black] (154) at (8.5, 0) {};
		\node [style=none] (156) at (6.75, 1) {};
		\node [style=none] (158) at (5.5, 1) {};
		\node [style=none] (161) at (6.25, -1) {};
		\node [style=none] (162) at (10, 0) {};
		\node [style=basic box] (163) at (5.75, -1) {};
		\node [style=none] (164) at (5.25, -0.75) {};
		\node [style=none] (165) at (5.25, -1.25) {};
		\node [style=none] (166) at (2, -1.25) {};
		\node [style=none] (167) at (2, 0) {};
	\end{pgfonlayer}
	\begin{pgfonlayer}{edgelayer}
		\draw [in=105, out=0, looseness=0.75] (134.center) to (132);
		\draw [in=180, out=60, looseness=0.75] (131) to (136.center);
		\draw [in=180, out=0] (137.center) to (135.center);
		\draw [in=-105, out=0, looseness=0.75] (139.center) to (132);
		\draw [in=-180, out=-60, looseness=0.75] (131) to (142.center);
		\draw (144.center) to (143.center);
		\draw (145.center) to (131);
		\draw (132) to (140.center);
		\draw [dashed] (148.center) to (151.center);
		\draw [dashed] (151.center) to (150.center);
		\draw [dashed] (150.center) to (149.center);
		\draw [dashed] (149.center) to (148.center);
		\draw [in=105, out=0, looseness=0.75] (156.center) to (154);
		\draw [in=180, out=60] (153) to (158.center);
		\draw [in=-105, out=0, looseness=0.75] (161.center) to (154);
		\draw [in=-180, out=-60, looseness=0.75] (153) to (164.center);
		\draw (166.center) to (165.center);
		\draw (167.center) to (153);
		\draw (154) to (162.center);
		\draw (158.center) to (156.center);
	\end{pgfonlayer}
\end{tikzpicture}
\end{equation*}
As seen in this example, string diagram rewriting is performed modulo certain structural laws, which reflect the categorical structure in which the string diagrams live. However, from a practical viewpoint, this form of rewriting is not really feasible, as each rewrite step would require us to inspect all string diagrams equivalent to a given one looking for redexes.

This issue can be solved via an interpretation of string diagrams as certain hypergraphs, and of string diagram rewriting as \emph{double-pushout rewriting} (DPO)~\cite{dpoapproach} of such hypergraphs. We refer to \cite{pt1,pt2,pt3} for a systematic introduction to this approach. In a nutshell, the benefit of working with such an interpretation is that an equivalence class of string diagrams corresponds to just one hypergraph, meaning that our search for redexes is drastically simplified. However, there is a mismatch: if we want to rewrite string diagrams in a SMC, then soundness is only ensured by adopting a restricted notion of DPO rewriting, called \emph{convex} DPO rewriting~\cite{BonchiGKSZ16}. Conversely, if we want to work with arbitrary DPO rewriting steps, then the corresponding notion of string diagram rewriting does not rewrite only modulo the laws of SMCs, but requires a special commutative \emph{Frobenius algebra} on each object of the category. Recall that a Frobenius algebra consists of a commutative monoid and a commutative comonoid, interacting with each other via the so-called Frobenius law~\cite{Carboni1987}.

When modelling a certain class of systems with string diagrams, assuming that such Frobenius structure exists is not always reasonable, or desirable. A first class of such examples are matrix-like semantic structures, which are axiomatised by bialgebra equations---see \emph{e.g.}~\cite{Zanasi15} for a survey. It is known that if the monoid and the comonoid both obey the Frobenius and the bialgebra laws, then the equational theory trivialises, \emph{cf.}~\cite[Ex. 4.3]{FongZ18}. A second important class are semantic structures for probability theory, which usually feature a commutative comonoid structure, but no Frobenius equations --- introducing Frobenius structure amounts to allowing unnormalised probabilities, \emph{cf.}~\cite{JacobsKZ21,Fritz_2020}. These categories, sometimes called~\emph{CD-categories}, also play a special role in the study of algebraic theories, because they model the cartesian handling of variables~\cite{BonchiSZ18}.

All these models motivate the study of rewriting for intermediate structures between plain symmetric monoidal and those equipped with Frobenius algebras. More specifically, we focus on string diagrams in categories where each object comes with a \emph{commutative monoid} structure. From a rewriting viewpoint, this case is particularly significant because symmetries in a SMC may always create redexes for the commutativity axiom of the monoid  multiplication, yielding a non-terminating rewrite system:
\[\begin{tikzpicture}
	\begin{pgfonlayer}{nodelayer}
		\node [style=none] (1) at (5.5, 0.25) {};
		\node [style=none] (2) at (3.75, 0.75) {};
		\node [style=none] (3) at (2.75, 0.75) {};
		\node [style=none] (4) at (2.75, -0.25) {};
		\node [style=none] (5) at (3.75, 0.75) {};
		\node [style=none] (6) at (3.75, -0.25) {};
		\node [style=none] (7) at (2.5, 0.75) {};
		\node [style=none] (8) at (2.5, -0.25) {};
		\node [style=none] (9) at (4, 0.75) {};
		\node [style=none] (10) at (3.75, 0.75) {};
		\node [style=none] (11) at (4, -0.25) {};
		\node [style=none] (13) at (4, -0.25) {};
		\node [style=none] (15) at (15.25, 0.25) {};
		\node [style=none] (16) at (14, 0.75) {};
		\node [style=none] (17) at (13, 0.75) {};
		\node [style=none] (18) at (13, -0.25) {};
		\node [style=none] (19) at (14, 0.75) {};
		\node [style=none] (20) at (14, -0.25) {};
		\node [style=none] (21) at (12.75, 0.75) {};
		\node [style=none] (22) at (12.75, -0.25) {};
		\node [style=none] (23) at (14.5, 0.25) {};
		\node [style=none] (24) at (14, 0.75) {};
		\node [style=none] (25) at (14.5, 0.25) {};
		\node [style=none] (27) at (14.5, 0.25) {};
		\node [style=none] (28) at (12.75, 0.75) {};
		\node [style=none] (29) at (11.75, 0.75) {};
		\node [style=none] (30) at (11.75, -0.25) {};
		\node [style=none] (31) at (12.75, 0.75) {};
		\node [style=none] (32) at (12.75, -0.25) {};
		\node [style=none] (33) at (11.5, 0.75) {};
		\node [style=none] (34) at (11.5, -0.25) {};
		\node [style=none] (35) at (12.75, 0.75) {};
		\node [style=none] (37) at (9, 0.25) {};
		\node [style=none] (38) at (7.75, 0.75) {};
		\node [style=none] (39) at (7.75, 0.75) {};
		\node [style=none] (40) at (7.75, -0.25) {};
		\node [style=none] (41) at (8.75, 0.25) {};
		\node [style=none] (42) at (7.75, 0.75) {};
		\node [style=none] (43) at (8, -0.25) {};
		\node [style=none] (44) at (8.75, 0.25) {};
		\node [style=none] (45) at (8, -0.25) {};
		\node [style=none] (46) at (6.25, 0.25) {$\Rightarrow$};
		\node [style=none] (47) at (10.25, 0.25) {$=$};
		\node [style=none] (48) at (16.25, 0.25) {$\Rightarrow$};
		\node [style=none] (50) at (19.75, 0.25) {};
		\node [style=none] (51) at (18.5, 0.75) {};
		\node [style=none] (52) at (17.5, 0.75) {};
		\node [style=none] (53) at (17.5, -0.25) {};
		\node [style=none] (54) at (18.5, 0.75) {};
		\node [style=none] (55) at (18.5, -0.25) {};
		\node [style=none] (56) at (17.25, 0.75) {};
		\node [style=none] (57) at (17.25, -0.25) {};
		\node [style=none] (58) at (19.25, 0.25) {};
		\node [style=none] (59) at (18.5, 0.75) {};
		\node [style=none] (60) at (19.25, 0.25) {};
		\node [style=none] (62) at (19.25, 0.25) {};
		\node [style=none] (63) at (20.75, 0.25) {$\Rightarrow$};
		\node [style=none] (64) at (23.75, 0.25) {$\dots$};
		\node [style=small black dot] (66) at (19.25, 0.25) {};
		\node [style=none] (67) at (19.75, 0.25) {};
		\node [style=small black dot] (68) at (14.5, 0.25) {};
		\node [style=none] (69) at (15.25, 0.25) {};
		\node [style=small black dot] (70) at (8.75, 0.25) {};
		\node [style=none] (71) at (9.5, 0.25) {};
		\node [style=none] (73) at (3.75, 0.75) {};
		\node [style=none] (74) at (3.75, 0.75) {};
		\node [style=none] (75) at (3.75, -0.25) {};
		\node [style=none] (76) at (4.75, 0.25) {};
		\node [style=none] (77) at (3.75, 0.75) {};
		\node [style=none] (78) at (4, -0.25) {};
		\node [style=none] (79) at (4.75, 0.25) {};
		\node [style=none] (80) at (4, -0.25) {};
		\node [style=small black dot] (81) at (4.75, 0.25) {};
		\node [style=none] (82) at (13.75, -0.75) {};
	\end{pgfonlayer}
	\begin{pgfonlayer}{edgelayer}
		\draw [in=-180, out=0, looseness=1.25] (3.center) to (6.center);
		\draw [in=180, out=0, looseness=1.25] (4.center) to (5.center);
		\draw (7.center) to (3.center);
		\draw (8.center) to (4.center);
		\draw (10.center) to (9.center);
		\draw (6.center) to (11.center);
		\draw [in=-180, out=0, looseness=1.25] (17.center) to (20.center);
		\draw [in=180, out=0, looseness=1.25] (18.center) to (19.center);
		\draw (21.center) to (17.center);
		\draw (22.center) to (18.center);
		\draw [bend left] (24.center) to (23.center);
		\draw [bend right, looseness=1.25] (20.center) to (25.center);
		\draw [in=-180, out=0, looseness=1.25] (29.center) to (32.center);
		\draw [in=180, out=0, looseness=1.25] (30.center) to (31.center);
		\draw (33.center) to (29.center);
		\draw (34.center) to (30.center);
		\draw [in=150, out=0, looseness=1.50] (42.center) to (41.center);
		\draw (40.center) to (43.center);
		\draw [bend right, looseness=0.50] (45.center) to (44.center);
		\draw [in=-180, out=0, looseness=1.25] (52.center) to (55.center);
		\draw [in=180, out=0, looseness=1.25] (53.center) to (54.center);
		\draw (56.center) to (52.center);
		\draw (57.center) to (53.center);
		\draw [bend left=15, looseness=1.25] (59.center) to (58.center);
		\draw [bend right=15, looseness=1.25] (55.center) to (60.center);
		\draw (66) to (67.center);
		\draw (68) to (69.center);
		\draw (70) to (71.center);
		\draw [in=150, out=0, looseness=1.50] (77.center) to (76.center);
		\draw (75.center) to (78.center);
		\draw [bend right, looseness=0.50] (80.center) to (79.center);
		\draw (81) to (1.center);
	\end{pgfonlayer}
\end{tikzpicture}\]
Therefore, rather than taking commutativity as a rewrite rule, we need to find an alternative representation of string diagrams (and of string diagram rewriting) that is invariant modulo the axioms of commutative monoids (and the laws of SMCs), which is the focus of this paper. Our contribution is two-fold:
\begin{itemize}
    \item we identify which class of hypergraphs provides an adequate interpretation of string diagrams in a SMC with commutative monoid structure, and organise them into a SMC. This characterisation will take the form of an isomorphism between the SMC of string diagrams and the SMC of hypergraphs.\footnote{Simultaneously to a preprint of our work~\cite{arXivversion}, a preprint of~\cite{Fritz}, showing a result closely related to this first of our contributions, also appeared on ArXiv. We comment on their relation in~Section~\ref{sec:conclusions}.}
    \item We identify which notion of double-pushout hypergraph rewriting interprets string diagram rewriting modulo the axioms of commutative monoids in a sound and complete way.
\end{itemize}
Note that all of the theory developed in this work can be easily dualised to obtain a framework for rewriting modulo commutative \emph{comonoid} structure, which justifies the title and makes it relevant also for the aforementioned CD-categories.

\noindent \textit{Synopsis. } Section~\ref{sec:cospanmonoids} recalls background on string diagrams and hypergraphs. Section~\ref{sec:characterisation} shows the hypergraph characterisation of string diagrams with a chosen commutative monoid structure. Section~\ref{sec:rewriting} shows how string diagram rewriting may be characterised in terms of DPO hypergraph rewriting. We summarise our work and suggest future directions in Section~\ref{sec:conclusions}. 

\medskip

This work extends the conference paper~\cite{MilosavljevicPZ23}. It contains additional details, examples, and all the missing proofs, some of which have also been simplified. Additionally, whereas~\cite{MilosavljevicPZ23} only considers single-sorted theories (props), here we develop our approach in the more general case of multi-sorted theories (coloured props).

\section{Preliminaries}\label{sec:cospanmonoids}

We recall some basic definitions, using the same terminology as \cite{pt1}. For a systematic treatment of coloured props, we refer the reader to~\cite{HackneyColouredPROPs15}, and for an introduction to string diagrams to~\cite{piedeleu2023introduction}.

\begin{defi}[Theories] A \emph{symmetric monoidal theory} is a tuple $(\Sigma, \col, E)$, where $\col$ is a finite set of \emph{colours}, $\Sigma$ is a monoidal signature and $E$ is a set of equations. The signature $\Sigma$ is a set of operations $o: w \rightarrow v $ with a fixed arity $w \in {\col}^{\star}$ and coarity $v \in {\col}^{\star}$, and $E$ is a set of equations, i.e. pairs $\langle l,r \rangle$ of $\Sigma$-terms $l,r \colon v \rightarrow w$ with the same arity and coarity (note that we use $\epsilon$ below to denote the empty word). $\Sigma$-terms are freely obtained by combining operations in $\Sigma$, a \emph{unit} ${id}_c \colon c\to c$ for each $c \in \col$ and a \emph{symmetry} $\sigma_{c,d} \colon cd\to dc$ for each $c,d\in \col$, by sequential ($;$) and parallel ($\tns$) composition. That means, given terms $a \colon w_1 \to w_2$, $b \colon w_2\to w_3$, $a' \colon v_1\to v_2$, one constructs new terms $a ; b \colon w_1\to w_3$ and $a \tns a' \colon w_1 v_1 \to w_2 v_2$.
\begin{figure}
\centering
    \begin{gather*}
(a; b) ; c \equiv a; (b; c),\ id_w; a\equiv a\equiv a; id_v,\
(a \oplus b) \oplus c\equiv a\oplus(b \oplus c), \ id_\emptyword \oplus a\equiv a \equiv a \oplus
id_\emptyword,\\
id_w \oplus id_v \equiv id_{wv},\quad \sigma_{w,v}; \sigma_{v,w} \equiv id_{wv}, \quad
(a \oplus id_w); \sigma_{w,v} \equiv \sigma_{w,u} ; (id_w \oplus a)\\(a; c) \oplus (b;d) \equiv (a\oplus b) ; (c \oplus d),\quad
(\sigma_{w,v}\oplus id_u) ; (id_v \oplus \sigma_{w,u} ) \equiv \sigma_{w,vu},
\end{gather*}
    \caption{Laws of symmetric monoidal categories, for morphisms of a $\mathcal{C}$-coloured prop.}
    \label{fig:laws-smc}
\end{figure}
\end{defi}
\begin{defi}[Props]\label{def:props}
A \emph{$\col$-coloured prop} is a symmetric monoidal category (SMC) where the set of objects is ${\col}^{\star}$ and the monoidal product $\tns$ on objects is word concatenation. $\col$-coloured props form a category $\PROP{\col}$ with morphisms the identity-on-objects symmetric monoidal functors. Similarly, we can form a category $\PROP{}$ where objects are $\col$-coloured props of any colour $\col$ and morphisms are strict symmetric monoidal functors. 
\end{defi}

Coloured props on a singleton colour set $\{\bullet\}$ are often referred to simply as props. The set of objects of a prop is $\{\bullet\}^{\star}$, which may regarded as the set $\mathbb{N}$ of natural numbers, with unit $0$ and monoidal product given by addition.

Given an SMT $(\Sigma, \col, E)$, one can freely construct a $\col$-coloured prop $\syntax{\Sigma , \col, E}$ with morphisms the $\Sigma$-terms quotiented by the laws of symmetric monoidal categories (Figure~\ref{fig:laws-smc}) and by the equations in $E$. We write $\syntax{\Sigma , \col}$ for empty $E$.  Details of this construction can be found in~\cite[Appendix A]{baez2018props} or~\cite[Chapter 2]{Zanasi15}.

We shall adopt the graphical notation of string diagrams~\cite{JOYAL199155} for the morphisms of $\syntax{\Sigma,\col}$. A morphism $a \from w_1 \to w_2$ is pictured as $\diagbox{a}{w_1}{w_2}$. 
Compositions via $;$ and $\tns$ are drawn respectively as horizontal and vertical juxtaposition, that means, $a ; b$ is drawn $\begin{tikzpicture}
	\begin{pgfonlayer}{nodelayer}
		\node [style=basic box] (10) at (-1.25, 0) {$a$};
		\node [style=none] (11) at (-2.5, 0) {};
		\node [style=basic box] (19) at (0.75, 0) {$b$};
		\node [style=none] (20) at (2, 0) {};
		\node [style=none] (21) at (-2.25, 0.25) {\scriptsize $w_1$};
		\node [style=none] (22) at (1.75, 0.25) {\scriptsize $w_3$};
		\node [style=none] (23) at (-0.15, 0.25) {\scriptsize $w_2$};
	\end{pgfonlayer}
	\begin{pgfonlayer}{edgelayer}
		\draw (11.center) to (10);
		\draw (10) to (19);
		\draw (19) to (20.center);
	\end{pgfonlayer}
\end{tikzpicture}$ and $a_1 \tns a_2$ is drawn $\begin{tikzpicture}
	\begin{pgfonlayer}{nodelayer}
		\node [style=basic box] (10) at (0, 0.5) {\scriptsize $a$};
		\node [style=none] (11) at (-1.5, 0.5) {};
		\node [style=none] (20) at (1.5, 0.5) {};
		\node [style=none] (21) at (-1.25, 0.75) {\scriptsize $w_1$};
		\node [style=none] (22) at (1.25, 0.75) {\scriptsize $v_1$};
		\node [style=basic box] (24) at (0, -0.75) {\scriptsize $a'$};
		\node [style=none] (25) at (1.5, -0.75) {};
		\node [style=none] (27) at (-1.5, -0.75) {};
		\node [style=none] (28) at (-1.25, -0.5) {\scriptsize $w_2$};
		\node [style=none] (29) at (1.25, -0.5) {\scriptsize $v_2$};
	\end{pgfonlayer}
	\begin{pgfonlayer}{edgelayer}
		\draw (11.center) to (10);
		\draw (24) to (25.center);
		\draw (27.center) to (24);
		\draw (10) to (20.center);
	\end{pgfonlayer}
\end{tikzpicture}$. There are specific diagrams for the symmetric monoidal structure, namely $\idx{w}$ for the identity $id_w \from w \to w$, and $\begin{tikzpicture}
	\begin{pgfonlayer}{nodelayer}
		\node [style=none] (0) at (-0.75, -0.5) {};
		\node [style=none] (1) at (-0.75, 0.25) {};
		\node [style=none] (2) at (0.75, -0.5) {};
		\node [style=none] (3) at (0.75, 0.25) {};
		\node [style=none] (4) at (0.75, -0.25) {\scriptsize $w$};
		\node [style=none] (5) at (0.75, 0.5) {\scriptsize $v$};
		\node [style=none] (6) at (-0.75, -0.25) {\scriptsize $v$};
		\node [style=none] (7) at (-0.75, 0.5) {\scriptsize $w$};
	\end{pgfonlayer}
	\begin{pgfonlayer}{edgelayer}
		\draw [in=180, out=0] (1.center) to (2.center);
		\draw [in=180, out=0] (0.center) to (3.center);
	\end{pgfonlayer}
\end{tikzpicture}$ for the symmetry $\sigma_{w,v} \colon wv \to vw$, for $w, u \in \col^{\star}$. These are definable from the basic identities and symmetries for colours in $\col$ using the pasting rules for $;$ and $\tns$ (see~\cite{piedeleu2023introduction}).

\begin{exa}\label{ex:perm}
The initial object in $\PROP{\col}$ is the $\col$-coloured prop $\perm{\col}$ whose morphisms $w \to v$ are permutations of $w$ into $v$ (thus morphisms exist only when the word $v$ is an anagram of the word $w$). $\perm{\col}$ is freely generated by the monoidal theory $(\emptyset,\col,\emptyset)$.
\end{exa}

\begin{rem}\label{rmk:coproducts} Coproducts in $\PROP{\col}$ and in $\PROP{}$ behave a bit differently, and it is convenient for later developments to recall both constructions. In $\PROP{}$, the coproduct $\cat + \cat'$ of a $\col$-coloured prop $\cat$ and a $\col'$-coloured prop $\cat'$ is constructed just as in the category $\SMCAT$ of symmetric monoidal categories, by taking $(\col \uplus \col')^{\star}$ as set of objects and combinations of $\cat$- and $\cat'$-morphisms by $;$ and $\oplus$ as morphisms. In case $\cat = \syntax{(\Sigma,\col, E)}$ and $\cat'= \syntax{(\Sigma',\col', E')}$ are freely generated by SMTs, then $\cat + \cat'= \syntax{(\Sigma \uplus \Sigma',\col \uplus \col', E \uplus E')}$. This is just a mild generalisation of the analogous observation for (single-coloured) props in \cite[Prop.~2.8]{Zanasi15}.

Coproducts in $\PROP{\col}$ are constructed similarly, the fundamental difference being that all props are $\col$-coloured, and when taking a coproduct $\cat + \cat'$ we thus identify the sets of objects of $\cat$ and of $\cat'$. Formally, $\cat + \cat'$ can be defined as a certain pushout object $\cat +_{\scriptscriptstyle{\perm{\col}}} \cat'$ in $\PROP{}$:
 \begin{equation*}\label{eq:pushout}
 \vcenter{
\xymatrix@R=20pt@C=40pt{
 \ar[d]_{!_1} \perm{\col}  \ar[r]^-{!_2} \drcorner & \cat \ar[d]^-{}\\
\ar[r]_-{} \cat' & {\cat +_{\scriptscriptstyle\perm{\col} } \cat'}
}
}
\end{equation*}
where the maps $!_1$ and $!_2$ are given by initiality of $\perm{\col}$ in $\PROP{\col}$ (see Example \ref{ex:perm}). Intuitively, in $\cat +_{\scriptscriptstyle{\perm{\col}}} \cat'$ one identifies the `copy' of $\perm{\col}$ in $\cat$ with the one in $\cat'$, meaning that objects, identity and symmetry morphisms are identified. 
\end{rem}

\begin{exa}[Monoids and Functions] \label{cmonf} 
The $\col$-coloured prop $\mathbf{CMon}_\col$ of \emph{commutative $\col$-monoids} is particularly relevant to our development. It is freely generated from the signature containing $\mu_c \from cc \to c$ (multiplication) depicted as $\begin{tikzpicture}
	\begin{pgfonlayer}{nodelayer}
		\node [style=black] (0) at (0.75, 0) {};
		\node [style=none] (1) at (1.75, 0) {};
		\node [style=none] (2) at (0, 0.5) {};
		\node [style=none] (3) at (0, -0.5) {};
		\node [style=none] (6) at (-0.25, 0.5) {};
		\node [style=none] (8) at (1.5, 0.25) {\scriptsize $c$};
		\node [style=none] (10) at (-0.25, 0.75) {\scriptsize $c$};
		\node [style=none] (11) at (-0.25, -0.5) {};
		\node [style=none] (12) at (-0.25, -0.25) {\scriptsize $c$};
	\end{pgfonlayer}
	\begin{pgfonlayer}{edgelayer}
		\draw (0) to (1.center);
		\draw [bend left] (2.center) to (0);
		\draw [in=0, out=-120] (0) to (3.center);
		\draw [in=180, out=0] (6.center) to (2.center);
		\draw [in=180, out=0] (11.center) to (3.center);
	\end{pgfonlayer}
\end{tikzpicture}$,  and $\eta_c \from \emptyword \to c$ (unit) depicted as $\begin{tikzpicture}
	\begin{pgfonlayer}{nodelayer}
		\node [style=black] (0) at (0.75, 0) {};
		\node [style=none] (1) at (1.75, 0) {};
		\node [style=none] (8) at (1.5, 0.25) {\scriptsize $c$};
	\end{pgfonlayer}
	\begin{pgfonlayer}{edgelayer}
		\draw (0) to (1.center);
	\end{pgfonlayer}
\end{tikzpicture}
$ for each colour $c\in\col$, and the following equations, expressing associativity, unitality and commutativity of these operations, respectively:
\begin{equation*}
\begin{tikzpicture}
	\begin{pgfonlayer}{nodelayer}
		\node [style=black] (0) at (0.75, 0) {};
		\node [style=none] (1) at (1.75, 0) {};
		\node [style=none] (2) at (-0.25, -0.75) {};
		\node [style=black] (3) at (-0.25, 0.5) {};
		\node [style=none] (4) at (-1.25, 1) {};
		\node [style=none] (5) at (-1.25, 0) {};
		\node [style=none] (6) at (-1.25, -0.75) {};
		\node [style=none] (7) at (-1.25, 1.25) {\scriptsize $c$};
		\node [style=none] (8) at (1.5, 0.25) {\scriptsize $c$};
		\node [style=none] (9) at (-1.25, 0.25) {\scriptsize $c$};
		\node [style=none] (10) at (-1.25, -0.5) {\scriptsize $c$};
	\end{pgfonlayer}
	\begin{pgfonlayer}{edgelayer}
		\draw (0) to (1.center);
		\draw [bend right] (2.center) to (0);
		\draw [bend left] (4.center) to (3);
		\draw [bend right] (5.center) to (3);
		\draw [in=0, out=120] (0) to (3);
		\draw (6.center) to (2.center);
	\end{pgfonlayer}
\end{tikzpicture} = \begin{tikzpicture}
	\begin{pgfonlayer}{nodelayer}
		\node [style=black] (0) at (0.75, 0.25) {};
		\node [style=none] (1) at (1.75, 0.25) {};
		\node [style=none] (2) at (-0.25, 1) {};
		\node [style=black] (3) at (-0.25, -0.25) {};
		\node [style=none] (4) at (-1.25, -0.75) {};
		\node [style=none] (5) at (-1.25, 0.25) {};
		\node [style=none] (6) at (-1.25, 1) {};
		\node [style=none] (7) at (-1.25, -0.5) {\scriptsize $c$};
		\node [style=none] (8) at (1.5, 0.5) {\scriptsize $c$};
		\node [style=none] (9) at (-1.25, 0.5) {\scriptsize $c$};
		\node [style=none] (10) at (-1.25, 1.25) {\scriptsize $c$};
	\end{pgfonlayer}
	\begin{pgfonlayer}{edgelayer}
		\draw (0) to (1.center);
		\draw [bend left] (2.center) to (0);
		\draw [bend right] (4.center) to (3);
		\draw [bend left] (5.center) to (3);
		\draw [in=0, out=-120] (0) to (3);
		\draw (6.center) to (2.center);
	\end{pgfonlayer}
\end{tikzpicture}\qquad \quad \begin{tikzpicture}
	\begin{pgfonlayer}{nodelayer}
		\node [style=black] (0) at (0.75, 0) {};
		\node [style=none] (1) at (1.75, 0) {};
		\node [style=none] (2) at (-0.25, -0.5) {};
		\node [style=black] (3) at (-0.25, 0.5) {};
		\node [style=none] (6) at (-1.25, -0.5) {};
		\node [style=none] (8) at (1.5, 0.25) {\scriptsize $c$};
		\node [style=none] (10) at (-1.25, -0.25) {\scriptsize $c$};
	\end{pgfonlayer}
	\begin{pgfonlayer}{edgelayer}
		\draw (0) to (1.center);
		\draw [bend right] (2.center) to (0);
		\draw [in=0, out=120] (0) to (3);
		\draw (6.center) to (2.center);
	\end{pgfonlayer}
\end{tikzpicture} = \idx{c} =  \begin{tikzpicture}
	\begin{pgfonlayer}{nodelayer}
		\node [style=black] (0) at (0.75, 0) {};
		\node [style=none] (1) at (1.75, 0) {};
		\node [style=none] (2) at (-0.25, 0.5) {};
		\node [style=black] (3) at (-0.25, -0.5) {};
		\node [style=none] (6) at (-1.25, 0.5) {};
		\node [style=none] (8) at (1.5, 0.25) {\scriptsize $c$};
		\node [style=none] (10) at (-1.25, 0.75) {\scriptsize $c$};
	\end{pgfonlayer}
	\begin{pgfonlayer}{edgelayer}
		\draw (0) to (1.center);
		\draw [bend left] (2.center) to (0);
		\draw [in=0, out=-120] (0) to (3);
		\draw (6.center) to (2.center);
	\end{pgfonlayer}
\end{tikzpicture} \qquad \quad \begin{tikzpicture}
	\begin{pgfonlayer}{nodelayer}
		\node [style=black] (0) at (0.75, 0) {};
		\node [style=none] (1) at (1.75, 0) {};
		\node [style=none] (2) at (0, 0.5) {};
		\node [style=none] (3) at (0, -0.5) {};
		\node [style=none] (6) at (-1.25, -0.5) {};
		\node [style=none] (8) at (1.5, 0.25) {\scriptsize $c$};
		\node [style=none] (10) at (-1.25, -0.25) {\scriptsize $c$};
		\node [style=none] (11) at (-1.25, 0.5) {};
		\node [style=none] (12) at (-1.25, 0.75) {\scriptsize $c$};
	\end{pgfonlayer}
	\begin{pgfonlayer}{edgelayer}
		\draw (0) to (1.center);
		\draw [bend left] (2.center) to (0);
		\draw [in=0, out=-120] (0) to (3.center);
		\draw [in=180, out=0] (6.center) to (2.center);
		\draw [in=180, out=0] (11.center) to (3.center);
	\end{pgfonlayer}
\end{tikzpicture} = \begin{tikzpicture}
	\begin{pgfonlayer}{nodelayer}
		\node [style=black] (0) at (0.75, 0) {};
		\node [style=none] (1) at (1.75, 0) {};
		\node [style=none] (2) at (0, 0.5) {};
		\node [style=none] (3) at (0, -0.5) {};
		\node [style=none] (6) at (-0.25, 0.5) {};
		\node [style=none] (8) at (1.5, 0.25) {\scriptsize $c$};
		\node [style=none] (10) at (-0.25, 0.75) {\scriptsize $c$};
		\node [style=none] (11) at (-0.25, -0.5) {};
		\node [style=none] (12) at (-0.25, -0.25) {\scriptsize $c$};
	\end{pgfonlayer}
	\begin{pgfonlayer}{edgelayer}
		\draw (0) to (1.center);
		\draw [bend left] (2.center) to (0);
		\draw [in=0, out=-120] (0) to (3.center);
		\draw [in=180, out=0] (6.center) to (2.center);
		\draw [in=180, out=0] (11.center) to (3.center);
	\end{pgfonlayer}
\end{tikzpicture}
\end{equation*}
Then, we obtain a monoid structure over composite objects $uw$, given inductively by
\[\begin{tikzpicture}
	\begin{pgfonlayer}{nodelayer}
		\node [style=black] (0) at (0.75, 0.75) {};
		\node [style=none] (1) at (1.75, 0.75) {};
		\node [style=none] (2) at (-0.5, 1.25) {};
		\node [style=none] (3) at (-0.25, 0.25) {};
		\node [style=none] (6) at (-1.75, 1.25) {};
		\node [style=none] (8) at (1.5, 1) {\scriptsize $u$};
		\node [style=none] (10) at (-1.5, 1.5) {\scriptsize $u$};
		\node [style=none] (11) at (-1.75, -0.5) {};
		\node [style=black] (13) at (0.75, -0.75) {};
		\node [style=none] (14) at (1.75, -0.75) {};
		\node [style=none] (15) at (-0.25, -0.25) {};
		\node [style=none] (16) at (-0.5, -1.25) {};
		\node [style=none] (17) at (-1.75, 0.5) {};
		\node [style=none] (18) at (1.5, -0.5) {\scriptsize $w$};
		\node [style=none] (20) at (-1.75, -1.25) {};
		\node [style=none] (21) at (-1.5, -0.25) {\scriptsize $u$};
		\node [style=none] (22) at (-1.5, 0.75) {\scriptsize $w$};
		\node [style=none] (23) at (-1.5, -1) {\scriptsize $w$};
	\end{pgfonlayer}
	\begin{pgfonlayer}{edgelayer}
		\draw (0) to (1.center);
		\draw [bend left] (2.center) to (0);
		\draw [in=0, out=-120] (0) to (3.center);
		\draw [in=180, out=0] (6.center) to (2.center);
		\draw [in=180, out=0] (11.center) to (3.center);
		\draw (13) to (14.center);
		\draw [bend left] (15.center) to (13);
		\draw [in=0, out=-120] (13) to (16.center);
		\draw [in=180, out=0] (17.center) to (15.center);
		\draw [in=180, out=0] (20.center) to (16.center);
	\end{pgfonlayer}
\end{tikzpicture}\qquad\quad\qquad \begin{tikzpicture}
	\begin{pgfonlayer}{nodelayer}
		\node [style=black] (0) at (0.75, 0.5) {};
		\node [style=none] (1) at (1.75, 0.5) {};
		\node [style=none] (8) at (1.5, 0.75) {\scriptsize $u$};
		\node [style=black] (13) at (0.75, -0.5) {};
		\node [style=none] (14) at (1.75, -0.5) {};
		\node [style=none] (18) at (1.5, -0.25) {\scriptsize $w$};
	\end{pgfonlayer}
	\begin{pgfonlayer}{edgelayer}
		\draw (0) to (1.center);
		\draw (13) to (14.center);
	\end{pgfonlayer}
\end{tikzpicture}\]
Notice that this is a coproduct of coloured props (in $\PROP{}$, \emph{cf.} Remark~\ref{rmk:coproducts}) with $\mathbf{CMon}_{\mathcal{C}} = \Sigma_{c\in \mathcal{C}} \mathbf{CMon}_{c}$.

An important particular case is the (single-colour) prop $\mathbf{CMon} : = \mathbf{CMon}_{\{\bullet\}}$ of commutative monoids. The importance of this prop stems from the fact that it presents the prop $\mathbb{F}$ of functions---see \emph{e.g.}~\cite{composing}. Recall that $\mathbb{F}$ has morphisms $f: m \rightarrow n$ the functions from the set $\{0,\dots,m-1\}$ to $\{0,\dots,n-1\}$, with the monoidal product on functions given by their disjoint union. 
\end{exa}

    \emph{Cospans} are another central concept of this paper. When interpreting string diagrams as hypergraphs, it is fundamental to record the information of what wires are available for composition on the left and right hand side of the diagram: this is achieved by considering cospans of hypergraphs, with the cospan structure indicating which nodes constitute the left and the right interface of the hypergraph.
    \begin{defi}[Cospan]\label{def:cospan}
    A \emph{cospan} in some category $\mathbb{C}$ is a pair $f:X\to A, g:Y\to A$ of morphisms of $\mathbb{C}$ with the same codomain, which we write as $\cspobj{X}{f}{A}{g}{Y}$. 
    An isomorphism between cospans $\cspobj{X}{f}{A}{g}{Y}$ and $\cspobj{X}{f'}{A'}{g'}{Y}$ is an isomorphism $h:A\to A'$  of $\mathbb{C}$ such that $h\circ f = f'$ and $h\circ g=g'$.
    \end{defi}
Cospans over categories with enough structure form a symmetric monoidal category.
     \begin{defi}[SMC of cospans] \label{defn:csp}
    Given a category $\mathbb{C}$ with finite colimits, let $\mathrm{Csp}(\mathbb{C})$ be the category with the same objects as $\mathbb{C}$ and morphisms $X \rightarrow Y$ isomorphism classes of \emph{cospans} $\cspobj{X}{}{A}{}{Y}$, for any object $A$ (called the \emph{carrier} of the cospan). Composition of cospans $\cspobj{X}{f}{A}{h}{Z}$ and $\cspobj{Z}{h}{B}{i}{Y}$ is defined by pushout of the span formed by the middle legs, \emph{i.e.}, it is the (isomorphism class of the) cospan $\cspobj{X}{f;p}{Q}{i;q}{Y}$ where $\cspobj{A}{p}{Q}{q}{B}$ is the pushout of $A \xleftarrow{h} Z {\xrightarrow{i}} B$.
$\mathrm{Csp}(\mathbb{C})$ is symmetric monoidal with the monoidal unit being the initial object $0 \in \mathbb{C}$ and the monoidal product given by the coproduct in $\mathbb{C}$ of the two maps of each cospan.
\end{defi}

\emph{Hypergraphs} \cite{hypergraphs} generalise graphs by replacing edges with ordered and directed hyperedges, which may have lists of source and target nodes instead of just individual ones. Hypergraphs and hypergraph homomorphisms form a category $\mathbf{Hyp}$.  As observed in~\cite{pt1}, this category may also be defined as a presheaf topos---this is particularly convenient for calculating (co)limits and to ensure that it is \emph{adhesive}~\cite{adh}, a fundamental property for DPO rewriting. For this reason, we define $\mathbf{Hyp}$ as the functor category $\mathbb{F}^\mathbf{I}$, where $\mathbf{I}$ has objects the pairs of natural numbers $(k, l) \in \mathbb{N} \times \mathbb{N}$ and an extra object $\star$, with $k+l$ morphisms from $(k,l)$ to $\star$, for all $k,l \in \mathbb{N}$. A hypergraph $G$ is therefore given by a set $G_\star$ of nodes, and sets $G_{k,l}$ of hyperedges for each $(k, l) \in \mathbb{N} \times \mathbb{N}$, with source maps $s_i : G_{k,l}\to G_\star$ for $1\leq i\leq k$ and target maps $t_j:G_{k,l}\to G_\star$, $1\leq j\leq l$. A monoidal signature $(\Sigma,\col)$ yields a directed hypergraph $G_{(\Sigma,\col)}$ with a node for each $c \in \col$ and a hyperedge for every $\Sigma$-operation $o \colon w \rightarrow v$, whose source and target lists of nodes are given respectively by the arity $w \in \col^{\star}$  and the coarity $v \in \col^{\star}$ of $o$. We can use this observation to define the category of $(\Sigma,\col)$-labelled hypergraphs as follows.
\begin{defi}
The slice category $\mathbf{Hyp} \downarrow G_{(\Sigma,\col)}$ is called the category of $(\Sigma,\col)$-labelled hypergraphs and denoted by $\hyp$.
\end{defi}

\begin{rem}\label{rmk:monogamous} 
As proven in~\cite{pt1}, morphisms in a prop freely generated by a signature $(\Sigma,\col)$ may be faithfully interpreted as certain cospans of $(\Sigma,\col)$-labelled hypergraphs, where the domain of the cospan legs are discrete hypergraphs (\emph{i.e.}, sets), used to represent the left and right interfaces of the  string diagram---see Example~\ref{ex:monogamous} below. 

That sets can be seen as discrete hypergraphs extends to a faithful, coproduct-preserving functor $D : \mathbb{F} \rightarrow \hyp$ mapping every object $i \in \mathrm{Ob}(\mathbb{F})=\mathbb{N}$ to the hypergraph with set of nodes $i=\{0,\dots,i-1\}$ and mapping each function to the induced hypergraph homomorphism. The \emph{prop} $\csphyp$ used to interpret string diagrams in~\cite{pt1} is defined as the full subcategory of $\mathrm{Csp}(\hyp)$ (\emph{cf.} Definition~\ref{defn:csp}) whose objects are discrete hypergraphs. Note that we will reserve the term \emph{discrete cospan} (of hypergraphs) for a cospan whose apex is also discrete.  
Finally, notice that, when $\Sigma=\emptyset$, the cospans of hypergraphs in $\mathrm{Csp}(\hyp)$ do not have any hyperedges and the resulting prop is equivalent to that of cospans of sets---we will use this fact in Section~\ref{sec:discrete-rm-cospans} below.
\end{rem}

\section{The Combinatorial Interpretation}\label{sec:characterisation}

When referring to ``string diagrams with a chosen commutative monoid structure'', we mean morphisms of the $\mathcal{C}$-coloured prop $\cprd$, the coproduct of the free $\mathcal{C}$-coloured props over signature $\Sigma$ and $\mathbf{CMon}_{\mathcal{C}}$. Intuitively, such morphisms are obtained by freely combining $\Sigma$-terms with terms of the theory of commutative monoids, then quotienting by the laws of symmetric monoidal categories and those of $\mathbf{CMon}$. The aim of this section is give a combinatorial characterisation of string diagrams in $\cprd$.
Specifically, we prove that the freely generated coloured props with a chosen commutative monoid structure are isomorphic to a category of cospans of hypergraphs with certain restrictions (Theorem~\ref{thm:iso} below). 

\subsection{Right-monogamous cospans}
As shown in \cite{pt2}, the standard interpretation of string diagrams in a prop as cospans of hypergraphs is not full. In order to characterise the image of the interpretation, it is necessary to restrict ourselves to a class of so-called \emph{acyclic} and \emph{monogamous} cospans. 
\begin{exa}\label{ex:monogamous}
The cospan below on the right interprets the diagram on the left. Intuitively, nodes represent the wires and hyperedges the operations of the corresponding diagram. We use blue frames to indicate the left and the right interface of a cospan, 
indexes to indicate how the cospan legs are defined, rounded rectangles to represent hyperedges, and black dots to represent vertices (with the colour inscribed inside when there is more than one). As explained in Remark~\ref{rmk:monogamous} above, its left and right interface are discrete hypergraphs, \emph{i.e.} sets. 
\[\begin{tikzpicture}
	\begin{pgfonlayer}{nodelayer}
		\node [style=none] (296) at (-0.5, 1) {};
		\node [style=none] (297) at (-0.5, -0.5) {};
		\node [style=none] (298) at (0.5, -0.5) {};
		\node [style=none] (299) at (0.5, 1) {};
		\node [style=none] (300) at (0.5, -0.25) {};
		\node [style=none] (301) at (-0.5, -0.25) {};
		\node [style=none] (302) at (-0.5, 0.75) {};
		\node [style=none] (303) at (-2, 0.75) {};
		\node [style=none] (306) at (2.5, 0.75) {};
		\node [style=none] (307) at (0, 0.25) {$f$};
		\node [style=none] (309) at (1, -1.25) {};
		\node [style=none] (312) at (-2, -1.25) {};
		\node [style=none] (314) at (0.5, 0.75) {};
		\node [style=none] (316) at (-1.75, 1) {\scriptsize $c_0$};
		\node [style=none] (317) at (1.5, 1) {\scriptsize $c_2$};
		\node [style=none] (319) at (-1.75, -1) {\scriptsize $c_1$};
		\node [style=none] (320) at (2.5, 1.25) {};
		\node [style=none] (321) at (2.5, 0.25) {};
		\node [style=none] (322) at (3.5, 0.25) {};
		\node [style=none] (323) at (3.5, 1.25) {};
		\node [style=none] (328) at (5, 0.75) {};
		\node [style=none] (329) at (3, 0.75) {$e$};
		\node [style=none] (330) at (3.5, 0.75) {};
		\node [style=none] (332) at (4.75, 1) {\scriptsize $c_2$};
		\node [style=none] (333) at (5, -0.25) {};
		\node [style=none] (335) at (1, -0.75) {};
		\node [style=none] (336) at (1, -1.75) {};
		\node [style=none] (337) at (2, -1.75) {};
		\node [style=none] (338) at (2, -0.75) {};
		\node [style=none] (339) at (1.5, -1.25) {$d$};
		\node [style=none] (340) at (4.75, 0) {\scriptsize $c_1$};
	\end{pgfonlayer}
	\begin{pgfonlayer}{edgelayer}
		\draw (297.center) to (296.center);
		\draw (296.center) to (299.center);
		\draw (299.center) to (298.center);
		\draw (298.center) to (297.center);
		\draw [in=180, out=0] (303.center) to (302.center);
		\draw (312.center) to (309.center);
		\draw (314.center) to (306.center);
		\draw (321.center) to (320.center);
		\draw (320.center) to (323.center);
		\draw (323.center) to (322.center);
		\draw (322.center) to (321.center);
		\draw (330.center) to (328.center);
		\draw (300.center) to (333.center);
		\draw (336.center) to (335.center);
		\draw (335.center) to (338.center);
		\draw (338.center) to (337.center);
		\draw (337.center) to (336.center);
	\end{pgfonlayer}
\end{tikzpicture} \qquad \mapsto \qquad \begin{tikzpicture}
	\begin{pgfonlayer}{nodelayer}
		\node [style=none] (0) at (-8.25, 2.75) {};
		\node [style=none] (1) at (-8.25, -2.5) {};
		\node [style=none] (2) at (-6.5, -2.5) {};
		\node [style=none] (3) at (-6.5, 2.75) {};
		\node [style=none] (4) at (7.25, 2.5) {};
		\node [style=none] (5) at (7.25, -2.5) {};
		\node [style=none] (6) at (8.75, -2.5) {};
		\node [style=none] (7) at (8.75, 2.5) {};
		\node [style=small black dot] (9) at (-7.5, -0.25) {$c_2$};
		\node [style=small black dot] (10) at (-7.5, -1.75) {$c_2$};
		\node [style=none] (14) at (-2.25, 1.5) {};
		\node [style=none] (15) at (-2.25, 0.5) {};
		\node [style=none] (16) at (-1.25, 0.5) {};
		\node [style=none] (17) at (-1.25, 1.5) {};
		\node [style=none] (18) at (-1.25, 0.5) {};
		\node [style=none] (20) at (-2.25, -0.75) {};
		\node [style=none] (21) at (-2.25, -1.75) {};
		\node [style=none] (22) at (-1.25, -1.75) {};
		\node [style=none] (23) at (-1.25, -0.75) {};
		\node [style=none] (24) at (-1.25, -0.75) {};
		\node [style=small black dot] (25) at (4.5, 1.25) {$c_2$};
		\node [style=none] (26) at (2, 1.75) {};
		\node [style=none] (27) at (2, 0.75) {};
		\node [style=none] (28) at (3, 0.75) {};
		\node [style=none] (29) at (3, 1.75) {};
		\node [style=none] (30) at (3, 1.25) {};
		\node [style=small black dot] (31) at (8, 1) {$c_2$};
		\node [style=small black dot] (32) at (0.25, 1.25) {$c_2$};
		\node [style=none] (33) at (-1.25, 1.25) {};
		\node [style=none] (34) at (2, 1.25) {};
		\node [style=small black dot] (35) at (-3.5, -1.25) {};
		\node [style=small black dot] (36) at (-3.5, -1.25) {$c_1$};
		\node [style=none] (37) at (-2.25, -1.25) {};
		\node [style=none] (38) at (-2.25, 1) {};
		\node [style=small black dot] (39) at (-3.5, 1) {$c_0$};
		\node [style=none] (42) at (-7.5, 2.25) {$n_0$};
		\node [style=none] (43) at (-7.5, 0.5) {$n_1$};
		\node [style=none] (44) at (-7.5, -1) {$n_2$};
		\node [style=none] (47) at (8, 1.75) {$m_0$};
		\node [style=none] (48) at (-5.75, 0) {};
		\node [style=none] (49) at (-4.75, 0) {};
		\node [style=none] (50) at (-4.75, 0) {};
		\node [style=none] (51) at (5.5, 0) {};
		\node [style=none] (52) at (6.5, 0) {};
		\node [style=none] (53) at (-5.75, 0) {};
		\node [style=none] (54) at (-4.75, 0) {};
		\node [style=none] (55) at (4.5, 2) {$n_2,m_0$};
		\node [style=none] (56) at (0.25, 2) {$n_1$};
		\node [style=none] (58) at (-3.5, -0.5) {$n_4$};
		\node [style=none] (63) at (-3.5, 1.75) {$n_0$};
		\node [style=none] (64) at (-1.75, 1) {$f$};
		\node [style=none] (65) at (2.5, 1.25) {$e$};
		\node [style=none] (66) at (-1.75, -1.25) {$d$};
		\node [style=none] (67) at (-1.25, -1.75) {};
		\node [style=small black dot] (68) at (-7.5, 1.5) {$c_0$};
		\node [style=small black dot] (71) at (8, -1.5) {$c_1$};
		\node [style=none] (72) at (8, -0.75) {$m_1$};
		\node [style=none] (73) at (0.5, 0) {$m_1$};
		\node [style=small black dot] (74) at (0.25, -0.75) {};
		\node [style=small black dot] (75) at (0.25, -0.75) {$c_1$};
		\node [style=none] (76) at (-1.25, 0.75) {};
	\end{pgfonlayer}
	\begin{pgfonlayer}{edgelayer}
		\draw [style=hadamard edge] (0.center) to (3.center);
		\draw [style=hadamard edge] (2.center) to (3.center);
		\draw [style=hadamard edge] (0.center) to (1.center);
		\draw [style=hadamard edge] (1.center) to (2.center);
		\draw [style=hadamard edge] (4.center) to (7.center);
		\draw [style=hadamard edge] (6.center) to (7.center);
		\draw [style=hadamard edge] (4.center) to (5.center);
		\draw [style=hadamard edge] (5.center) to (6.center);
		\draw [style=gray edge] (15.center) to (14.center);
		\draw [style=gray edge, bend left=90, looseness=0.75] (14.center) to (17.center);
		\draw [style=gray edge] (17.center) to (16.center);
		\draw [style=gray edge, in=270, out=270, looseness=0.75] (16.center) to (15.center);
		\draw [style=gray edge] (21.center) to (20.center);
		\draw [style=gray edge, bend left=90, looseness=0.75] (20.center) to (23.center);
		\draw [style=gray edge] (23.center) to (22.center);
		\draw [style=gray edge, in=270, out=270, looseness=0.75] (22.center) to (21.center);
		\draw [style=gray edge] (27.center) to (26.center);
		\draw [style=gray edge, bend left=90, looseness=0.75] (26.center) to (29.center);
		\draw [style=gray edge] (29.center) to (28.center);
		\draw [style=gray edge, in=270, out=270, looseness=0.75] (28.center) to (27.center);
		\draw [style=gray edge, in=180, out=0, looseness=0.75] (30.center) to (25);
		\draw [style=gray edge] (33.center) to (32);
		\draw [style=gray edge] (32) to (34.center);
		\draw [style=gray edge] (36) to (37.center);
		\draw [style=gray edge] (39) to (38.center);
		\draw [style=pointer edge] (53.center) to (54.center);
		\draw [style=pointer edge] (52.center) to (51.center);
		\draw [style=gray edge] (75) to (76.center);
	\end{pgfonlayer}
\end{tikzpicture}\]
Notice that every node is the source and target of a single hyperedge. This is the requirement we will need to relax below, in order to accommodate commutative monoids.
\end{exa}
In order to prove our result for coloured props with a chosen commutative monoid structure, we relax this notion to \emph{right-monogamous} cospans, which we now introduce.

\begin{defi}[Degree of a node \cite{pt2}]
The \emph{in-degree} of a node $v$ in hypergraph $H$ is the number of pairs
$(h, i)$ where $h$ is a hyperedge with $v$ as its $i^{th}$ target. Similarly, the \emph{out-degree} of $v$ is the number
of pairs $(h, j)$ where h is a hyperedge with $v$ as its $j^{th}$ source.
\end{defi}

\begin{defi}[Terminal node]\label{def:terminal}
We say that a node $v$ of a hypergraph $H$ is \emph{terminal} if its out-degree is $0$, i.e., if there are no hyperedges of $H$ with source~v.
\end{defi}
Given $\cspobj{v}{f}{H}{g}{w}$ in $\csp(\hyp)$,  we call inputs of $H$ the set $\mathrm{in(H)}$, defined as the image of $f$ and outputs, the set $\mathrm{out(H)}$ defined as the image of $g$.
\begin{defi}[Right-monogamy]
We say that a cospan $\cspobj{v}{f}{H}{g}{w}$ is \emph{right-monogamous} if $g$ is mono and $\mathrm{out}(H)$ is the set of terminal nodes of $H$.



 \end{defi}

Compared to monogamy~\cite{pt2}, right-monogamy does not impose any requirement on $f$, and only constraints the out-degree of nodes (not the in-degree).


Acyclicity is a standard condition which forbids (directed) loops in a hypergraph and was already present in~\cite[Definition 20]{pt2} to characterise string diagrams for plain symmetric monoidal categories. We also need it here.
\begin{defi}[Acyclicity]
Given a hypergraph G and two nodes or hyperedges $a$ and $b$,  a \emph{path} from $a$ to $b$ in G is an alternating list $p= [p_1, . . . , p_n], p_1 = a, p_n = b$ of hyperedges and nodes such that for all hyperedges $p_i$, the nodes $p_{i-1}$ and $p_{i+1}$ are a source and target for $p_i$ when
they are defined (i.e. when $i>1$ and $i< n$, respectively). A hypergraph is said to be \emph{acyclic} if it has no path containing the same node twice. Similarly, we say that a cospan $\cspobj{v}{}{G}{}{w}$ is acyclic if $G$ is acyclic.
\end{defi}
Equivalently to the previous definition, we can define acyclicity in terms of hyperedges, since the existence of a path containing the same node twice is equivalent to the the existence of a path containing the same hyperedge twice. 

\begin{exa}
\label{ex:right-mono-cospan}
The cospan depicted below is right-monogamous and acyclic.
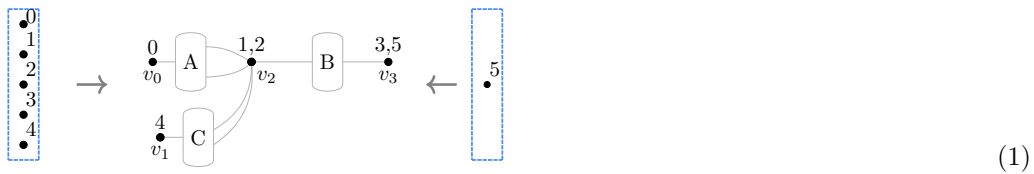
\begin{equation}\label{eq:right-mono-cospan}
\begin{tikzpicture}
	\begin{pgfonlayer}{nodelayer}
		\node [style=none] (0) at (-8.5, 2.75) {};
		\node [style=none] (1) at (-8.5, -2.5) {};
		\node [style=none] (2) at (-6.5, -2.5) {};
		\node [style=none] (3) at (-6.5, 2.75) {};
		\node [style=none] (4) at (7.5, 2.5) {};
		\node [style=none] (5) at (7.5, -2.5) {};
		\node [style=none] (6) at (9, -2.5) {};
		\node [style=none] (7) at (9, 2.5) {};
		\node [style=small black dot] (9) at (-7.75, 1) {$c_2$};
		\node [style=small black dot] (10) at (-7.75, 0) {$c_2$};
		\node [style=small black dot] (11) at (-7.75, -1) {$c_3$};
		\node [style=small black dot] (12) at (-7.75, -2) {$c_1$};
		\node [style=small black dot] (13) at (0.25, 1) {};
		\node [style=none] (14) at (-2.25, 1.5) {};
		\node [style=none] (15) at (-2.25, 0.5) {};
		\node [style=none] (16) at (-1.25, 0.5) {};
		\node [style=none] (17) at (-1.25, 1.5) {};
		\node [style=none] (18) at (-1.25, 0.5) {};
		\node [style=small black dot] (19) at (0.25, 1) {};
		\node [style=none] (20) at (-2.25, -0.75) {};
		\node [style=none] (21) at (-2.25, -1.75) {};
		\node [style=none] (22) at (-1.25, -1.75) {};
		\node [style=none] (23) at (-1.25, -0.75) {};
		\node [style=none] (24) at (-1.25, -0.75) {};
		\node [style=small black dot] (25) at (4.75, 1) {$c_3$};
		\node [style=none] (26) at (2.25, 1.5) {};
		\node [style=none] (27) at (2.25, 0.5) {};
		\node [style=none] (28) at (3.25, 0.5) {};
		\node [style=none] (29) at (3.25, 1.5) {};
		\node [style=none] (30) at (3.25, 1) {};
		\node [style=small black dot] (31) at (8.25, 0) {$c_3$};
		\node [style=small black dot] (32) at (0.25, 1) {$c_2$};
		\node [style=none] (33) at (-1.25, 1.5) {};
		\node [style=none] (34) at (2.25, 1) {};
		\node [style=small black dot] (35) at (-3.5, -1.25) {};
		\node [style=small black dot] (36) at (-3.5, -1.25) {$c_1$};
		\node [style=none] (37) at (-2.25, -1.25) {};
		\node [style=none] (38) at (-2.25, 1) {};
		\node [style=small black dot] (39) at (-3.5, 1) {$c_0$};
		\node [style=none] (41) at (-7, 1.25) {};
		\node [style=none] (42) at (-7, 2.25) {$n_0$};
		\node [style=none] (43) at (-7, 1.25) {$n_1$};
		\node [style=none] (44) at (-7, 0.25) {$n_2$};
		\node [style=none] (45) at (-7, -0.75) {$n_3$};
		\node [style=none] (46) at (-7, -1.75) {$n_4$};
		\node [style=none] (47) at (8.25, 0.75) {$m_0$};
		\node [style=none] (48) at (-5.75, 0) {};
		\node [style=none] (49) at (-4.75, 0) {};
		\node [style=none] (50) at (-4.75, 0) {};
		\node [style=none] (51) at (5.75, 0) {};
		\node [style=none] (52) at (6.75, 0) {};
		\node [style=none] (53) at (-5.75, 0) {};
		\node [style=none] (54) at (-4.75, 0) {};
		\node [style=none] (55) at (4.75, 1.75) {$n_3,m_0$};
		\node [style=none] (56) at (0.25, 1.75) {$n_1,n_2$};
		\node [style=none] (58) at (-3.5, -0.5) {$n_4$};
		\node [style=none] (63) at (-3.5, 1.75) {$n_0$};
		\node [style=none] (64) at (-1.75, 1) {$d$};
		\node [style=none] (65) at (2.75, 1) {$f$};
		\node [style=none] (66) at (-1.75, -1.25) {$e$};
		\node [style=none] (67) at (-1.25, -1.75) {};
		\node [style=small black dot] (68) at (-7.75, 2) {$c_0$};
	\end{pgfonlayer}
	\begin{pgfonlayer}{edgelayer}
		\draw [style=hadamard edge] (0.center) to (3.center);
		\draw [style=hadamard edge] (2.center) to (3.center);
		\draw [style=hadamard edge] (0.center) to (1.center);
		\draw [style=hadamard edge] (1.center) to (2.center);
		\draw [style=hadamard edge] (4.center) to (7.center);
		\draw [style=hadamard edge] (6.center) to (7.center);
		\draw [style=hadamard edge] (4.center) to (5.center);
		\draw [style=hadamard edge] (5.center) to (6.center);
		\draw [style=gray edge] (15.center) to (14.center);
		\draw [style=gray edge, bend left=90, looseness=0.75] (14.center) to (17.center);
		\draw [style=gray edge] (17.center) to (16.center);
		\draw [style=gray edge, in=270, out=270, looseness=0.75] (16.center) to (15.center);
		\draw [style=gray edge, in=-135, out=0, looseness=0.75] (18.center) to (13);
		\draw [style=gray edge] (21.center) to (20.center);
		\draw [style=gray edge, bend left=90, looseness=0.75] (20.center) to (23.center);
		\draw [style=gray edge] (23.center) to (22.center);
		\draw [style=gray edge, in=270, out=270, looseness=0.75] (22.center) to (21.center);
		\draw [style=gray edge] (24.center) to (19);
		\draw [style=gray edge] (27.center) to (26.center);
		\draw [style=gray edge, bend left=90, looseness=0.75] (26.center) to (29.center);
		\draw [style=gray edge] (29.center) to (28.center);
		\draw [style=gray edge, in=270, out=270, looseness=0.75] (28.center) to (27.center);
		\draw [style=gray edge, in=180, out=0, looseness=0.75] (30.center) to (25);
		\draw [style=gray edge, in=150, out=0] (33.center) to (32);
		\draw [style=gray edge] (32) to (34.center);
		\draw [style=gray edge] (36) to (37.center);
		\draw [style=gray edge] (39) to (38.center);
		\draw [style=pointer edge] (53.center) to (54.center);
		\draw [style=pointer edge] (52.center) to (51.center);
		\draw [style=gray edge, bend right] (67.center) to (32);
	\end{pgfonlayer}
\end{tikzpicture}
\end{equation}
\end{exa}
\begin{prop}
\label{prodac}
Let $\cspobj{u}{}{G}{}{v}$, $\cspobj{v}{}{H}{}{w}$, $\cspobj{v_1}{}{G_1}{}{w_1}$ and $\cspobj{v_2}{}{G_2}{}{w_2}$ be right-monogamous acyclic cospans in $\csp(\hyp)$. Then
\begin{itemize}
\item Identities and symmetries in $\csp(\hyp)$ are right-monogamous and acyclic;
\item $(\cspobj{u}{}{G}{}{v}) ; (\cspobj{v}{}{H}{}{w})$ is right-monogamous acyclic;
\item  $(\cspobj{v_1}{}{G_1}{}{w_1}) \oplus (\cspobj{v_2}{}{G_2}{}{w_2})$ is right-monogamous acyclic.
\end{itemize}
\end{prop}
\begin{proof}
Entirely analogous to the monogamous case, proven in~\cite[Lemmas 15-17]{pt2}.
\end{proof}
Thus, acyclic right-monogamous cospans form a coloured sub-prop of $\csphyp$, which we write $\rmacsphyp$.

\subsection{Commutative monoids and discrete right-monogamous cospans}\label{sec:discrete-rm-cospans}
The notion of right-monogamy is justified by its connection to commutative monoids, crystallised in the following result. 
\begin{prop}\label{cmoniso}
$\mathbf{CMon}_{\mathcal{C}} \cong \rmacspemptyhyp$.
\end{prop}
For the single-colour case, the fundamental observation is that the prop of right-monogamous acyclic cospans of $\emptyset$-labelled hypergraphs is isomorphic to $\mathbb{F}$. Indeed, as explained in Remark~\ref{rmk:monogamous}, $\mathrm{Csp}(\hyp)$ is equivalent to that of cospans of sets, so that \emph{right-monogamous} cospans of this category coincide with cospans of the form $\cspobj{m}{f}{n}{id}{n}$ and can thus be thought of simply as maps of finite sets, \emph{i.e.}, morphisms in $\mathbb{F}$. We prove this in Lemma~\ref{lem:right-monogamous-functions} below. Paired with the fact that $\mathbf{CMon} \cong \mathbb{F}$ (\emph{cf.} Example~\ref{cmonf}), this will show that $\mathbf{CMon} \cong \mathrm{RMACsp}_{\disc}\left(\mathbf{Hyp}_{\emptyset,\{\bullet\}}\right)$. The general (multicolour) result then follows from decomposing $\rmacspemptyhyp$ into a coproduct of single-colour props, in Lemma~\ref{lem:hypergraph-coproduct}.

Note that, since cospans of sets can also be seen as discrete cospans of hypergraphs (\emph{cf.} Remark~\ref{rmk:monogamous}), we can define right-monogamous cospans of sets in the same way.
\begin{lem}\label{lem:right-monogamous-functions}
\label{isocspf}
Given a right-monogamous cospan $\cspobj{m}{f}{n}{g}{n}$ in $\cspf$, there exists a unique cospan $\cspobj{m}{f^*}{n}{id}{n}$ isomorphic to it.
\end{lem}
\begin{proof}
Since $g$ is an isomorphism (a permutation on $n$), we can denote its inverse by $g^{-1}$. We define $f^*$ as $f; g^{-1} = g^{-1} \circ f$, which makes the following diagram commute:
\begin{equation}\label{eq:cospansiso}
\begin{tikzcd}
                                     & n \arrow[dd, "g^{-1}"] &                                     \\
m \arrow[ru, "f"] \arrow[rd, "f^*"'] &                        & n \arrow[lu, "g"'] \arrow[ld, "id"] \\
                                     & n                      &                                    
\end{tikzcd}
\end{equation}
Since $g^{-1}$ is also an isomorphism, the two cospans in~\eqref{eq:cospansiso} are indeed isomorphic. For uniqueness, suppose there is another cospan $\cspobj{m}{h}{n}{id}{n}$ isomorphic to $\cspobj{m}{f}{n}{g}{n}$. But, then, there must exist $\psi$ such that the following diagram commutes:
\begin{equation*}
\begin{tikzcd}
                                     & n \arrow[dd, "\psi"] &                                      \\
m \arrow[ru, "f^*"] \arrow[rd, "h"'] &                      & n \arrow[lu, "id"'] \arrow[ld, "id"] \\
                                     & n                    &                                     
\end{tikzcd}
\end{equation*}
Because $id; \psi = id$ we have $\psi = id$ and thus $h=f^*; \psi =f^* ; id = f^*$.
\end{proof}
The following states that $\rmacspemptyhyp$ is the coproduct of $|\mathcal{C}|$ copies of (isomorphic) single-colour props (\emph{cf.} Remark~\ref{rmk:coproducts}).
\begin{lem}\label{lem:hypergraph-coproduct}
$\rmacspemptyhyp$ is isomorphic to the coproduct $\displaystyle\sum_{c\in\mathcal{C}}\mathrm{RMACsp}_{\disc}\!\left(\!\mathbf{Hyp}_{\emptyset,\{c\}}\!\right)$ in the category $\PROP{\col}$.
\end{lem}
\begin{proof}
We want to show that $\rmacspemptyhyp$  satisfies the universal property of the coproduct $\sum_{c\in\mathcal{C}}\mathrm{RMACsp}_{\disc}\left(\mathbf{Hyp}_{\emptyset,\{c\}}\right)$. First, notice that, for each $c\in\mathcal{C}$, there is an obvious faithful prop morphism $[\cdot]_c\from \mathrm{RMACsp}_{\disc}\left(\mathbf{Hyp}_{\emptyset,\{c\}}\right)\to \rmacspemptyhyp$ mapping each cospan to itself.  To prove the universal property of the coproduct, given a $\mathcal{C}$-coloured prop $\mathbb{A}$ and prop morphisms $\alpha_c$ for all $c\in\mathcal{C}$ as in the diagram below
\begin{equation}\label{eq:cospan-discrete}
\begin{tikzcd}
{\mathrm{RMACsp}_{\disc}\left(\mathbf{Hyp}_{\emptyset,\{c\}}\right)} \arrow[rr, "{[\cdot]_c}"] \arrow[rrd, "\alpha_c"] &  & \rmacspemptyhyp \arrow[d, "\exists !\beta", dashed] \\
                                                                                                                             &  & \mathbb{A}                                          
\end{tikzcd}
\end{equation}
we want to show that there exists a unique $\beta$ such that the diagram~\eqref{eq:cospan-discrete} commutes for any $c\in\mathcal{C}$. First, notice that every cospan $\cspobj{v}{f}{H}{g}{w}$ of $\rmacspemptyhyp$ is isomorphic to one that is the disjoint sum of cospans of each colour: writing $n=|\mathcal{C}|$, we can always find (unique) isomorphisms $\pi\from v_1\dots v_n\to v$ and $\theta\from w\to w_1\dots w_n$ that rearrange the words $v$ and $w$ such that each $v_i$ or $w_i$ are single colour words, and such that  
\[(\cspobj{v_1\dots v_n}{f\circ \pi}{H}{g\circ \theta}{w_1\dots w_n}) = (\cspobj{v_1\dots v_n}{f\circ \pi}{\sum_{c\in\mathcal{C}} H_c}{g\circ \theta}{w_1\dots w_n}) \]
where each $H_c$ has sets of nodes of the single colour $c$. Here, we are using the fact that, in any finite commutative free monoid, we can rearrange a word $w$ as the concatenation $v^{k_1}_1...v^{k_n}_n$ with $n$ the cardinality of the set of generators. Thus, we have
\[(\cspobj{v_1\dots v_n}{f\circ \pi}{H}{g\circ \theta}{w_1\dots w_n}) = \sum_{c\in\mathcal{C}}( \cspobj{v_c}{f_c}{ H_c}{g_c}{w_c}) \]
where the legs of the last cospans are simply the restrictions of $f\circ \pi$ and $g\circ \theta$ to each $v_c$ or $w_c$.
 
This decomposition, the requirements that $\beta$ be a prop morphism, and that diagram~\eqref{eq:cospan-discrete} commutes, fully determine $\beta$: 
\begin{align*}
\beta(\cspobj{v}{f}{H}{g}{w}) &= \beta\big(\pi^{-1};(\cspobj{v_1\dots v_n}{f\circ \pi}{H}{g\circ \theta}{w_1\dots w_n});\theta^{-1}\big)\\
&=\pi^{-1} ; \beta\left(\sum_{c\in\mathcal{C}} \left(\cspobj{v_c}{f_c}{ H_c}{g_c}{w_c}\right)\right); \theta^{-1}\\
&=\pi^{-1} ; \bigotimes_{c\in\mathcal{C}} \beta\left(\cspobj{v_c}{f_c}{ H_c}{g_c}{w_c}\right); \theta^{-1}\\
&=\pi^{-1} ; \bigotimes_{c\in\mathcal{C}} \alpha_c\left(\cspobj{v_c}{f_c}{ H_c}{g_c}{w_c}\right); \theta^{-1}
\end{align*}
where we use the same notation for the symmetries $\pi,\theta$ in all coloured props involved (since they exist in all of them, and any prop morphism has to preserve them), and where the product denotes the monoidal product in $\mathbb{A}$. Note that the definition of $\beta$ does not actually depend on the choice of $\pi,\theta$, since these are just used to reorder the interface before applying their inverse to put them back in the same order. Finally, for any colour $c$, $\beta([\cspobj{v}{f}{H}{g}{w}]_c)  = \pi^{-1};\alpha_c(\cspobj{v}{f\circ \pi}{H}{g\circ \theta}{w});\theta^{-1} = \alpha_c(\cspobj{v}{f}{H}{g}{w})$, since $\alpha_c$ is a prop morphism. So the diagram above does commute for any colour $c$. 
\end{proof}

We are now ready to prove the main result of the section.

\begin{proof}[Proof of Proposition~\ref{cmoniso}]
The first step is to notice that the coloured version follows from the single-colour case and the facts that 
$\mathbf{CMon}_{\mathcal{C}}$ is isomorphic to the coproduct of $\mathcal{C}$ copies of $\mathbf{Cmon}$ and $\rmacspemptyhyp$ is isomorphic to the coproduct $\sum_{c\in\mathcal{C}}\mathrm{RMACsp}_{\disc}\!\left(\mathbf{Hyp}_{\emptyset,\{\bullet\}}\right)$ by Lemma~\ref{lem:hypergraph-coproduct}. Thus, we only have to show that $\mathbf{CMon} \cong \mathrm{RMACsp}_{\disc}\left(\mathbf{Hyp}_{\emptyset,\{c\}}\right)$ for an arbitrary single colour $c$. But the morphisms of $\mathrm{RMACsp}_{\disc}\left(\mathbf{Hyp}_{\emptyset,\{c\}}\right)$ are just discrete cospans, \emph{i.e.} cospans of sets (since there are no hyperedges). Thus $\mathrm{RMACsp}_{\disc}\left(\mathbf{Hyp}_{\emptyset,\{c\}}\right)\cong \rmacspf$.

Recall that we also know that $\mathbf{CMon} \cong \mathbb{F}$ (see Example~\ref{cmonf}). Therefore, it suffices to show that there exists an isomorphism $H: \mathbb{F} \rightarrow \rmacspf$ to prove the desired result. We construct this isomorphism explicitly in the following way:
\begin{itemize}
    \item $H(m)=m$ for all $m \in \mathrm{Ob}(\mathbb{F})$
    \item $H(f) = \cspobj{m}{f}{n}{id}{n}$, for all morphisms $f: m \rightarrow n$  in $\mathbb{F}$
\end{itemize}
First, $H$ is well-defined, as any cospan in its image is clearly right-monogamous. Furthermore, $H$ is identity-on-objects, and it maps the identity $id_m\from m \to m$ to the identity cospan $\cspobj{m}{id}{m}{id}{m}$, and the symmetry $\sigma_m^n\from m+n\to n+m$ to the cospan  $\cspobj{m+n}{\sigma_{m,n}}{n+m}{id}{n+m}$. Finally, the composition $(\cspobj{m}{f}{n}{id}{n}); (\cspobj{n}{g}{t}{id}{t})$ is obtained by taking a pushout, which gives $\cspobj{m}{g \circ f }{t}{id}{t}$;
the monoidal product $(\cspobj{m_1}{f_1}{n_1}{id}{n_1})\bigoplus (\cspobj{m_2}{f_2}{n_2}{id}{n_2})$ is obtained by taking a coproduct, which is simply a disjoint union in $\mathbb{F}$, giving $(\cspobj{m_1+m_2}{f_1+f_2}{n_1+n_2}{id}{n_1+n_2})$. Thus $H$ preserves composition and monoidal product, and 
is a morphism of props. We now show it is full and faithful. 

Suppose $H(f_1)=H(f_2)$ for $f_1: m_1 \rightarrow n_1$ and $f_2: m_2 \rightarrow n_2$. Then $\cspobj{m_1}{f_1}{n_1}{id}{n_1}$  and $\cspobj{m_2}{f_2}{n_2}{id}{n_2}$ are isomorphic and $m_1 = m_2$ and $n_1=n_2$. Furthermore, we have a commutative diagram
\[
\begin{tikzcd}
                                     & n_1 \arrow[dd, "\psi"] &                                      \\
m_1 \arrow[ru, "f_1"] \arrow[rd, "f_2"'] &                      & n_1 \arrow[lu, "id"'] \arrow[ld, "id"] \\
                                     & n_1                    &                                     
\end{tikzcd}
\]
where $\psi$ is an iso. Hence $id ; \psi = id$ and therefore $\psi=id$ and $f_1=f_2$. This shows that $H$ is faithful.

Finally, every $\cspobj{m}{f}{n}{g}{n}$ in $\rmacspf$ is isomorphic to $\cspobj{m}{f^*}{n}{id}{n}$, for some $f^*$ by Lemma~\ref{isocspf}, giving $H(f^*) = \cspobj{m}{f^*}{n}{id}{n}$. Therefore, $H$ is full.
\end{proof}
\subsection{The general case.}
Our next goal, and the core result of this section, is extending Proposition~\ref{cmoniso} to the case where $\Sigma$ is non-empty, \emph{i.e.}, an isomorphism between $\cprd$ and $\rmacsphyp$. This will allow us to refer to $\rmacsphyp$ as the combinatorial characterisation of string diagrams in $\cprd$, and study their rewriting as DPO-rewriting in $\rmacsphyp$ in Section~\ref{sec:rewriting}.

In order to relate $\cprd$ and $\rmacsphyp$, we will use a strategy analogous to the one used in \cite{pt2} for theories with plain symmetric monoidal structure. In essence, we want to show that $\rmacsphyp$ has the universal property of the coproduct. Consider
$$\cspobj{\mathbf{S}_{\Sigma,\mathcal{C}}}{\sfun}{\rmacsphyp}{|\cdot|}{\mathbf{CMon}}$$
where 
\begin{itemize}
\item $\sfun: \mathbf{S}_{\Sigma,\mathcal{C}} \rightarrow \rmacsphyp$ is the $\mathcal{C}$-coloured prop morphism
defined in~\cite{pt2} (monogamous cospans are, in particular, right-monogamous); on each generating operation $o\from c_1\dots c_k \to d_1\dots d_l$ of $\Sigma$ (where $c_i,b_j\in\mathcal{C}$) it is given by
\[\begin{tikzpicture}
	\begin{pgfonlayer}{nodelayer}
		\node [style=none] (6) at (-7.25, 0) {};
		\node [style=none] (7) at (-6.25, 0) {};
		\node [style=none] (8) at (-6.25, 0) {};
		\node [style=none] (9) at (-0.25, 0) {};
		\node [style=none] (10) at (0.75, 0) {};
		\node [style=none] (11) at (-7.25, 0) {};
		\node [style=none] (12) at (-6.25, 0) {};
		\node [style=none] (25) at (-7.25, 0) {};
		\node [style=none] (26) at (-6.25, 0) {};
		\node [style=none] (27) at (-6.25, 0) {};
		\node [style=none] (67) at (-11.25, 0) {$\mapsto$};
		\node [style=none] (78) at (-16.75, 1.25) {\scriptsize $c_1$};
		\node [style=none] (82) at (-11.25, 0.75) {\scriptsize $\sfun$};
		\node [style=none] (83) at (-3.75, 0.5) {};
		\node [style=none] (84) at (-3.75, -0.5) {};
		\node [style=none] (85) at (-2.75, -0.5) {};
		\node [style=none] (86) at (-2.75, 0.5) {};
		\node [style=none] (87) at (-3.75, 0.5) {};
		\node [style=small black dot] (88) at (-5.25, 1) {$c_1$};
		\node [style=none] (89) at (-3.25, 0) {$o$};
		\node [style=none] (90) at (-2.75, -0.5) {};
		\node [style=small black dot] (91) at (-1.25, -1) {$d_l$};
		\node [style=small black dot] (92) at (-5.25, -1) {$c_k$};
		\node [style=small black dot] (93) at (-1.25, 1) {$d_1$};
		\node [style=none] (94) at (-5.25, 0.25) {$\vdots$};
		\node [style=none] (95) at (-1.25, 0.25) {$\vdots$};
		\node [style=none] (96) at (-9.25, 2) {};
		\node [style=none] (97) at (-9.25, -2) {};
		\node [style=none] (98) at (-7.75, -2) {};
		\node [style=none] (99) at (-7.75, 2) {};
		\node [style=small black dot] (100) at (-8.5, 1) {$c_1$};
		\node [style=small black dot] (101) at (-8.5, -1) {$c_k$};
		\node [style=none] (102) at (-8.5, 0) {$\vdots$};
		\node [style=small black dot] (103) at (2, -1) {$d_l$};
		\node [style=small black dot] (104) at (2, 1) {$d_1$};
		\node [style=none] (105) at (2, 0) {$\vdots$};
		\node [style=none] (106) at (1.25, 2) {};
		\node [style=none] (107) at (1.25, -2) {};
		\node [style=none] (108) at (2.75, -2) {};
		\node [style=none] (109) at (2.75, 2) {};
		\node [style=none] (110) at (-15.5, 1) {};
		\node [style=none] (111) at (-15.5, -1) {};
		\node [style=none] (112) at (-14.5, -1) {};
		\node [style=none] (113) at (-14.5, 1) {};
		\node [style=none] (114) at (-15.5, 1) {};
		\node [style=none] (116) at (-15, 0) {$o$};
		\node [style=none] (117) at (-14.5, -1) {};
		\node [style=none] (121) at (-16.25, 0.25) {$\vdots$};
		\node [style=none] (122) at (-14, 0.25) {$\vdots$};
		\node [style=none] (123) at (-17, 0.75) {};
		\node [style=none] (124) at (-15.5, -0.75) {};
		\node [style=none] (125) at (-14.5, -0.75) {};
		\node [style=none] (126) at (-14.5, 0.75) {};
		\node [style=none] (127) at (-15.5, 0.75) {};
		\node [style=none] (128) at (-13.25, 0.75) {};
		\node [style=none] (129) at (-13.25, -0.75) {};
		\node [style=none] (130) at (-17, -0.75) {};
		\node [style=none] (131) at (-16.75, -0.25) {\scriptsize $c_k$};
		\node [style=none] (132) at (-13.25, 1.25) {\scriptsize $d_1$};
		\node [style=none] (133) at (-13.25, -0.25) {\scriptsize $d_l$};
	\end{pgfonlayer}
	\begin{pgfonlayer}{edgelayer}
		\draw [style=pointer edge] (11.center) to (12.center);
		\draw [style=pointer edge] (10.center) to (9.center);
		\draw [style=gray edge] (84.center) to (83.center);
		\draw [style=gray edge, bend left=90, looseness=0.75] (83.center) to (86.center);
		\draw [style=gray edge] (86.center) to (85.center);
		\draw [style=gray edge, in=270, out=270, looseness=0.75] (85.center) to (84.center);
		\draw [style=gray edge] (88) to (87.center);
		\draw [style=gray edge] (91) to (90.center);
		\draw [style=hadamard edge] (96.center) to (99.center);
		\draw [style=hadamard edge] (98.center) to (99.center);
		\draw [style=hadamard edge] (96.center) to (97.center);
		\draw [style=hadamard edge] (97.center) to (98.center);
		\draw [style=hadamard edge] (106.center) to (109.center);
		\draw [style=hadamard edge] (108.center) to (109.center);
		\draw [style=hadamard edge] (106.center) to (107.center);
		\draw [style=hadamard edge] (107.center) to (108.center);
		\draw (111.center) to (110.center);
		\draw (110.center) to (113.center);
		\draw (113.center) to (112.center);
		\draw (112.center) to (111.center);
		\draw (123.center) to (127.center);
		\draw (126.center) to (128.center);
		\draw (125.center) to (129.center);
		\draw (130.center) to (124.center);
		\draw [style=gray edge] (92) to (84.center);
		\draw [style=gray edge] (86.center) to (93);
	\end{pgfonlayer}
\end{tikzpicture}\]
\item $|\cdot| : \mathbf{CMon}_{\mathcal{C}} \rightarrow \rmacsphyp$ is defined by composing the isomorphism of Proposition~\ref{cmoniso} with the obvious morphism $\rmacspemptyhyp \to \rmacsphyp$. We can describe it explicitly, by interpreting the generators of $\mathbf{CMon}_{\mathcal{C}}$ as follows, for each $c\in\mathcal{C}$:
\[
 \begin{tikzpicture}
	\begin{pgfonlayer}{nodelayer}
		\node [style=none] (0) at (-3.5, 1) {};
		\node [style=none] (1) at (-3.5, -1) {};
		\node [style=none] (2) at (-2, -1) {};
		\node [style=none] (3) at (-2, 1) {};
		\node [style=none] (6) at (-7.75, 0) {};
		\node [style=none] (7) at (-6.75, 0) {};
		\node [style=none] (8) at (-6.75, 0) {};
		\node [style=none] (9) at (-5, 0) {};
		\node [style=none] (10) at (-4, 0) {};
		\node [style=none] (11) at (-7.75, 0) {};
		\node [style=none] (12) at (-6.75, 0) {};
		\node [style=none] (14) at (-9.75, 1) {};
		\node [style=none] (15) at (-9.75, -1) {};
		\node [style=none] (16) at (-8.25, -1) {};
		\node [style=none] (17) at (-8.25, 1) {};
		\node [style=small black dot] (18) at (-9, 0.5) {$c$};
		\node [style=small black dot] (19) at (-9, -0.5) {$c$};
		\node [style=small black dot] (20) at (-2.75, 0) {$c$};
		\node [style=none] (25) at (-7.75, 0) {};
		\node [style=none] (26) at (-6.75, 0) {};
		\node [style=none] (27) at (-6.75, 0) {};
		\node [style=small black dot] (35) at (-6, 0) {$c$};
		\node [style=none] (36) at (12, 1) {};
		\node [style=none] (37) at (12, -1) {};
		\node [style=none] (38) at (13.5, -1) {};
		\node [style=none] (39) at (13.5, 1) {};
		\node [style=none] (40) at (8, 0) {};
		\node [style=none] (41) at (9, 0) {};
		\node [style=none] (42) at (9, 0) {};
		\node [style=none] (43) at (10.5, 0) {};
		\node [style=none] (44) at (11.5, 0) {};
		\node [style=none] (45) at (8, 0) {};
		\node [style=none] (46) at (9, 0) {};
		\node [style=none] (47) at (6.25, 1) {};
		\node [style=none] (48) at (6.25, -1) {};
		\node [style=none] (49) at (7.5, -1) {};
		\node [style=none] (50) at (7.5, 1) {};
		\node [style=small black dot] (52) at (12.75, 0) {$c$};
		\node [style=none] (54) at (8, 0) {};
		\node [style=none] (55) at (9, 0) {};
		\node [style=none] (56) at (9, 0) {};
		\node [style=small black dot] (57) at (9.75, 0) {$c$};
		\node [style=none] (58) at (-12.75, 0) {};
		\node [style=none] (59) at (-14, 0.5) {};
		\node [style=none] (60) at (-14, 0.5) {};
		\node [style=none] (61) at (-14, -0.5) {};
		\node [style=none] (62) at (-13, 0) {};
		\node [style=none] (63) at (-14, 0.5) {};
		\node [style=none] (64) at (-13.75, -0.5) {};
		\node [style=none] (65) at (-13, 0) {};
		\node [style=none] (66) at (-13.75, -0.5) {};
		\node [style=none] (67) at (-11, 0) {$\mapsto$};
		\node [style=small black dot] (68) at (-13, 0) {};
		\node [style=none] (69) at (-12.25, 0) {};
		\node [style=none] (70) at (3, 0) {};
		\node [style=none] (71) at (2.75, 0) {};
		\node [style=none] (72) at (2.75, 0) {};
		\node [style=none] (73) at (4.75, 0) {$\mapsto$};
		\node [style=small black dot] (74) at (2.75, 0) {};
		\node [style=none] (75) at (3.5, 0) {};
		\node [style=none] (78) at (-12.5, 0.25) {\scriptsize $c$};
		\node [style=none] (79) at (-13.75, 0.75) {\scriptsize $c$};
		\node [style=none] (80) at (-13.75, -0.25) {\scriptsize $c$};
		\node [style=none] (81) at (3.25, 0.25) {\scriptsize $c$};
		\node [style=none] (82) at (-11, 0.75) {\scriptsize $\cfun$};
		\node [style=none] (83) at (4.75, 0.75) {\scriptsize $\cfun$};
	\end{pgfonlayer}
	\begin{pgfonlayer}{edgelayer}
		\draw [style=hadamard edge] (0.center) to (3.center);
		\draw [style=hadamard edge] (2.center) to (3.center);
		\draw [style=hadamard edge] (0.center) to (1.center);
		\draw [style=hadamard edge] (1.center) to (2.center);
		\draw [style=pointer edge] (11.center) to (12.center);
		\draw [style=pointer edge] (10.center) to (9.center);
		\draw [style=hadamard edge] (14.center) to (17.center);
		\draw [style=hadamard edge] (16.center) to (17.center);
		\draw [style=hadamard edge] (14.center) to (15.center);
		\draw [style=hadamard edge] (15.center) to (16.center);
		\draw [style=hadamard edge] (36.center) to (39.center);
		\draw [style=hadamard edge] (38.center) to (39.center);
		\draw [style=hadamard edge] (36.center) to (37.center);
		\draw [style=hadamard edge] (37.center) to (38.center);
		\draw [style=pointer edge] (45.center) to (46.center);
		\draw [style=pointer edge] (44.center) to (43.center);
		\draw [style=hadamard edge] (47.center) to (50.center);
		\draw [style=hadamard edge] (49.center) to (50.center);
		\draw [style=hadamard edge] (47.center) to (48.center);
		\draw [style=hadamard edge] (48.center) to (49.center);
		\draw [in=150, out=0, looseness=1.50] (63.center) to (62.center);
		\draw (61.center) to (64.center);
		\draw [bend right, looseness=0.50] (66.center) to (65.center);
		\draw (68) to (69.center);
		\draw (74) to (75.center);
	\end{pgfonlayer}
\end{tikzpicture}
\]
where the legs of each cospan are the only possible maps of the appropriate type.
\end{itemize}
To prove our main result we will need to rely on the faithfulness of the prop morphism $\sfun: \mathbf{S}_{\Sigma,\mathcal{C}} \rightarrow \rmacsphyp$. In other words, we need to rely on the fact that plain string diagrams are mapped faithfully to cospans of hypergraphs and that they are precisely the monogamous acyclic such cospans. 
\begin{prop}\label{prop:monogamous-interpret-faithful}
$\sfun: \mathbf{S}_{\Sigma,\mathcal{C}} \rightarrow \rmacsphyp$ is a faithful $\mathcal{C}$-coloured prop morphism whose image consists precisely of the monogamous acyclic cospans.
\end{prop}
\begin{proof}
See Appendix~\ref{app}.
\end{proof}
Note \cite[Corollary 3.4]{Zanasi17} and \cite[Proposition 3.4]{BonchiGKSZ16} provide results analogue to our Proposition~\ref{prop:monogamous-interpret-faithful}. However, the argument used in those proofs is incomplete, as pointed out in~\cite{Fritz}. This is why we supply a different argument, based on a similar characterisation by Joyal and Street~\cite[Theorem 2.3]{JOYAL199155}. Their formalisation of string diagrams is in terms of certain \emph{graphs} instead of hypergraphs, but the result is equivalent---we elaborate on the correspondence in Appendix~\ref{app}.

We will need an analogous result for the combinatorial interpretation of $\mathbf{CMon}_{\mathcal{C}}$.
\begin{prop}\label{prop:right-monogamous-interpret-faithful}
$|\cdot| \from \mathbf{CMon}_{\mathcal{C}}\rightarrow \rmacsphyp$ is a faithful $\mathcal{C}$-coloured prop morphism whose image are precisely the discrete right-monogamous acyclic cospans.
\end{prop}
\begin{proof}
It is the composition of the isomorphism of Proposition~\ref{cmoniso} and the obvious morphism $\rmacspemptyhyp \to \rmacsphyp$, which is faithful.
\end{proof}
To show that $\rmacsphyp$ has the universal property of the coproduct, the fundamental step is investigating how right-monogamous acyclic cospans  can be factorised into a composite cospan $\mathcal{M}_0;\mathcal{D}_0; \dots ; \mathcal{M}_l;\mathcal{D}_l$ that alternates between monogamous acyclic cospans, \emph{i.e.}, in the image $\sfun \from \mathbf{S}_{\Sigma,\mathcal{C}} \rightarrow \rmacsphyp$, and discrete right-monogamous acyclic cospans, \emph{i.e.}, in the image of  $|\cdot| \from \mathbf{CMon}_{\mathcal{C}} \rightarrow \rmacsphyp$.



As we saw, the hypergraphs that correspond to plain string diagrams, \emph{i.e.} morphisms of $\syntax{\Sigma , \col}$ are monogamous: nodes are precisely the target and source of one hyperedge. The commutative monoid structure relaxes this requirement for targets. For our last decomposition, we would like to identify nodes that can only appear in the hypergraph representation of diagrams that contain some occurrence of the commutative monoid structure (multiplication or unit), that is, nodes that do not simply represent plain wires. The following definition formalises this idea.
\begin{defi}[Left-amonogamous nodes]
Let $\cspobj{v}{}{G}{}{w}$ be a right-monogamous acyclic cospan. We say that a node in $G$ is \emph{\bad} if:
\begin{itemize}
    \item it is in $\mathrm{in(G)}$ and its in-degree is not equal to 0, or
    \item it is not in $\mathrm{in(G)}$ and its in-degree is not equal to 1.
\end{itemize}
\end{defi}
\begin{exa}\label{ex:left-amonogamous}
In the cospan~\eqref{eq:right-mono-cospan}, the nodes with colours $c_2$ and $c_3$ are left-amonogamous, while others are~not.
\end{exa}
As explained above, we want to factorise right-monogamous acyclic cospans into a composite $\mathcal{M}_0;\mathcal{D}_0; \dots ; \mathcal{M}_l;\mathcal{D}_l$, alternating between monogamous acyclic cospans and discrete right-monogamous acyclic cospans. The index of each factor roughly indicates the maximum number of left-amonogamous nodes on a path preceding it. To make this idea precise, we require the two notions below, for nodes (corresponding to discrete right-monogamous cospans) and for hyperedges (corresponding to monogamous cospans).
\begin{defi}[Order of nodes and level of hyperedges]
\label{def:levels}
Let $\cspobj{v}{}{G}{}{w}$ be a right-monogamous acyclic cospan.
We define the \emph{order} of a node $n$ as the largest number of left-amonogamous nodes preceding it (including itself) on a path leading to $n$. 
The \emph{level} of an hyperedge $h$ is the minimum of
\begin{itemize}
    \item the order of any node in $out(G)$ that is its successor, or
    \item the order of any left-amonogamous node that is its target, minus 1.
\end{itemize}
\end{defi}
\begin{exa}
In the cospan~\eqref{eq:right-mono-cospan}, hyperedges $A$ and $C$ are level-0 hyperedges, and hyperedge $B$ is a level-1 hyperedge.
\end{exa}

Recall that we want to obtain a factorisation of any cospan into an alternating composition of discrete right-monogamous cospans---corresponding to diagrams with no generating boxes from the chosen signature---and monogamous cospans---corresponding to plain string diagrams over the chosen signature. We will do this by induction on the maximum level of hyperedges, effectively stripping the necessary cospans (discrete right-monogamous and monogamous) at each level as we move from left to right. 

The following lemma will be used at each induction step: here, we require the decomposition to not only alternate between monogamous and discrete right-monogamous cospans, but to also keep track of the order of terminal nodes. Diagrammatically, we want a decomposition of a right-monogamous cospan into the following form:
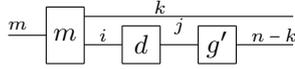
\[\begin{tikzpicture}
	\begin{pgfonlayer}{nodelayer}
		\node [style=none] (229) at (1, 0.25) {};
		\node [style=none] (230) at (1, -0.75) {};
		\node [style=none] (231) at (2, -0.75) {};
		\node [style=none] (232) at (2, 0.25) {};
		\node [style=none] (233) at (2, -0.25) {};
		\node [style=none] (235) at (3.5, 0.5) {};
		\node [style=none] (236) at (-3, 0.75) {};
		\node [style=none] (237) at (-3, -0.75) {};
		\node [style=none] (238) at (-2, -0.75) {};
		\node [style=none] (239) at (-2, 0.75) {};
		\node [style=none] (241) at (-3, 0) {};
		\node [style=none] (242) at (-2, 0.5) {};
		\node [style=none] (245) at (-4, 0) {};
		\node [style=none] (246) at (3.5, -0.25) {};
		\node [style=none] (250) at (-1, 0.25) {};
		\node [style=none] (251) at (-1, -0.75) {};
		\node [style=none] (252) at (0, -0.75) {};
		\node [style=none] (253) at (0, 0.25) {};
		\node [style=none] (256) at (-1, -0.25) {};
		\node [style=none] (257) at (-2, -0.25) {};
		\node [style=none] (262) at (-0.5, -0.25) {$d$};
		\node [style=none] (267) at (-3.75, 0.25) {\scriptsize $v$};
		\node [style=none] (270) at (3, 0) {\scriptsize $w_2$};
		\node [style=none] (271) at (0, 0.75) {\scriptsize $w_1$};
		\node [style=none] (272) at (0, -0.25) {};
		\node [style=none] (273) at (1, -0.25) {};
		\node [style=none] (276) at (-1.5, 0) {\scriptsize $u_1$};
		\node [style=none] (278) at (1.5, -0.25) {$g'$};
		\node [style=none] (279) at (-2.5, 0) {$m$};
		\node [style=none] (280) at (0.5, 0) {\scriptsize $u_2$};
	\end{pgfonlayer}
	\begin{pgfonlayer}{edgelayer}
		\draw (230.center) to (229.center);
		\draw (229.center) to (232.center);
		\draw (232.center) to (231.center);
		\draw (231.center) to (230.center);
		\draw (237.center) to (236.center);
		\draw (236.center) to (239.center);
		\draw (239.center) to (238.center);
		\draw (238.center) to (237.center);
		\draw (245.center) to (241.center);
		\draw (233.center) to (246.center);
		\draw (251.center) to (250.center);
		\draw (250.center) to (253.center);
		\draw (253.center) to (252.center);
		\draw (252.center) to (251.center);
		\draw [in=180, out=0] (257.center) to (256.center);
		\draw (242.center) to (235.center);
		\draw (272.center) to (273.center);
	\end{pgfonlayer}
\end{tikzpicture}\]
with $m$ corresponding to a monogamous cospan, and $d$ to a discrete right-monogamous one, and $g'$ is the rest of the decomposition. Here, the $w_1$-labelled wires correspond to the terminal order-$0$ nodes of the overall composite diagram. Keeping track of where terminal nodes of each order are located is an important technical complication that will be needed to prove that the map that we construct out of $\rmacsphyp$ is a monoidal functor which satisfies the universal property of the coproduct $\cprd$. Recall that, following Proposition~\ref{cmoniso}, we refer abusively to permutations $\pi: w\to w$ below as cospans, assuming implicitly that we mean the cospan $\cspobj{w}{\pi}{w}{id}{w}$. Finally, we use the term \emph{in-connection} of a node $x$ in a given cospan $\cspobj{v}{f}{G}{g}{w}$, for any hyperedge $h$ that has $x$ as target in $G$, or for any boundary node of $v$ that is mapped to $x$ by $f$. In diagrammatic terms, nodes without any in-connections correspond to occurrences of the unit $\begin{tikzpicture}
	\begin{pgfonlayer}{nodelayer}
		\node [style=black] (0) at (0.75, 0) {};
		\node [style=none] (1) at (1.75, 0) {};
		\node [style=none] (8) at (1.5, 0.25) {\scriptsize $c$};
	\end{pgfonlayer}
	\begin{pgfonlayer}{edgelayer}
		\draw (0) to (1.center);
	\end{pgfonlayer}
\end{tikzpicture}
$. 
\begin{lem}[level-0 decomposition]
\label{lem:level-0}
Let $\mathcal{G} = (\cspobj{v}{f}{G}{g}{w})$ be a right-monogamous acyclic cospan whose order-$0$ terminal nodes are the first $|w_1|$ nodes of $w$. Then there exists a decomposition of $\mathcal{G}$ as $\mathcal{M};(id_{w_1}\oplus (\mathcal{D};\mathcal{G}'))$,
where (A) $\mathcal{M}$ is monogamous acyclic and contains precisely all the level-0 hyperedges; (B) $\mathcal{D}$ is discrete right-monogamous and contains precisely all order-1 left-amonogamous nodes; (C) $\mathcal{G}'$ is right-monogamous acyclic and has no left-amonogamous nodes without any in-connections.

Moreover, any two such factorisations differ only by permutations of the terminal nodes of the factors, i.e., if $\mathcal{G}=\mathcal{M};(id_{w_1}\oplus (\mathcal{D};\mathcal{G}'))=\mathcal{M}';(id_{w_1}\oplus (\mathcal{D}';\mathcal{G}''))$, then there exists permutations $\pi,\theta$ such that $\mathcal{M}' = \mathcal{M};\pi$, $\pi;(id_{w_1}\oplus (\mathcal{D}';\mathcal{G}')) = id_{w_1}\oplus (\mathcal{D};\mathcal{G}')$, and $\mathcal{D}' = \mathcal{D};\theta$, $\theta;\mathcal{G}''=\mathcal{G}'$.
\end{lem}
\begin{proof}
Let $\mathcal{M} = (\cspobj{v}{f}{M}{h}{w_1 u_1})$ first, where
\begin{itemize}
\item we define $M$ to be the hypergraph whose hyperedges are exactly the level-0 hyperedges of $G$; its nodes are all monogamous sources and targets of level-0 hyperedges, as well as a new node called $f(x)$ for each letter $x$ in $v$  such that $f(x)$ is not the source of any level-0 hyperedge (\emph{i.e.}, a copy of each node in $v$ whose image is left-amonogamous), and a new node called $(e,j)$ for every pair of a hyperedge $e$ with a degree-1 left-amonogamous node as its $j$-th target (whose colour is the same as that target). 
\item For $e$ some hyperedge of $M$ (\emph{i.e.}, some level-0 hyperedge of $G$), the sources of $e$ in $M$ are the same as those of $e$ in $G$, and its $j$-th target is the same as the $j$-th target of $e$ in $G$ if that target is monogamous, or simply the new node $(e,j)$ otherwise.
\item To define the right leg $h$ of the cospan, we establish an arbitrary ordering of the $|u_1|$ terminal nodes of $M$ which are not terminal in $G$. From this ordering, we can construct a map $h\from w_1 u_1 \to M$ which coincides with $g$ on $w_1$ (identified with the terminal nodes of $G$ by assumption). Note that the cospan $\mathcal{M}$ is unique up to this choice of ordering, and any permutation of these nodes will give a valid decomposition, as in the statement of the lemma.
\end{itemize}
We now construct $\mathcal{D} = (\cspobj{u_1}{p}{D}{q}{u_2})$, where 
\begin{itemize}
\item $D$ is the discrete hypergraph containing only the order-1 left amonogamous nodes of $G$ (and no hyperedges, by definition). Fixing some ordering of these nodes, we obtain the word $u_2$.
\item The left leg $p$ of the cospan is uniquely defined (up to the chosen ordering above) to connect all left-amonogamous nodes of order-1 in $G$ to the corresponding nodes in $M$. The right leg $q$ of the same cospan is uniquely defined by the ordering chosen above. Once again, this is unique up to some permutation, as stated in the lemma.
\end{itemize}
Finally, let $\mathcal{G}' = (\cspobj{u_2}{f'}{G'}{g'}{w_2})$ be the last cospan, which we define below.
\begin{itemize}
\item $G'$ has all remaining hyperedges; its nodes are those of $G$ which are not in $M$, plus the nodes of $D$.
\item Its source and target maps are restrictions of those of $G$ to the remaining hyperedges. 
\item The left leg $f'$ of the cospan $\mathcal{G}'$ is the inverse of $q$ (which is not only monic, but one-to-one, by construction), and the right leg is the restriction of $g$ to the terminal nodes of $G'$.
\end{itemize} 
Note that $\mathcal{G}'$ is right-monogamous because $\mathcal{G}$ was. Moreover, it does not have any left-amonogamous nodes without any in-connections (occurrences of the black unit, in diagrammatic terms) because these are left-amonogamous nodes of order-1, which are all in $\mathcal{D}$, by construction.
\end{proof}
\begin{exa}\label{ex:level-0}
Taking the level-0 decomposition of the cospan appearing in Example~\ref{ex:right-mono-cospan} gives the following three cospans:
\[\scalebox{0.8}{\begin{tikzpicture}
	\begin{pgfonlayer}{nodelayer}
		\node [style=none] (0) at (-8.75, 3) {};
		\node [style=none] (1) at (-8.75, -2.75) {};
		\node [style=none] (2) at (-6.75, -2.75) {};
		\node [style=none] (3) at (-6.75, 3) {};
		\node [style=none] (4) at (24, 2.5) {};
		\node [style=none] (5) at (24, -2.5) {};
		\node [style=none] (6) at (25.5, -2.5) {};
		\node [style=none] (7) at (25.5, 2.5) {};
		\node [style=small black dot] (9) at (-8, 1) {$c_2$};
		\node [style=small black dot] (10) at (-8, 0) {$c_2$};
		\node [style=small black dot] (11) at (-8, -1) {$c_3$};
		\node [style=small black dot] (12) at (-8, -2) {$c_1$};
		\node [style=none] (14) at (-2.5, 2.5) {};
		\node [style=none] (15) at (-2.5, 1.5) {};
		\node [style=none] (16) at (-1.5, 1.5) {};
		\node [style=none] (17) at (-1.5, 2.5) {};
		\node [style=none] (20) at (-2.5, -2) {};
		\node [style=none] (21) at (-2.5, -3) {};
		\node [style=none] (22) at (-1.5, -3) {};
		\node [style=none] (23) at (-1.5, -2) {};
		\node [style=small black dot] (31) at (24.75, 0) {$c_3$};
		\node [style=small black dot] (35) at (-3.75, -2.5) {};
		\node [style=small black dot] (36) at (-3.75, -2.5) {$c_1$};
		\node [style=none] (37) at (-2.5, -2.5) {};
		\node [style=none] (38) at (-2.5, 2) {};
		\node [style=small black dot] (39) at (-3.75, 2) {$c_0$};
		\node [style=none] (41) at (-7.25, 1.25) {};
		\node [style=none] (42) at (-7.25, 2.25) {$k_0$};
		\node [style=none] (43) at (-7.25, 1.25) {$k_1$};
		\node [style=none] (44) at (-7.25, 0.25) {$k_2$};
		\node [style=none] (45) at (-7.25, -0.75) {$k_3$};
		\node [style=none] (46) at (-7.25, -1.75) {$k_4$};
		\node [style=none] (47) at (24.75, 0.75) {$n_0$};
		\node [style=none] (48) at (-6, 0) {};
		\node [style=none] (49) at (-5, 0) {};
		\node [style=none] (50) at (-5, 0) {};
		\node [style=none] (51) at (22.25, 0) {};
		\node [style=none] (52) at (23.25, 0) {};
		\node [style=none] (53) at (-6, 0) {};
		\node [style=none] (54) at (-5, 0) {};
		\node [style=none] (58) at (-3.75, -1.75) {$k_4$};
		\node [style=none] (63) at (-3.75, 2.75) {$k_0$};
		\node [style=none] (64) at (-2, 2) {$d$};
		\node [style=none] (66) at (-2, -2.5) {$e$};
		\node [style=small black dot] (68) at (-8, 2) {$c_0$};
		\node [style=small black dot] (69) at (16.5, 0) {};
		\node [style=small black dot] (73) at (16.5, 0) {};
		\node [style=small black dot] (77) at (20.75, 0) {$c_3$};
		\node [style=none] (78) at (18.25, 0.5) {};
		\node [style=none] (79) at (18.25, -0.5) {};
		\node [style=none] (80) at (19.25, -0.5) {};
		\node [style=none] (81) at (19.25, 0.5) {};
		\node [style=none] (82) at (19.25, 0) {};
		\node [style=small black dot] (83) at (16.5, 0) {$c_2$};
		\node [style=none] (85) at (18.25, 0) {};
		\node [style=none] (86) at (20.75, 0.75) {$m_0,n_0$};
		\node [style=none] (87) at (16.5, 0.75) {$m_0$};
		\node [style=none] (88) at (18.75, 0) {$f$};
		\node [style=small black dot] (91) at (0, 2.75) {$c_2$};
		\node [style=small black dot] (92) at (0, 1.5) {$c_2$};
		\node [style=small black dot] (93) at (0, -2) {$c_2$};
		\node [style=small black dot] (94) at (0, -3.25) {$c_2$};
		\node [style=none] (95) at (4, 4) {};
		\node [style=none] (96) at (4, -3.75) {};
		\node [style=none] (97) at (6.25, -3.75) {};
		\node [style=none] (98) at (6.25, 4) {};
		\node [style=none] (105) at (5.5, 3.5) {$l_0$};
		\node [style=none] (106) at (5.5, 2.5) {$l_1$};
		\node [style=none] (107) at (5.5, 1.5) {$l_2$};
		\node [style=none] (108) at (5.5, 0.5) {$l_3$};
		\node [style=none] (109) at (5.5, -0.5) {$l_4$};
		\node [style=none] (111) at (2.5, 0) {};
		\node [style=none] (112) at (3.5, 0) {};
		\node [style=small black dot] (113) at (0, 0.25) {$c_2$};
		\node [style=small black dot] (114) at (0, -1) {$c_2$};
		\node [style=none] (116) at (1.25, 0.5) {$k_1,l_3$};
		\node [style=none] (117) at (1.25, -0.5) {$k_2,l_4$};
		\node [style=small black dot] (118) at (4.75, 3) {$c_3$};
		\node [style=small black dot] (119) at (4.75, 2) {$c_2$};
		\node [style=small black dot] (120) at (4.75, -1) {$c_2$};
		\node [style=small black dot] (121) at (4.75, -2) {$c_2$};
		\node [style=small black dot] (122) at (4.75, 1) {$c_2$};
		\node [style=small black dot] (123) at (4.75, 0) {$c_2$};
		\node [style=none] (124) at (5.5, -1.5) {$l_5$};
		\node [style=small black dot] (125) at (0, 4) {$c_3$};
		\node [style=none] (126) at (1, 4.5) {$k_3,l_0$};
		\node [style=none] (127) at (0.5, 3.25) {$l_1$};
		\node [style=none] (128) at (0.5, 2) {$l_2$};
		\node [style=none] (129) at (0.75, -1.75) {$l_5$};
		\node [style=none] (130) at (0.75, -2.75) {$l_6$};
		\node [style=small black dot] (131) at (4.75, -3) {$c_2$};
		\node [style=none] (132) at (5.5, -2.5) {$l_6$};
		\node [style=none] (133) at (6.75, 0) {};
		\node [style=none] (134) at (7.75, 0) {};
		\node [style=none] (135) at (7.75, 0) {};
		\node [style=none] (136) at (6.75, 0) {};
		\node [style=none] (137) at (7.75, 0) {};
		\node [style=none] (138) at (11.75, 2.5) {};
		\node [style=none] (139) at (11.75, -2.5) {};
		\node [style=none] (140) at (13.75, -2.5) {};
		\node [style=none] (141) at (13.75, 2.5) {};
		\node [style=small black dot] (142) at (12.75, 0.5) {$c_2$};
		\node [style=none] (143) at (12.75, -0.5) {$m_1$};
		\node [style=none] (144) at (10.25, 0) {};
		\node [style=none] (145) at (11.25, 0) {};
		\node [style=none] (146) at (14.5, 0) {};
		\node [style=none] (147) at (15.5, 0) {};
		\node [style=none] (148) at (15.5, 0) {};
		\node [style=none] (149) at (14.5, 0) {};
		\node [style=none] (150) at (15.5, 0) {};
		\node [style=small black dot] (151) at (9, -1.75) {};
		\node [style=small black dot] (152) at (9, -1.75) {};
		\node [style=small black dot] (153) at (9, -1.75) {$c_2$};
		\node [style=none] (154) at (9, -1) {$l_1,\dots,l_6,m_1$};
		\node [style=none] (155) at (9, 1.5) {$l_0,m_0$};
		\node [style=small black dot] (156) at (9, 0.75) {$c_3$};
		\node [style=none] (157) at (12.75, 1.25) {$m_0$};
		\node [style=small black dot] (158) at (12.75, -1.25) {$c_3$};
		\node [style=none] (159) at (-7.5, -4.5) {};
		\node [style=none] (160) at (-1.75, -5) {};
		\node [style=none] (161) at (4.25, -4.5) {};
		\node [style=none] (162) at (-1.75, -5.75) {$\mathcal{M}$};
		\node [style=none] (163) at (6, -4.5) {};
		\node [style=none] (164) at (9, -5) {};
		\node [style=none] (165) at (12.25, -4.5) {};
		\node [style=none] (166) at (9, -5.75) {$\mathcal{D}$};
		\node [style=none] (167) at (13.5, -4.5) {};
		\node [style=none] (168) at (18.75, -5) {};
		\node [style=none] (169) at (24.5, -4.5) {};
		\node [style=none] (170) at (18.75, -5.75) {$\mathcal{G}'$};
	\end{pgfonlayer}
	\begin{pgfonlayer}{edgelayer}
		\draw [style=hadamard edge] (0.center) to (3.center);
		\draw [style=hadamard edge] (2.center) to (3.center);
		\draw [style=hadamard edge] (0.center) to (1.center);
		\draw [style=hadamard edge] (1.center) to (2.center);
		\draw [style=hadamard edge] (4.center) to (7.center);
		\draw [style=hadamard edge] (6.center) to (7.center);
		\draw [style=hadamard edge] (4.center) to (5.center);
		\draw [style=hadamard edge] (5.center) to (6.center);
		\draw [style=gray edge] (15.center) to (14.center);
		\draw [style=gray edge, bend left=90, looseness=0.75] (14.center) to (17.center);
		\draw [style=gray edge] (17.center) to (16.center);
		\draw [style=gray edge, in=270, out=270, looseness=0.75] (16.center) to (15.center);
		\draw [style=gray edge] (21.center) to (20.center);
		\draw [style=gray edge, bend left=90, looseness=0.75] (20.center) to (23.center);
		\draw [style=gray edge] (23.center) to (22.center);
		\draw [style=gray edge, in=270, out=270, looseness=0.75] (22.center) to (21.center);
		\draw [style=gray edge] (36) to (37.center);
		\draw [style=gray edge] (39) to (38.center);
		\draw [style=pointer edge] (53.center) to (54.center);
		\draw [style=pointer edge] (52.center) to (51.center);
		\draw [style=gray edge] (79.center) to (78.center);
		\draw [style=gray edge, bend left=90, looseness=0.75] (78.center) to (81.center);
		\draw [style=gray edge] (81.center) to (80.center);
		\draw [style=gray edge, in=270, out=270, looseness=0.75] (80.center) to (79.center);
		\draw [style=gray edge, in=180, out=0, looseness=0.75] (82.center) to (77);
		\draw [style=gray edge] (83) to (85.center);
		\draw (22.center) to (94);
		\draw (23.center) to (93);
		\draw (16.center) to (92);
		\draw (17.center) to (91);
		\draw [style=hadamard edge] (95.center) to (98.center);
		\draw [style=hadamard edge] (97.center) to (98.center);
		\draw [style=hadamard edge] (95.center) to (96.center);
		\draw [style=hadamard edge] (96.center) to (97.center);
		\draw [style=pointer edge] (112.center) to (111.center);
		\draw [style=pointer edge] (136.center) to (137.center);
		\draw [style=hadamard edge] (138.center) to (141.center);
		\draw [style=hadamard edge] (140.center) to (141.center);
		\draw [style=hadamard edge] (138.center) to (139.center);
		\draw [style=hadamard edge] (139.center) to (140.center);
		\draw [style=pointer edge] (145.center) to (144.center);
		\draw [style=pointer edge] (149.center) to (150.center);
		\draw [in=90, out=-90, looseness=0.25] (159.center) to (160.center);
		\draw [in=-90, out=90, looseness=0.25] (160.center) to (161.center);
		\draw [in=90, out=-90, looseness=0.25] (163.center) to (164.center);
		\draw [in=-90, out=90, looseness=0.25] (164.center) to (165.center);
		\draw [in=90, out=-90, looseness=0.25] (167.center) to (168.center);
		\draw [in=-90, out=90, looseness=0.25] (168.center) to (169.center);
	\end{pgfonlayer}
\end{tikzpicture}}\]
We see that the first cospan corresponds to $\mathcal{M}$ in the notation of Lemma~\ref{lem:level-0}: it contains precisely the two level-0 hyperedges ($d$ and $e$). Note that this cospan is monogamous, requiring the creation of new nodes for each target of $d$ and $e$, as well as new nodes to split the amonogamous nodes into multiple occurrences of monogamous ones. The second component corresponds to $\mathcal{D}$ in the lemma's notation and merges nodes $6$-$11$ together into the single order-1 node of the original cospan. Finally, the third cospan corresponds to $\mathcal{G}'$ and contains the only remaining hyperedge of order $>0$ and node of order $>1$. 
\end{exa}
Note that the non-uniqueness of the decomposition comes from two distinct sources: 1) arbitrary ordering of nodes on the boundaries of cospans, and 2) the commutativity of the monoid multiplication which implies that it can absord any permutation of the wires that it merges. We can now iterate level-0 decomposition to factorise any cospan of hypergraph into successive levels.
\begin{lem}[Factorisation into levels]
\label{mononandright}
Any right-monogamous acyclic cospan $\mathcal{G}=\cspobj{v}{}{G}{}{w}$ can be factored into $\mathcal{M}_0 ; \big(id_{u_0}\oplus(\mathcal{D}_0;\dots ; \mathcal{M}_l;(id_{u_l}\oplus\mathcal{D}_l)\dots )\big);\pi$ for some permutation $\pi$ and where, for each $i$, (A) $\mathcal{M}_i$ is monogamous acyclic and contains precisely the level $i$ hyperedges of $\mathcal{G}$, and (B) $\mathcal{D}_i$ is discrete right-monogamous, contains all order $i+1$ left-amonogamous nodes and all order-$i$ terminal nodes of $\mathcal{G}$.
\end{lem}
\begin{proof}
First, let $u_i$ be the set of order-$i$ terminal nodes of $\mathcal{G}$. We then define the permutation $\pi$ to be a reordering of $w$ into $u_0u_1\dots u_l w'$, where the order-$k$ terminal nodes correspond to $u_k$. 

We can now prove the lemma using induction on the highest order of left-amonogamous nodes in $\mathcal{G};\pi^{-1}$, using Lemma~\ref{lem:level-0} (level-0 decomposition).

For the base case we note that any right-monogamous acyclic cospan without any left-amonogamous nodes is simply monogamous acyclic.

For the induction hypothesis, we assume that the statement holds for all the right-monogamous acyclic cospans with the maximum order of left-amonogamous nodes strictly less than $r$, where $r$ is a positive integer.
For the inductive case, suppose that $\mathcal{G};\pi^{-1}$ is a cospan whose highest order of left-amonogamous nodes is $r$. Then, by Lemma \ref{lem:level-0} (level-0  decomposition), it can be factored into
$\mathcal{G};\pi^{-1}=\mathcal{M}_0;(id_{u_0}\oplus(\mathcal{D}_0;\mathcal{G}'))$
where the first cospan is monogamous acyclic and contains all level 0 hyperedges of $\mathcal{G}$, and $\mathcal{D}_0$ is discrete right monogamous with all order $1$ left-amonogamous nodes of $\mathcal{G};\pi^{-1}$ and $u_0$ all order $0$ terminal nodes of $\mathcal{G};\pi^{-1}$.
Now, every node in $\mathcal{G}'$ corresponding to an order $i$ left-amonogamous node in $\mathcal{G};\pi^{-1}$, is now an order $i-1$ left-amonogamous. Thus, the highest order of left-amonogamous nodes in $\mathcal{G}'$ is $r-1$ and, by the induction hypothesis, $\mathcal{G}'$ can be factored into $\mathcal{M}_1 ; \big(id_{u_1}\oplus(\mathcal{D}_1;\dots ; \mathcal{M}_l;(id_{u_l}\oplus\mathcal{D}_l)\dots )\big)$ as in the statement of the lemma. Therefore, the composite $\mathcal{M}_0 ; \big(id_{u_0}\oplus(\mathcal{D}_0;\dots ; \mathcal{M}_l;(id_{u_l}\oplus\mathcal{D}_l)\dots )\big) =\mathcal{M}_0;(id_{u_0}\oplus (\mathcal{D}_0;\mathcal{G}')) = \mathcal{G};\pi^{-1}$ satisfies conditions \emph{(A)} and \emph{(B)} of the lemma and $\mathcal{M}_0 ; \big(id_{u_0}\oplus(\mathcal{D}_0;\dots ; \mathcal{M}_l;(id_{u_l}\oplus\mathcal{D}_l)\dots )\big);\pi$ is the factorisation we are looking for.
\end{proof}
We will also need the following simpler form of the factorisation into levels, which matches closely the leading intuition of a factorisation into an alternating composition of monogamous and discrete right-monogamous cospans. 
\begin{cor}\label{cor:factorisation-levels}
Any right-monogamous acyclic cospan $\mathcal{G}=\cspobj{v}{}{G}{}{w}$ can be factorised into an alternating sequence of monogamous cospans and discrete right-monogamous cospans, \emph{i.e.}, as $\mathcal{M}_0 ; \mathcal{D}_0;\dots ; \mathcal{M}_l;\mathcal{D}_l$. 

Moreover any two such factorisations differ only by permutations of the terminal nodes of each factor, i.e., if $\mathcal{M}_0 ; \mathcal{D}_0\dots ; \mathcal{M}_l;\mathcal{D}_l=\mathcal{M}'_0 ; \mathcal{D}'_0;\dots ; \mathcal{M}'_l;\mathcal{D}'_l$, there exists permutations $\pi_i,\theta_i$ such that $\mathcal{M}'_i = \mathcal{M}_i;\pi_i$, $\pi\mathcal{D}'_i = \mathcal{D}_i$ and $\mathcal{D}'_i = \mathcal{D}_i;\theta_i$, $\theta_i{M}'_{i+1} = \mathcal{M}_{i+1}$.
\end{cor}
\begin{proof}
Since identities can be seen as monogamous cospans or discrete right-monogamous, and a permutation can be seen as discrete right-monogamous cospan, if we can get a factorisation of $\mathcal{G}$ into levels as in Lemma~\ref{mononandright}, we also obtain a factorisation as in the statement of this lemma.

Finally, we can prove by induction, using the second part of the statement of Lemma~\ref{lem:level-0} that any two such factorisations differ only by some permutation of the factors.
\end{proof}

\noindent We are now able to conclude with our characterisation theorem.

\begin{thm}\label{thm:iso} There exists an isomorphism $\iso{\cdot}:\cprd \rightarrow \rmacsphyp$.
\end{thm}
\begin{proof}
Let us define $\iso{\cdot}$ as copairing of the faithful $\mathcal{C}$-coloured prop morphisms $\sfun : \mathbf{S}_{\Sigma,\mathcal{C}} \rightarrow \rmacsphyp$ and $\cfun: \mathbf{CMon}_{\mathcal{C}} \rightarrow \rmacsphyp$. It suffices to show that the prop $\rmacsphyp$ satisfies the universal property of the coproduct $\cprd$:
\[
\begin{tikzcd}
\mathbf{S}_{\Sigma,\mathcal{C}} \arrow[rr, "\sfun"] \arrow[rrd, "\alpha"'] &  & \rmacsphyp \arrow[d, "\exists ! \gamma", dashed] &  & \mathbf{CMon}_{\mathcal{C}} \arrow[ll, "\cfun"'] \arrow[lld, "\beta"] \\
 &  & \mathbb{A} &  &
\end{tikzcd}
\]
Given a $\mathcal{C}$-coloured prop $\mathbb{A}$ and $\mathcal{C}$-coloured prop morphisms $\alpha: \mathbf{S}_{\Sigma,\mathcal{C}} \rightarrow \mathbb{A}$ , $\beta: \mathbf{CMon}_{\mathcal{C}} \rightarrow \mathbb{A}$, we need to prove there exists a unique prop-morphism $\gamma : \rmacsphyp \rightarrow \mathbb{A}$, such that the diagram above commutes. Now, since coloured prop morphisms are identity-on-objects functors, it is sufficient to consider what happens to the morphisms.
The diagram above needs to commute, so for any morphism $s$ in $\mathbf{S}_{\Sigma,\mathcal{C}}$ and for any morphism $c$ in $\mathbf{CMon}_{\mathcal{C}}$ we want $\gamma(\sfun[s]) = \alpha(s)$ and $\gamma(\cfun[c]) = \beta(c)$. But, by Corollary~\ref{cor:factorisation-levels}, any cospan $\mathcal{G}$ in $\rmacsphyp$ can be factorised as an alternating sequence of monogamous cospans and discrete right-monogamous cospans, \emph{i.e.}, as $\mathcal{G}=\mathcal{M}_0 ; \mathcal{D}_0\dots ; \mathcal{M}_l;\mathcal{D}_l$. Moreover, by Proposition~\ref{prop:monogamous-interpret-faithful}, we can find $s_i$ in $\mathbf{S}_\Sigma$ such that $\mathcal{M}_i=\sfun[s_i]$. Similarly, by Proposition~\ref{prop:right-monogamous-interpret-faithful}, we can find $c_i$ in $\mathbf{CMon}_{\mathcal{C}}$ such that $\mathcal{D}_i=\cfun[c_i]$. Then, to make the diagram above commute, let 
$\gamma(\mathcal{G}) = \gamma\left(\mathcal{M}_0 ; \mathcal{D}_0;\dots ; \mathcal{M}_l;\mathcal{D}_l\right) = \alpha(s_0)\beta(c_0)\dots \alpha(s_l)\beta(c_l)$. This defines $\gamma$ uniquely since both $\sfun$ and $\cfun$ are faithful, once again by Proposition~\ref{prop:monogamous-interpret-faithful} and Proposition~\ref{prop:right-monogamous-interpret-faithful} respectively. We now verify that $\gamma$ is well-defined, functorial, and monoidal.

\textit{Well-definedness.} Since the factorisation of $\mathcal{G}$ into levels is not unique, we need to show that $\gamma$ is well-defined, \emph{i.e.}, that any two such factorisations will define the same value of $\gamma(\mathcal{G})$. Consider another factorisation $\mathcal{G}=\mathcal{M}'_0 ; \mathcal{D}'_0;\dots ; \mathcal{M}'_l;\mathcal{D}'_l$ obtained from Corollary~\ref{cor:factorisation-levels}. Then there exists permutations $\pi_i,\theta_i$ such that $\mathcal{M}'_i = \mathcal{M}_i;\pi_i$, $\pi_i\mathcal{D}'_i = \mathcal{D}_i$ and $\mathcal{D}'_i = \mathcal{D}_i;\theta_i$, $\theta_i\mathcal{M}'_{i+1} = \mathcal{M}_{i+1}$. In addition, $\mathcal{M}'_i=\sfun[s'_i]$ for some $s'_i$ in $\mathbf{S}_\Sigma$, $\mathcal{D}'_i=\cfun[c'_i]$ for some $c'_i$ in $\mathbf{CMon}_{\mathcal{C}}$. To show well-definedness of $\gamma$, we will use the following facts:
\begin{enumerate}
\item since $\mathbf{S}_{\Sigma,\mathcal{C}},\mathbf{CMon}_{\mathcal{C}}$, and $\mathbb{A}$ are $\mathcal{C}$-coloured props, they all contain a copy of the prop of permutations so we will abuse notation slightly  and use the same names to refer to the same permutation in all of them;
\item prop morphisms preserve permutations so that $\alpha(\pi)=\beta(\pi)=\gamma(\pi)=\sfun[\pi]=\cfun[\pi]=\pi$ for any permutation $\pi$;
\item by definition of $\gamma$ it is clear that $\gamma(\mathcal{G};\pi) = \gamma(\mathcal{G});\pi$ and $\gamma(\pi;\mathcal{G}) = \pi;\gamma(\mathcal{G})$.
\end{enumerate}
Now, we have
\begin{align*}
\gamma(\mathcal{M}_0 ; \mathcal{D}_0;\dots ; \mathcal{M}_l;\mathcal{D}_l)&=
\gamma(\mathcal{M}_0) ; \gamma(\mathcal{D}_0);\dots ; \gamma(\mathcal{M}_l);\gamma(\mathcal{D}_l)\\
&= \gamma(\mathcal{M}_0) ; \gamma(\pi_0;\mathcal{D}'_0);\dots ; \gamma(\mathcal{M}_l);\gamma(\pi_l;\mathcal{D}'_l)\\
&= \gamma(\mathcal{M}_0) ; \pi_0;\gamma(\mathcal{D}'_0);\dots ; \gamma(\mathcal{M}_l);\pi_l;\gamma(\mathcal{D}'_l)\\
&= \gamma(\mathcal{M}_0; \pi_0);\gamma(\mathcal{D}'_0);\dots ; \gamma(\mathcal{M}_l;\pi_l);\gamma(\mathcal{D}'_l)\\
&=\gamma(\mathcal{M}'_0);\gamma(\mathcal{D}'_0);\dots ; \gamma(\mathcal{M}'_l);\gamma(\mathcal{D}'_l)
\\
&=\gamma(\mathcal{M}'_0 ; \mathcal{D}'_0;\dots ; \mathcal{M}'_l;\mathcal{D}'_l)
\end{align*}

\textit{Monoidal functoriality.} First, $\gamma$ preserves monoidal products, as the decomposition of a monoidal product is obtained by taking a monoidal product of monogamous acyclic cospans, and a monoidal product of discrete right monogamous cospans, for each level separately.
Second, consider two cospans $\mathcal{G}=\cspobj{w}{}{G}{}{v}$ and $\mathcal{H}=\cspobj{v}{}{G}{}{u}$. We can factorise $\mathcal{H}$ as $\mathcal{M}_0 ; \mathcal{D}_0;\dots ; \mathcal{M}_l;\mathcal{D}_l$. Hence, if we can show that $\gamma(\mathcal{G};\mathcal{M};\mathcal{D})=\gamma(\mathcal{G});\gamma(\mathcal{M};\mathcal{D})$, for $\mathcal{M}$ monogamous and $\mathcal{D}$ discrete right-monogamous, a simple induction will allow us to conclude that $\gamma(\mathcal{G};\mathcal{H})=\gamma(\mathcal{G});\gamma(\mathcal{H})$. In fact, to show the induction step, it is enough to show that $\gamma(\mathcal{G};(\mathcal{M}\oplus id_{u'}))=\gamma(\mathcal{G});\gamma(\mathcal{M}\oplus id_{u'})$ where $\mathcal{M}$ consists of a single hyperedge $h$, with $k$ source nodes and $l$ target nodes such that $|u'| = |u|-k+l$---we can recover the general case of all monogamous cospans by performing another induction on the number of hyperedges in $\mathcal{M}$. 

Now, we need to understand to what level in $\mathcal{G};(\mathcal{M}\oplus id_{u'})$ the single hyperedge $h$ of $\mathcal{M}$ belongs.
By the definition of the level of hyperedges (Definition~\ref{def:levels}), $h$ will belong to level $i$ of in $\mathcal{G};(\mathcal{M}\oplus id_{u'})$ if the node with the largest order in the first $k$ terminal nodes of $\mathcal{G}$ is $i$. If we assume without loss of generality (as we can always post-compose with a permutation to achieve this), that the terminal nodes of $\mathcal{G}$ are ordered by order size, this implies that the factorisation of $\mathcal{G};(\mathcal{M}\oplus id_{u'})$ into levels is $\mathcal{G}_{\leq i};(\mathcal{M}\oplus \mathcal{G}_{>i})$ where $\mathcal{G}_{>i}$ and $\mathcal{G}_{\leq i}$ are obtained from the factorisation $\mathcal{M}_0 ; \big(id_{u_0}\oplus(\mathcal{D}_0;\dots ; \mathcal{M}_l;(id_{u_l}\oplus\mathcal{D}_l)\dots )\big)$ of $\mathcal{G}$ (from Lemma~\ref{mononandright}) as follows: $\mathcal{G}_{\leq i}:= \mathcal{M}_0 ; \big(id_{u_0}\oplus(\mathcal{D}_0;\dots ; \mathcal{M}_i)\dots \big)$ and $\mathcal{G}_{>i}:= \mathcal{D}_i;\mathcal{M}_{i+1} ; \big(id_{u_{i+1}}\oplus(\mathcal{D}_{i+1};\dots ; \mathcal{M}_l;(id_{u_l}\oplus\mathcal{D}_l)\dots )\big)$. Note that, by construction, we have $\mathcal{G}=\mathcal{G}_{\leq i};(id_{u_i} \oplus \mathcal{G}_{>i})$.~Thus
\begin{align*}
\mathcal{G};(\mathcal{M}\oplus id_{u'})&=\mathcal{G}_{\leq i};(id_{u_i} \oplus \mathcal{G}_{>i});(\mathcal{M}\oplus id_{u'})\\
&=\mathcal{G}_{\leq i};\big((id_{u_i};\mathcal{M}) \oplus (\mathcal{G}_{>i};id_{u'})\big)\\
& = \mathcal{G}_{\leq i};(\mathcal{M}\oplus \mathcal{G}_{>i})
\end{align*}
by the interchange and unitality axioms of symmetric monoidal categories (see Fig.~\ref{fig:laws-smc}). 

The intuition now is that we are able to slide the hyperedge $h$ back to level $i$ into the decomposition of $\mathcal{G}$ and that the operation of sliding back---which only uses the monoidal product and composition with identities---is preserved by $\gamma$. This will be sufficient to prove functoriality of $\gamma$. We have
\begin{align*}
\gamma(\mathcal{G};(\mathcal{M}\oplus id_{u'})) &= \gamma(\mathcal{G}_{\leq i};(\mathcal{M}\oplus \mathcal{G}_{>i}))\\
&=\gamma(\mathcal{G}_{\leq i});\gamma(\mathcal{M}\oplus \mathcal{G}_{>i})\\
& = \gamma(\mathcal{G}_{\leq i});\big(\gamma(\mathcal{M})\oplus \gamma(\mathcal{G}_{>i})\big)\\
& = \gamma(\mathcal{G}_{\leq i});\big(\gamma(id_{u_i}) \oplus \gamma(\mathcal{G}_{>i})\big);\big(\gamma(\mathcal{M})\oplus \gamma(id_{u'})\big)
\end{align*}
where the second equality holds because $\mathcal{G}_{\leq i};(\mathcal{M}\oplus \mathcal{G}_{>i})$ is the factorisation of $\mathcal{G};\mathcal{M}$ through which we define $\gamma$; the third equality holds because $\gamma$ preserves monoidal products and the remaining equalities use the interchange and unitality laws of symmetric monoidal categories as above. Finally, by definition of $\gamma$, $\gamma(\mathcal{G}_{\leq i});(\gamma(id_{u_i}) \oplus \gamma(\mathcal{G}_{>i}))=\gamma(\mathcal{G})$ and, since $\gamma$ also preserves monoidal products, we can conclude that $\gamma(\mathcal{G};(\mathcal{M}\oplus id_{u'}))=\gamma(\mathcal{G});\gamma(\mathcal{M}\oplus id_{u'}))$ as we wanted to show.
\end{proof}

\section{Characterisation of String Diagram Rewriting}\label{sec:rewriting}

Now that we have a characterisation theorem for $\cprd$, we are ready to interpret rewriting modulo commutative monoid structure as DPO rewriting, and to show that such a correspondence is sound and complete. 

We first recall the notion of sub-diagram. 
Formally, a sub-diagram $c$ of some larger string diagram $d\from v\to w$ can be defined as a sub-term (modulo the laws of symmetric monoidal categories) of $d$. It is not difficult to show by induction that we can always find some $u\in\Sigma^*$ and diagrams $c_1,c_2$ such that $d=c_1;(id_u\oplus l);c_2$, that is, such that $d$ decomposes as
\begin{equation}\label{eq:sub-diagram}
\begin{tikzpicture}
	\begin{pgfonlayer}{nodelayer}
		\node [style=none] (67) at (3.75, 0) {};
		\node [style=none] (68) at (4.75, 0) {};
		\node [style=none] (69) at (3.75, -1) {};
		\node [style=none] (70) at (4.75, -1) {};
		\node [style=none] (71) at (4.25, -0.5) {$c$};
		\node [style=none] (72) at (2.75, -0.5) {};
		\node [style=none] (73) at (3.75, -0.5) {};
		\node [style=none] (74) at (4.75, -0.5) {};
		\node [style=none] (75) at (5.75, -0.5) {};
		\node [style=none] (76) at (5.75, -1) {};
		\node [style=none] (77) at (6.75, -1) {};
		\node [style=none] (78) at (6.75, 1) {};
		\node [style=none] (79) at (1.75, 1) {};
		\node [style=none] (80) at (1.75, -1) {};
		\node [style=none] (81) at (2.75, -1) {};
		\node [style=none] (82) at (2.75, 1) {};
		\node [style=none] (83) at (5.75, 1) {};
		\node [style=none] (85) at (6.75, 0) {};
		\node [style=none] (86) at (7.75, 0) {};
		\node [style=none] (87) at (1.75, 0) {};
		\node [style=none] (88) at (0.75, 0) {};
		\node [style=none] (89) at (7.5, 0.25) {\scriptsize $w$};
		\node [style=none] (90) at (1, 0.25) {\scriptsize $v$};
		\node [style=none] (91) at (3.25, 0) {\scriptsize $v'$};
		\node [style=none] (92) at (5.25, 0) {\scriptsize $w'$};
		\node [style=none] (93) at (-2.5, 0.75) {};
		\node [style=none] (94) at (-1.5, 0.75) {};
		\node [style=none] (95) at (-2.5, -0.75) {};
		\node [style=none] (96) at (-1.5, -0.75) {};
		\node [style=none] (97) at (-2, 0) {$d$};
		\node [style=none] (98) at (-3.25, 0) {};
		\node [style=none] (99) at (-2.5, 0) {};
		\node [style=none] (100) at (-1.5, 0) {};
		\node [style=none] (101) at (-0.75, 0) {};
		\node [style=none] (102) at (-3, 0.25) {\scriptsize $v$};
		\node [style=none] (103) at (-1, 0.25) {\scriptsize $w$};
		\node [style=none] (104) at (0, 0) {$=$};
		\node [style=none] (105) at (2.75, 0.5) {};
		\node [style=none] (106) at (5.75, 0.5) {};
		\node [style=none] (107) at (4.25, 1) {\scriptsize $u$};
		\node [style=none] (108) at (6.25, 0) {$c_2$};
		\node [style=none] (109) at (2.25, 0) {$c_1$};
	\end{pgfonlayer}
	\begin{pgfonlayer}{edgelayer}
		\draw (67.center) to (68.center);
		\draw (68.center) to (70.center);
		\draw (70.center) to (69.center);
		\draw (69.center) to (67.center);
		\draw (80.center) to (81.center);
		\draw (81.center) to (82.center);
		\draw (83.center) to (76.center);
		\draw (76.center) to (77.center);
		\draw (77.center) to (78.center);
		\draw (79.center) to (80.center);
		\draw (72.center) to (73.center);
		\draw (74.center) to (75.center);
		\draw (88.center) to (87.center);
		\draw (85.center) to (86.center);
		\draw (93.center) to (94.center);
		\draw (94.center) to (96.center);
		\draw (96.center) to (95.center);
		\draw (95.center) to (93.center);
		\draw (98.center) to (99.center);
		\draw (100.center) to (101.center);
		\draw (79.center) to (82.center);
		\draw (78.center) to (83.center);
		\draw (105.center) to (106.center);
	\end{pgfonlayer}
\end{tikzpicture}
\end{equation}
In fact, we could also take this decomposition as a definition of sub-diagrams. 
We can now recall the formal notion of string diagram rewriting.
\begin{defi}[String Diagram Rewriting Modulo Commutative Monoids] \label{def:stringdiagrew}
	Let $d,e \colon v \to w$ and $l, r \colon v' \to w'$ be pairs of morphisms in $\cprd$. We say that $d$ rewrites into $e$ modulo commutative monoid structure according to the rewrite rule $\R = \langle l, r \rangle$, notation $d \Rightarrow_\R e $, if, in $\cprd$, we have:
\begin{equation}\label{eq:stringdiadrew}\begin{tikzpicture}
	\begin{pgfonlayer}{nodelayer}
		\node [style=none] (0) at (2, 0) {};
		\node [style=none] (1) at (3, 0) {};
		\node [style=none] (2) at (2, -1) {};
		\node [style=none] (3) at (3, -1) {};
		\node [style=none] (4) at (2.5, -0.5) {$l$};
		\node [style=none] (5) at (1, -0.5) {};
		\node [style=none] (6) at (2, -0.5) {};
		\node [style=none] (7) at (3, -0.5) {};
		\node [style=none] (8) at (4, -0.5) {};
		\node [style=none] (9) at (4, -1) {};
		\node [style=none] (10) at (5, -1) {};
		\node [style=none] (11) at (5, 1) {};
		\node [style=none] (12) at (0, 1) {};
		\node [style=none] (13) at (0, -1) {};
		\node [style=none] (14) at (1, -1) {};
		\node [style=none] (15) at (1, 1) {};
		\node [style=none] (16) at (4, 1) {};
		\node [style=none] (17) at (5, 0) {};
		\node [style=none] (18) at (6, 0) {};
		\node [style=none] (19) at (0, 0) {};
		\node [style=none] (20) at (-1, 0) {};
		\node [style=none] (21) at (5.75, 0.25) {\scriptsize $w$};
		\node [style=none] (22) at (-0.75, 0.25) {\scriptsize $v$};
		\node [style=none] (23) at (1.5, 0) {\scriptsize $v'$};
		\node [style=none] (24) at (3.5, 0) {\scriptsize $w'$};
		\node [style=none] (25) at (-4.25, 0.75) {};
		\node [style=none] (26) at (-3.25, 0.75) {};
		\node [style=none] (27) at (-4.25, -0.75) {};
		\node [style=none] (28) at (-3.25, -0.75) {};
		\node [style=none] (29) at (-3.75, 0) {$d$};
		\node [style=none] (30) at (-5, 0) {};
		\node [style=none] (31) at (-4.25, 0) {};
		\node [style=none] (32) at (-3.25, 0) {};
		\node [style=none] (33) at (-2.5, 0) {};
		\node [style=none] (34) at (-4.75, 0.25) {\scriptsize $v$};
		\node [style=none] (35) at (-2.75, 0.25) {\scriptsize $w$};
		\node [style=none] (36) at (-1.75, 0) {$=$};
		\node [style=none] (37) at (1, 0.5) {};
		\node [style=none] (38) at (4, 0.5) {};
		\node [style=none] (39) at (2.5, 1) {\scriptsize $u$};
		\node [style=none] (40) at (4.5, 0) {$c_2$};
		\node [style=none] (41) at (0.5, 0) {$c_1$};
	\end{pgfonlayer}
	\begin{pgfonlayer}{edgelayer}
		\draw (0.center) to (1.center);
		\draw (1.center) to (3.center);
		\draw (3.center) to (2.center);
		\draw (2.center) to (0.center);
		\draw (13.center) to (14.center);
		\draw (14.center) to (15.center);
		\draw (16.center) to (9.center);
		\draw (9.center) to (10.center);
		\draw (10.center) to (11.center);
		\draw (12.center) to (13.center);
		\draw (5.center) to (6.center);
		\draw (7.center) to (8.center);
		\draw (20.center) to (19.center);
		\draw (17.center) to (18.center);
		\draw (25.center) to (26.center);
		\draw (26.center) to (28.center);
		\draw (28.center) to (27.center);
		\draw (27.center) to (25.center);
		\draw (30.center) to (31.center);
		\draw (32.center) to (33.center);
		\draw (12.center) to (15.center);
		\draw (11.center) to (16.center);
		\draw (37.center) to (38.center);
	\end{pgfonlayer}
\end{tikzpicture} \qquad \begin{tikzpicture}
	\begin{pgfonlayer}{nodelayer}
		\node [style=none] (0) at (2, 0) {};
		\node [style=none] (1) at (3, 0) {};
		\node [style=none] (2) at (2, -1) {};
		\node [style=none] (3) at (3, -1) {};
		\node [style=none] (4) at (2.5, -0.5) {$r$};
		\node [style=none] (5) at (1, -0.5) {};
		\node [style=none] (6) at (2, -0.5) {};
		\node [style=none] (7) at (3, -0.5) {};
		\node [style=none] (8) at (4, -0.5) {};
		\node [style=none] (9) at (4, -1) {};
		\node [style=none] (10) at (5, -1) {};
		\node [style=none] (11) at (5, 1) {};
		\node [style=none] (12) at (0, 1) {};
		\node [style=none] (13) at (0, -1) {};
		\node [style=none] (14) at (1, -1) {};
		\node [style=none] (15) at (1, 1) {};
		\node [style=none] (16) at (4, 1) {};
		\node [style=none] (17) at (5, 0) {};
		\node [style=none] (18) at (6, 0) {};
		\node [style=none] (19) at (0, 0) {};
		\node [style=none] (20) at (-1, 0) {};
		\node [style=none] (21) at (5.75, 0.25) {\scriptsize $w$};
		\node [style=none] (22) at (-0.75, 0.25) {\scriptsize $v$};
		\node [style=none] (23) at (1.5, 0) {\scriptsize $v'$};
		\node [style=none] (24) at (3.5, 0) {\scriptsize $w'$};
		\node [style=none] (25) at (-4.25, 0.75) {};
		\node [style=none] (26) at (-3.25, 0.75) {};
		\node [style=none] (27) at (-4.25, -0.75) {};
		\node [style=none] (28) at (-3.25, -0.75) {};
		\node [style=none] (29) at (-3.75, 0) {$e$};
		\node [style=none] (30) at (-5, 0) {};
		\node [style=none] (31) at (-4.25, 0) {};
		\node [style=none] (32) at (-3.25, 0) {};
		\node [style=none] (33) at (-2.5, 0) {};
		\node [style=none] (34) at (-4.75, 0.25) {\scriptsize $v$};
		\node [style=none] (35) at (-2.75, 0.25) {\scriptsize $w$};
		\node [style=none] (36) at (-1.75, 0) {$=$};
		\node [style=none] (37) at (1, 0.5) {};
		\node [style=none] (38) at (4, 0.5) {};
		\node [style=none] (39) at (2.5, 1) {\scriptsize $u$};
		\node [style=none] (40) at (4.5, 0) {$c_2$};
		\node [style=none] (41) at (0.5, 0) {$c_1$};
	\end{pgfonlayer}
	\begin{pgfonlayer}{edgelayer}
		\draw (0.center) to (1.center);
		\draw (1.center) to (3.center);
		\draw (3.center) to (2.center);
		\draw (2.center) to (0.center);
		\draw (13.center) to (14.center);
		\draw (14.center) to (15.center);
		\draw (16.center) to (9.center);
		\draw (9.center) to (10.center);
		\draw (10.center) to (11.center);
		\draw (12.center) to (13.center);
		\draw (5.center) to (6.center);
		\draw (7.center) to (8.center);
		\draw (20.center) to (19.center);
		\draw (17.center) to (18.center);
		\draw (25.center) to (26.center);
		\draw (26.center) to (28.center);
		\draw (28.center) to (27.center);
		\draw (27.center) to (25.center);
		\draw (30.center) to (31.center);
		\draw (32.center) to (33.center);
		\draw (12.center) to (15.center);
		\draw (11.center) to (16.center);
		\draw (37.center) to (38.center);
	\end{pgfonlayer}
\end{tikzpicture} 
\end{equation}
\end{defi}


In the plain symmetric monoidal case, not all sub-hypergraphs of the cospan representation of a string diagram $d$ correspond to sub-diagrams of $d$. Those that do have the additional property of being \emph{convex}~\cite{pt2}.
\begin{defi}[Convex sub-hypergraph]
A sub-hypergraph $H \subseteq G$ is \emph{convex} if, for any nodes $v, v'$
in $H$ and any path $p$ from $v$ to $v'$ in $G$, every hyperedge in $p$ is also in $H$.
\end{defi}
\begin{exa}
\label{ex:convexsub}
Taking Example \ref{ex:right-mono-cospan} as reference, the sub-hypergraph on the left below is convex, while the one on the right is not:
\[{\begin{tikzpicture}
	\begin{pgfonlayer}{nodelayer}
		\node [style=small black dot] (13) at (2, 0) {};
		\node [style=none] (14) at (-0.5, 0.5) {};
		\node [style=none] (15) at (-0.5, -0.5) {};
		\node [style=none] (16) at (0.5, -0.5) {};
		\node [style=none] (17) at (0.5, 0.5) {};
		\node [style=none] (18) at (0.5, -0.5) {};
		\node [style=small black dot] (19) at (2, 0) {};
		\node [style=small black dot] (32) at (2, 0) {$c_2$};
		\node [style=none] (33) at (0.5, 0.5) {};
		\node [style=none] (38) at (-0.5, 0) {};
		\node [style=small black dot] (39) at (-1.5, 0) {$c_0$};
		\node [style=none] (64) at (0, 0) {$d$};
	\end{pgfonlayer}
	\begin{pgfonlayer}{edgelayer}
		\draw [style=gray edge] (15.center) to (14.center);
		\draw [style=gray edge, bend left=90, looseness=0.75] (14.center) to (17.center);
		\draw [style=gray edge] (17.center) to (16.center);
		\draw [style=gray edge, in=270, out=270, looseness=0.75] (16.center) to (15.center);
		\draw [style=gray edge, in=-135, out=0, looseness=0.75] (18.center) to (13);
		\draw [style=gray edge, in=150, out=0] (33.center) to (32);
		\draw [style=gray edge] (39) to (38.center);
	\end{pgfonlayer}
\end{tikzpicture}}\qquad \qquad\qquad {\begin{tikzpicture}
	\begin{pgfonlayer}{nodelayer}
		\node [style=small black dot] (13) at (2, 0.75) {};
		\node [style=none] (14) at (-0.5, 1.25) {};
		\node [style=none] (15) at (-0.5, 0.25) {};
		\node [style=none] (16) at (0.5, 0.25) {};
		\node [style=none] (17) at (0.5, 1.25) {};
		\node [style=none] (18) at (0.5, 0.25) {};
		\node [style=small black dot] (19) at (2, 0.75) {};
		\node [style=small black dot] (32) at (2, 0.75) {$c_2$};
		\node [style=none] (33) at (0.5, 1.25) {};
		\node [style=small black dot] (36) at (-1.75, -0.75) {$c_1$};
		\node [style=none] (38) at (-0.5, 0.75) {};
		\node [style=small black dot] (39) at (-1.75, 0.75) {$c_0$};
		\node [style=none] (64) at (0, 0.75) {$d$};
	\end{pgfonlayer}
	\begin{pgfonlayer}{edgelayer}
		\draw [style=gray edge] (15.center) to (14.center);
		\draw [style=gray edge, bend left=90, looseness=0.75] (14.center) to (17.center);
		\draw [style=gray edge] (17.center) to (16.center);
		\draw [style=gray edge, in=270, out=270, looseness=0.75] (16.center) to (15.center);
		\draw [style=gray edge, in=-135, out=0, looseness=0.75] (18.center) to (13);
		\draw [style=gray edge, in=150, out=0] (33.center) to (32);
		\draw [style=gray edge] (39) to (38.center);
	\end{pgfonlayer}
\end{tikzpicture}}\]
\end{exa}

Moreover, unlike in the monogamous case, convex sub-hypergraphs of right-monogamous cospans do not uniquely identify a sub-diagram of the corresponding diagram. This is because specifying a sub-hypergraph does not uniquely fix the legs of the cospan corresponding to the sub-diagram, as the following example illustrates.
\begin{exa}\label{ex:weak-decomposition}
Consider the diagram below ($\col = \{c_0, c_1\}$) with its corresponding cospan representation:
\[\begin{tikzpicture}
	\begin{pgfonlayer}{nodelayer}
		\node [style=none] (229) at (2, 1) {};
		\node [style=none] (230) at (2, -0.75) {};
		\node [style=none] (231) at (3, -0.75) {};
		\node [style=none] (232) at (3, 1) {};
		\node [style=none] (233) at (3, 0) {};
		\node [style=none] (235) at (2, 0.75) {};
		\node [style=none] (236) at (-2.75, 1) {};
		\node [style=none] (237) at (-2.75, -0.75) {};
		\node [style=none] (238) at (-1.75, -0.75) {};
		\node [style=none] (239) at (-1.75, 1) {};
		\node [style=none] (241) at (-2.75, 0) {};
		\node [style=none] (242) at (-1.75, 0.75) {};
		\node [style=none] (245) at (-3.75, 0) {};
		\node [style=none] (246) at (4, 0) {};
		\node [style=none] (250) at (-0.5, -0.25) {};
		\node [style=none] (251) at (-0.5, -1.25) {};
		\node [style=none] (252) at (0.5, -1.25) {};
		\node [style=none] (253) at (0.5, -0.25) {};
		\node [style=none] (254) at (-1.75, 0) {};
		\node [style=none] (256) at (-0.5, -0.75) {};
		\node [style=none] (257) at (-1.75, -0.5) {};
		\node [style=small black dot] (258) at (1.25, -0.25) {};
		\node [style=none] (262) at (0, -0.75) {$d$};
		\node [style=none] (263) at (2, -0.25) {};
		\node [style=none] (264) at (0.5, -0.75) {};
		\node [style=none] (265) at (-3.5, 0.25) {\scriptsize $c_0$};
		\node [style=none] (266) at (0, 1) {\scriptsize $c_0$};
		\node [style=none] (267) at (-1, -0.25) {\scriptsize $c_0$};
		\node [style=none] (268) at (1.75, 0.25) {\scriptsize $c_1$};
		\node [style=none] (269) at (3.75, 0.25) {\scriptsize $c_1$};
		\node [style=none] (270) at (-1.25, 0.25) {\scriptsize $c_1$};
	\end{pgfonlayer}
	\begin{pgfonlayer}{edgelayer}
		\draw (230.center) to (229.center);
		\draw (229.center) to (232.center);
		\draw (232.center) to (231.center);
		\draw (231.center) to (230.center);
		\draw (237.center) to (236.center);
		\draw (236.center) to (239.center);
		\draw (239.center) to (238.center);
		\draw (238.center) to (237.center);
		\draw (245.center) to (241.center);
		\draw (233.center) to (246.center);
		\draw (251.center) to (250.center);
		\draw (250.center) to (253.center);
		\draw (253.center) to (252.center);
		\draw (252.center) to (251.center);
		\draw [in=180, out=0] (257.center) to (256.center);
		\draw [in=135, out=0] (254.center) to (258);
		\draw (258) to (263.center);
		\draw [in=-120, out=0] (264.center) to (258);
		\draw (242.center) to (235.center);
	\end{pgfonlayer}
\end{tikzpicture}\qquad\qquad\begin{tikzpicture}
	\begin{pgfonlayer}{nodelayer}
		\node [style=none] (0) at (-8, 1.25) {};
		\node [style=none] (1) at (-8, -0.75) {};
		\node [style=none] (2) at (-6.5, -0.75) {};
		\node [style=none] (3) at (-6.5, 1.25) {};
		\node [style=none] (4) at (6.5, 1.25) {};
		\node [style=none] (5) at (6.5, -0.75) {};
		\node [style=none] (6) at (8, -0.75) {};
		\node [style=none] (7) at (8, 1.25) {};
		\node [style=small black dot] (8) at (-7.25, 0) {$c_0$};
		\node [style=small black dot] (31) at (7.25, 0) {$c_1$};
		\node [style=none] (42) at (-7.25, 0.75) {$n_0$};
		\node [style=none] (47) at (7.25, 0.75) {$n_0$};
		\node [style=none] (48) at (-6, 0) {};
		\node [style=none] (49) at (-5, 0) {};
		\node [style=none] (50) at (-5, 0) {};
		\node [style=none] (51) at (5, 0) {};
		\node [style=none] (52) at (6, 0) {};
		\node [style=none] (53) at (-6, 0) {};
		\node [style=none] (54) at (-5, 0) {};
		\node [style=none] (77) at (-0.5, -0.5) {};
		\node [style=none] (78) at (-0.5, -1.5) {};
		\node [style=none] (79) at (0.5, -1.5) {};
		\node [style=none] (80) at (0.5, -0.5) {};
		\node [style=none] (84) at (-0.5, -1) {};
		\node [style=small black dot] (85) at (-1.5, -1) {$c_0$};
		\node [style=small black dot] (86) at (1.5, 0) {$c_1$};
		\node [style=none] (87) at (-3.25, 0.5) {};
		\node [style=none] (88) at (-3.25, -0.5) {};
		\node [style=none] (89) at (-2.25, -0.5) {};
		\node [style=none] (90) at (-2.25, 0.5) {};
		\node [style=none] (91) at (-2.25, -0.5) {};
		\node [style=small black dot] (92) at (-0.25, 0.75) {$c_0$};
		\node [style=none] (93) at (-3.25, 0) {};
		\node [style=small black dot] (94) at (-4.25, 0) {$c_0$};
		\node [style=none] (95) at (-4.25, 0.75) {$m_0$};
		\node [style=small black dot] (96) at (4.25, 0) {};
		\node [style=none] (97) at (2.25, 0.5) {};
		\node [style=none] (98) at (2.25, -0.5) {};
		\node [style=none] (99) at (3.25, -0.5) {};
		\node [style=none] (100) at (3.25, 0.5) {};
		\node [style=none] (101) at (3.25, 0) {};
		\node [style=small black dot] (102) at (4.25, 0) {$c_1$};
		\node [style=none] (103) at (2.25, 0) {};
		\node [style=none] (104) at (4.25, 0.75) {$n_0$};
		\node [style=none] (105) at (0.5, -1) {};
		\node [style=none] (106) at (2.25, 0.5) {};
		\node [style=none] (107) at (-2.25, 0.5) {};
		\node [style=none] (108) at (0, -1) {$d$};
		\node [style=none] (109) at (-2.25, 0) {};
	\end{pgfonlayer}
	\begin{pgfonlayer}{edgelayer}
		\draw [style=hadamard edge] (0.center) to (3.center);
		\draw [style=hadamard edge] (2.center) to (3.center);
		\draw [style=hadamard edge] (0.center) to (1.center);
		\draw [style=hadamard edge] (1.center) to (2.center);
		\draw [style=hadamard edge] (4.center) to (7.center);
		\draw [style=hadamard edge] (6.center) to (7.center);
		\draw [style=hadamard edge] (4.center) to (5.center);
		\draw [style=hadamard edge] (5.center) to (6.center);
		\draw [style=pointer edge] (53.center) to (54.center);
		\draw [style=pointer edge] (52.center) to (51.center);
		\draw [style=gray edge] (78.center) to (77.center);
		\draw [style=gray edge, bend left=90, looseness=0.75] (77.center) to (80.center);
		\draw [style=gray edge] (80.center) to (79.center);
		\draw [style=gray edge, in=270, out=270, looseness=0.75] (79.center) to (78.center);
		\draw [style=gray edge] (85) to (84.center);
		\draw [style=gray edge] (88.center) to (87.center);
		\draw [style=gray edge, bend left=90, looseness=0.75] (87.center) to (90.center);
		\draw [style=gray edge] (90.center) to (89.center);
		\draw [style=gray edge, in=270, out=270, looseness=0.75] (89.center) to (88.center);
		\draw [style=gray edge] (94) to (93.center);
		\draw [style=gray edge] (98.center) to (97.center);
		\draw [style=gray edge, bend left=90, looseness=0.75] (97.center) to (100.center);
		\draw [style=gray edge] (100.center) to (99.center);
		\draw [style=gray edge, in=270, out=270, looseness=0.75] (99.center) to (98.center);
		\draw [style=gray edge, in=-180, out=0, looseness=0.50] (101.center) to (96);
		\draw [style=gray edge] (105.center) to (86);
		\draw [style=gray edge] (86) to (103.center);
		\draw [style=gray edge] (92) to (106.center);
		\draw [style=gray edge] (107.center) to (92);
		\draw [style=gray edge] (91.center) to (85);
		\draw [style=gray edge] (109.center) to (86);
	\end{pgfonlayer}
\end{tikzpicture} \]
Given the convex sub-hypergraph $L:=\scalebox{0.9}{\begin{tikzpicture}
	\begin{pgfonlayer}{nodelayer}
		\node [style=none] (77) at (-0.5, 0.5) {};
		\node [style=none] (78) at (-0.5, -0.5) {};
		\node [style=none] (79) at (0.5, -0.5) {};
		\node [style=none] (80) at (0.5, 0.5) {};
		\node [style=none] (84) at (-0.5, 0) {};
		\node [style=small black dot] (85) at (-1.5, 0) {$c_0$};
		\node [style=none] (108) at (0, 0) {$d$};
		\node [style=none] (109) at (0.5, 0) {};
		\node [style=small black dot] (110) at (1.5, 0) {$c_1$};
	\end{pgfonlayer}
	\begin{pgfonlayer}{edgelayer}
		\draw [style=gray edge] (78.center) to (77.center);
		\draw [style=gray edge, bend left=90, looseness=0.75] (77.center) to (80.center);
		\draw [style=gray edge] (80.center) to (79.center);
		\draw [style=gray edge, in=270, out=270, looseness=0.75] (79.center) to (78.center);
		\draw [style=gray edge] (85) to (84.center);
		\draw [style=gray edge] (110) to (109.center);
	\end{pgfonlayer}
\end{tikzpicture}}$, there are several possible choices of cospans, depending on where we attach the second leg of the monoid multiplication that appears in the corresponding string diagram:
\[ \begin{tikzpicture}
	\begin{pgfonlayer}{nodelayer}
		\node [style=none] (0) at (-5.25, 1.25) {};
		\node [style=none] (1) at (-5.25, -0.75) {};
		\node [style=none] (2) at (-3.75, -0.75) {};
		\node [style=none] (3) at (-3.75, 1.25) {};
		\node [style=none] (4) at (3.75, 1.25) {};
		\node [style=none] (5) at (3.75, -0.75) {};
		\node [style=none] (6) at (5.25, -0.75) {};
		\node [style=none] (7) at (5.25, 1.25) {};
		\node [style=small black dot] (8) at (-4.5, 0) {$c_0$};
		\node [style=small black dot] (31) at (4.5, 0) {$c_1$};
		\node [style=none] (42) at (-4.5, 0.75) {$m_0$};
		\node [style=none] (47) at (4.5, 0.75) {$n_0$};
		\node [style=none] (48) at (-3.25, 0) {};
		\node [style=none] (49) at (-2.25, 0) {};
		\node [style=none] (50) at (-2.25, 0) {};
		\node [style=none] (51) at (2.25, 0) {};
		\node [style=none] (52) at (3.25, 0) {};
		\node [style=none] (53) at (-3.25, 0) {};
		\node [style=none] (54) at (-2.25, 0) {};
		\node [style=none] (77) at (-0.5, 0.5) {};
		\node [style=none] (78) at (-0.5, -0.5) {};
		\node [style=none] (79) at (0.5, -0.5) {};
		\node [style=none] (80) at (0.5, 0.5) {};
		\node [style=none] (84) at (-0.5, 0) {};
		\node [style=small black dot] (85) at (-1.5, 0) {$c_0$};
		\node [style=small black dot] (86) at (1.5, 0) {$c_1$};
		\node [style=none] (95) at (-1.5, 0.75) {$m_0$};
		\node [style=none] (104) at (1.5, 0.75) {$n_0$};
		\node [style=none] (105) at (0.5, 0) {};
		\node [style=none] (108) at (0, 0) {$d$};
	\end{pgfonlayer}
	\begin{pgfonlayer}{edgelayer}
		\draw [style=hadamard edge] (0.center) to (3.center);
		\draw [style=hadamard edge] (2.center) to (3.center);
		\draw [style=hadamard edge] (0.center) to (1.center);
		\draw [style=hadamard edge] (1.center) to (2.center);
		\draw [style=hadamard edge] (4.center) to (7.center);
		\draw [style=hadamard edge] (6.center) to (7.center);
		\draw [style=hadamard edge] (4.center) to (5.center);
		\draw [style=hadamard edge] (5.center) to (6.center);
		\draw [style=pointer edge] (53.center) to (54.center);
		\draw [style=pointer edge] (52.center) to (51.center);
		\draw [style=gray edge] (78.center) to (77.center);
		\draw [style=gray edge, bend left=90, looseness=0.75] (77.center) to (80.center);
		\draw [style=gray edge] (80.center) to (79.center);
		\draw [style=gray edge, in=270, out=270, looseness=0.75] (79.center) to (78.center);
		\draw [style=gray edge] (85) to (84.center);
		\draw [style=gray edge] (105.center) to (86);
	\end{pgfonlayer}
\end{tikzpicture} \quad \text{ or }\quad \begin{tikzpicture}
	\begin{pgfonlayer}{nodelayer}
		\node [style=none] (0) at (-6, 1.5) {};
		\node [style=none] (1) at (-6, -1.5) {};
		\node [style=none] (2) at (-3.75, -1.5) {};
		\node [style=none] (3) at (-3.75, 1.5) {};
		\node [style=none] (4) at (3.75, 1.25) {};
		\node [style=none] (5) at (3.75, -0.75) {};
		\node [style=none] (6) at (5.25, -0.75) {};
		\node [style=none] (7) at (5.25, 1.25) {};
		\node [style=small black dot] (8) at (-5.25, -0.75) {$c_0$};
		\node [style=small black dot] (31) at (4.5, 0) {$c_1$};
		\node [style=none] (42) at (-4.5, -0.25) {$m_0$};
		\node [style=none] (47) at (4.5, 0.75) {$n_0$};
		\node [style=none] (48) at (-3.25, 0) {};
		\node [style=none] (49) at (-2.25, 0) {};
		\node [style=none] (50) at (-2.25, 0) {};
		\node [style=none] (51) at (2.25, 0) {};
		\node [style=none] (52) at (3.25, 0) {};
		\node [style=none] (53) at (-3.25, 0) {};
		\node [style=none] (54) at (-2.25, 0) {};
		\node [style=none] (77) at (-0.5, 0.5) {};
		\node [style=none] (78) at (-0.5, -0.5) {};
		\node [style=none] (79) at (0.5, -0.5) {};
		\node [style=none] (80) at (0.5, 0.5) {};
		\node [style=none] (84) at (-0.5, 0) {};
		\node [style=small black dot] (85) at (-1.5, 0) {$c_0$};
		\node [style=small black dot] (86) at (1.5, 0) {$c_1$};
		\node [style=none] (95) at (-1.5, 0.75) {$m_0$};
		\node [style=none] (104) at (1.5, 0.75) {$m_1,n_0$};
		\node [style=none] (105) at (0.5, 0) {};
		\node [style=none] (108) at (0, 0) {$d$};
		\node [style=small black dot] (109) at (-5.25, 0.5) {$c_1$};
		\node [style=none] (110) at (-4.5, 1) {$m_1$};
	\end{pgfonlayer}
	\begin{pgfonlayer}{edgelayer}
		\draw [style=hadamard edge] (0.center) to (3.center);
		\draw [style=hadamard edge] (2.center) to (3.center);
		\draw [style=hadamard edge] (0.center) to (1.center);
		\draw [style=hadamard edge] (1.center) to (2.center);
		\draw [style=hadamard edge] (4.center) to (7.center);
		\draw [style=hadamard edge] (6.center) to (7.center);
		\draw [style=hadamard edge] (4.center) to (5.center);
		\draw [style=hadamard edge] (5.center) to (6.center);
		\draw [style=pointer edge] (53.center) to (54.center);
		\draw [style=pointer edge] (52.center) to (51.center);
		\draw [style=gray edge] (78.center) to (77.center);
		\draw [style=gray edge, bend left=90, looseness=0.75] (77.center) to (80.center);
		\draw [style=gray edge] (80.center) to (79.center);
		\draw [style=gray edge, in=270, out=270, looseness=0.75] (79.center) to (78.center);
		\draw [style=gray edge] (85) to (84.center);
		\draw [style=gray edge] (105.center) to (86);
	\end{pgfonlayer}
\end{tikzpicture}\]
As we have said, this situation differs from the plain symmetric monoidal case~\cite{pt2}, where $\cspobj{v'}{}{L}{}{w'}$ is unique, given $L$. With commutative monoids, the non-uniqueness comes from having to choose whether we include the monoid structure nodes in the cospan  $\cspobj{v'}{}{L}{}{w'}$ or in the context (the surrounding cospans in the decomposition). 
\end{exa}


As studied in~\cite{pt1,pt2}, rewriting of string diagrams may be interpreted as DPO rewriting of the corresponding hypergraphs. The relevant notion is the one of DPO rewriting `with interfaces' (originally used for a single interface in~\cite{EhrigKonig04}, and adapted for two interfaces in~\cite{BonchiGKSZ16}), which ensures preservation of the interfaces described by the cospan structure.

\begin{defi}[DPO Rewriting (with interfaces)] \label{def:DPO}
Consider a DPO rewrite rule $\R = L \xleftarrow{[a_1,a_2]} v'w' \xrightarrow{[b_1,b_2]} R$ given by cospans $v' \xrightarrow{a_1} L \xleftarrow{a_2} w'$ and $v' \xrightarrow{b_1} R \xleftarrow{b_2} w'$ in $\hyp$. 

\noindent\begin{minipage}{.65\textwidth}
We say that the cospan $v \xrightarrow{q_1} D \xleftarrow{q_2} w$ rewrites to $v \xrightarrow{p_1} E \xleftarrow{p_2} w$ with rule $\R$, written $(v \rightarrow D \leftarrow w) \rew_\R (v \xrightarrow{} E \xleftarrow{} w)$,  if there is a cospan $v'w' \to C \leftarrow vw$ (called the pushout complement) making the diagram on the right commute with the two squares being pushouts.
\end{minipage}
\begin{minipage}{.35\textwidth}
\[\xymatrix@R=15pt@C=20pt{
L \ar[d]_{f}   &  v'w' \ar[d]
 \ar@{}[dl]|(.8){\text{\large $\urcorner$}}
 \ar@{}[dr]|(.8){\text{\large $\ulcorner$}}
 \ar[l]_{[a_1,a_2]} \ar[r]^{[b_1,b_2]}  & R \ar[d] \\
 D &  C \ar[l] \ar[r]  & E \\
&  vw \ar[u] \ar[ur]_{[p_1,p_2]}  \ar[ul]^{[q_1,q_2]}
}
\]
\end{minipage}
\end{defi}
\newlength{\origcolumnsep}
\setlength{\origcolumnsep}{\columnsep}

However, unless string diagram rewriting happens modulo the laws of Frobenius algebras, not all DPO rewrites are sound for string diagram rewriting: some pushout complements may yield as outcome of the rewriting hypergraphs that are not in the image of any string diagram~\cite{pt2}. To avoid these situations,~\cite{pt2} introduced the notion of boundary complements and convex matching. The former guarantees  that inputs can only be connected to outputs and vice-versa, while the latter are matches that do not contain directed paths from outputs to inputs, \emph{i.e.}, monomorphisms  whose image is convex. However,
\setlength{\columnsep}{-90pt}
\begin{wrapfigure}[8]{r}[-10pt]{.35\textwidth}  
    \vspace*{-10pt}
    \[ \xymatrix@C=50pt@R=20pt{
            L \ar[d]_f \ar@{}[dr]|{(\dagger)} & {v'w'} \ar[l]_{a = [a_1,a_2]}
                  \ar[d]^{c = [c_1,c_2]}
            \ar@{}[dl]|(.8){\text{\large $\urcorner$}}
             \\
           G & L^\perp \ar[l]^{g} \\
           & vw \ar[ul]^{[b_1,b_2]} \ar@{-->}[u]_{[d_1,d_2]}
        }
    \]
\end{wrapfigure}
these notions were designed for monogamous hypergraphs, and string diagram rewriting modulo symmetric monoidal structure. In order to capture the correct notion of DPO rewriting for right-monogamous hypergraphs, and rewriting modulo commutative monoid structure, we need to relax the first slightly to that of \emph{weak boundary complements}.
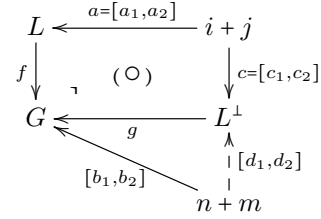
\setlength{\columnsep}{\origcolumnsep}

\begin{defi}[Weak boundary complement]\label{def:weak-boundary-complement}
For right-monogamous acyclic cospans $\cspobj{v'}{a_1}{L}{a_2}{w'}$ and $\cspobj{v}{b_1}{G}{b_2}{w}$ and a morphism $f : L \rightarrow G$, a pushout complement as on the right above is called a weak boundary complement if: (A) given two nodes in $L$ that are mapped  to the same node in $G$ by $f$, they must be in the image of $a_2$; (B) $c_1$ is mono; (C) no two nodes are both in the image of $c_1$ and $c_2$; (D) there exist $d_1: v \rightarrow L^\bot$ and $d_2: w \rightarrow L^\bot$ making the above diagram commute and such that $\cspobj{vw'}{[c_2,d_1]}{L^\bot}{[c_1,d_2]}{wv'}$ is right-monogamous.
\end{defi}
Intuitively, the complement $L^\bot$ is $G$ with an $L$-shaped hole, which we can picture as follows:
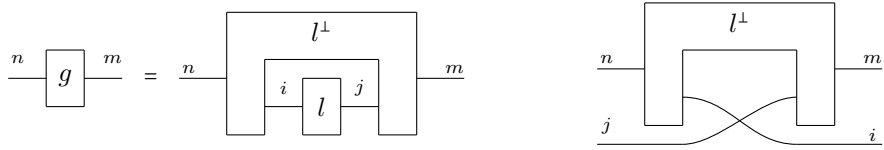
\[\begin{tikzpicture}
	\begin{pgfonlayer}{nodelayer}
		\node [style=none] (67) at (-4.75, 0) {};
		\node [style=none] (68) at (-3.75, 0) {};
		\node [style=none] (69) at (-4.75, -1.5) {};
		\node [style=none] (70) at (-3.75, -1.5) {};
		\node [style=none] (71) at (-4.25, -0.75) {$l$};
		\node [style=none] (72) at (-5.75, -0.75) {};
		\node [style=none] (73) at (-4.75, -0.75) {};
		\node [style=none] (74) at (-3.75, -0.75) {};
		\node [style=none] (75) at (-2.75, -0.75) {};
		\node [style=none] (76) at (-2.75, -1.5) {};
		\node [style=none] (77) at (-1.75, -1.5) {};
		\node [style=none] (78) at (-1.75, 1.75) {};
		\node [style=none] (79) at (-6.75, 1.75) {};
		\node [style=none] (80) at (-6.75, -1.5) {};
		\node [style=none] (81) at (-5.75, -1.5) {};
		\node [style=none] (82) at (-5.75, 0.5) {};
		\node [style=none] (83) at (-2.75, 0.5) {};
		\node [style=none] (84) at (-4.25, 1.25) {$l^\bot$};
		\node [style=none] (85) at (-1.75, 0) {};
		\node [style=none] (86) at (-0.5, 0) {};
		\node [style=none] (87) at (-6.75, 0) {};
		\node [style=none] (88) at (-8, 0) {};
		\node [style=none] (89) at (-0.75, 0.25) {\scriptsize $w$};
		\node [style=none] (90) at (-7.75, 0.25) {\scriptsize $v$};
		\node [style=none] (91) at (-5.25, -0.25) {\scriptsize $v'$};
		\node [style=none] (92) at (-3.25, -0.25) {\scriptsize $w'$};
		\node [style=none] (93) at (-11.5, 0.75) {};
		\node [style=none] (94) at (-10.5, 0.75) {};
		\node [style=none] (95) at (-11.5, -0.75) {};
		\node [style=none] (96) at (-10.5, -0.75) {};
		\node [style=none] (97) at (-11, 0) {$g$};
		\node [style=none] (98) at (-12.5, 0) {};
		\node [style=none] (99) at (-11.5, 0) {};
		\node [style=none] (100) at (-10.5, 0) {};
		\node [style=none] (101) at (-9.5, 0) {};
		\node [style=none] (102) at (-12.25, 0.5) {\scriptsize $v$};
		\node [style=none] (103) at (-9.75, 0.5) {\scriptsize $w$};
		\node [style=none] (104) at (-8.75, 0) {$=$};
		\node [style=none] (110) at (5.25, -0.5) {};
		\node [style=none] (111) at (8.25, -1.75) {};
		\node [style=none] (112) at (5.25, -1.75) {};
		\node [style=none] (113) at (8.25, -0.5) {};
		\node [style=none] (114) at (8.25, -1.25) {};
		\node [style=none] (115) at (9.25, -1.25) {};
		\node [style=none] (116) at (9.25, 2) {};
		\node [style=none] (117) at (4.25, 2) {};
		\node [style=none] (118) at (4.25, -1.25) {};
		\node [style=none] (119) at (5.25, -1.25) {};
		\node [style=none] (120) at (5.25, 0.75) {};
		\node [style=none] (121) at (8.25, 0.75) {};
		\node [style=none] (122) at (6.75, 1.5) {$l^\bot$};
		\node [style=none] (123) at (9.25, 0.25) {};
		\node [style=none] (124) at (10.5, 0.25) {};
		\node [style=none] (125) at (4.25, 0.25) {};
		\node [style=none] (126) at (3, 0.25) {};
		\node [style=none] (127) at (10.25, 0.5) {\scriptsize $w$};
		\node [style=none] (128) at (3.25, 0.5) {\scriptsize $v$};
		\node [style=none] (129) at (10.25, -1.25) {\scriptsize $v'$};
		\node [style=none] (130) at (3.25, -1.25) {\scriptsize $w'$};
		\node [style=none] (131) at (10.5, -1.75) {};
		\node [style=none] (132) at (3, -1.75) {};
	\end{pgfonlayer}
	\begin{pgfonlayer}{edgelayer}
		\draw (67.center) to (68.center);
		\draw (68.center) to (70.center);
		\draw (70.center) to (69.center);
		\draw (69.center) to (67.center);
		\draw (80.center) to (81.center);
		\draw (81.center) to (82.center);
		\draw (82.center) to (83.center);
		\draw (83.center) to (76.center);
		\draw (76.center) to (77.center);
		\draw (77.center) to (78.center);
		\draw (78.center) to (79.center);
		\draw (79.center) to (80.center);
		\draw (72.center) to (73.center);
		\draw (74.center) to (75.center);
		\draw (88.center) to (87.center);
		\draw (85.center) to (86.center);
		\draw (93.center) to (94.center);
		\draw (94.center) to (96.center);
		\draw (96.center) to (95.center);
		\draw (95.center) to (93.center);
		\draw (98.center) to (99.center);
		\draw (100.center) to (101.center);
		\draw (118.center) to (119.center);
		\draw (119.center) to (120.center);
		\draw (120.center) to (121.center);
		\draw (121.center) to (114.center);
		\draw (114.center) to (115.center);
		\draw (115.center) to (116.center);
		\draw (116.center) to (117.center);
		\draw (117.center) to (118.center);
		\draw [in=180, out=0] (110.center) to (111.center);
		\draw [in=-180, out=0, looseness=0.75] (112.center) to (113.center);
		\draw (126.center) to (125.center);
		\draw (123.center) to (124.center);
		\draw (132.center) to (112.center);
		\draw (111.center) to (131.center);
	\end{pgfonlayer}
\end{tikzpicture}\]
where $g:v\to w$, $l: v' \to w'$, and $l^\bot : vw'\to wv'$ are diagrams for the cospans $\cspobj{v}{}{G}{}{w}$,$\cspobj{v'}{}{L}{}{w'}$, and $\cspobj{vw'}{}{L^\bot}{}{wv'}$ respectively, \emph{i.e.} such that $\iso{g}=(\cspobj{v}{}{G}{}{w})$,$\iso{l}=\cspobj{v'}{}{L}{}{w'}$, and $\iso{l^\bot}=\cspobj{vw'}{}{L^\bot}{}{wv'}$. (Recall that $\iso{\cdot} \colon \cprd \rightarrow \rmacsphyp$ is the isomorphism established by Theorem~\ref{thm:iso}; we will use it quite liberally from now on in order to manipulate cospans as string diagrams when convenient).
Boundary complements restrict the shape that these can take. Let us explain the conditions of Definition~\ref{def:weak-boundary-complement} in plainer language.
\begin{itemize}
\item Condition \emph{(A)} allows matches to occur in a diagram $G$ that contains the sub-diagram $L$ potentially with some nodes identified, \emph{i.e.} wires connected by the monoid multiplication (see Example~\ref{ex:match} below. However, these can only occur as terminal nodes, that is, in the image of $a_2$, the right boundary of the subdiagram $L$.
\item Plain boundary complements~\cite{pt2} require $c_1,c_2$ to be jointly monic. This enforces two distinct properties: it prevents nodes from the left and right boundaries of the match to be identified, and it prevents nodes from within each of the two boundary sides to be identified. Here, we need to relax the second condition to allow nodes in the right boundary of the match to be identified. This is what conditions \emph{(B)} and \emph{(C)} give us.
\item Condition \emph{(D)} forces the boundary of the complement, both with the subdiagram $L$ and those of the larger diagram $G$, to be right-monogamous. In other words, we want the cospan $\cspobj{vw'}{[c_2,d_1]}{L^\bot}{[c_1,d_2]}{wv'}$  depicted above to be right-monogamous.
\end{itemize}


The last ingredient we require is the same as in~\cite{pt2}: we require the match to be \emph{convex}. 

\begin{defi}[Convex matching~\cite{pt2}]
$f : L\rightarrow G$ in $\hyp$ is a convex  match if its image is a convex sub-hypergraph of G.
\end{defi}

\begin{exa}\label{ex:match}
Consider the diagram below
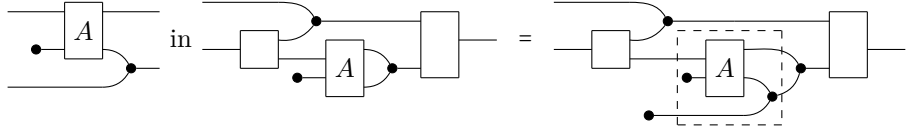
\[\begin{tikzpicture}
	\begin{pgfonlayer}{nodelayer}
		\node [style=none] (296) at (-0.5, 1) {};
		\node [style=none] (297) at (-0.5, -0.5) {};
		\node [style=none] (298) at (0.5, -0.5) {};
		\node [style=none] (299) at (0.5, 1) {};
		\node [style=none] (300) at (0.5, -0.25) {};
		\node [style=none] (301) at (-0.5, -0.25) {};
		\node [style=none] (302) at (-0.5, 0.75) {};
		\node [style=none] (303) at (-2, 0.75) {};
		\node [style=small black dot] (304) at (1.25, -0.75) {};
		\node [style=none] (306) at (2, 0.75) {};
		\node [style=none] (307) at (0, 0.25) {$d$};
		\node [style=none] (309) at (0.5, -1.25) {};
		\node [style=small black dot] (311) at (-1.25, -0.25) {};
		\node [style=none] (312) at (-2, -1.25) {};
		\node [style=none] (314) at (0.5, 0.75) {};
		\node [style=none] (315) at (2, -0.75) {};
		\node [style=none] (316) at (-1.75, 1) {\scriptsize $c_0$};
		\node [style=none] (317) at (1.75, 1) {\scriptsize $c_1$};
		\node [style=none] (318) at (1.75, -0.5) {\scriptsize $c_1$};
		\node [style=none] (319) at (-1.75, -1) {\scriptsize $c_1$};
		\node [style=none] (320) at (-1, 0.25) {\scriptsize $c_0$};
	\end{pgfonlayer}
	\begin{pgfonlayer}{edgelayer}
		\draw (297.center) to (296.center);
		\draw (296.center) to (299.center);
		\draw (299.center) to (298.center);
		\draw (298.center) to (297.center);
		\draw [in=180, out=0] (303.center) to (302.center);
		\draw [in=120, out=0] (300.center) to (304);
		\draw [in=-120, out=0] (309.center) to (304);
		\draw (311) to (301.center);
		\draw (312.center) to (309.center);
		\draw (314.center) to (306.center);
		\draw (304) to (315.center);
	\end{pgfonlayer}
\end{tikzpicture}\quad \text{ in }\quad \begin{tikzpicture}
	\begin{pgfonlayer}{nodelayer}
		\node [style=none] (229) at (-2.75, 0.75) {};
		\node [style=none] (230) at (-2.75, -1) {};
		\node [style=none] (231) at (-1.75, -1) {};
		\node [style=none] (232) at (-1.75, 0.75) {};
		\node [style=none] (233) at (-1.75, 0) {};
		\node [style=none] (235) at (-2.75, 0.5) {};
		\node [style=none] (236) at (-7.75, 0.25) {};
		\node [style=none] (237) at (-7.75, -0.75) {};
		\node [style=none] (238) at (-6.75, -0.75) {};
		\node [style=none] (239) at (-6.75, 0.25) {};
		\node [style=none] (241) at (-7.75, -0.25) {};
		\node [style=none] (242) at (-6.75, 0) {};
		\node [style=none] (245) at (-8.75, -0.25) {};
		\node [style=none] (246) at (-0.75, 0) {};
		\node [style=none] (250) at (-5.25, 0) {};
		\node [style=none] (251) at (-5.25, -1.5) {};
		\node [style=none] (252) at (-4.25, -1.5) {};
		\node [style=none] (253) at (-4.25, 0) {};
		\node [style=none] (254) at (-4.25, -0.25) {};
		\node [style=none] (255) at (-5.25, -1.25) {};
		\node [style=none] (256) at (-5.25, -0.5) {};
		\node [style=none] (257) at (-6.75, -0.5) {};
		\node [style=small black dot] (258) at (-3.5, -0.75) {};
		\node [style=none] (262) at (-4.75, -0.75) {$d$};
		\node [style=none] (263) at (-2.75, -0.75) {};
		\node [style=none] (264) at (-4.25, -1.25) {};
		\node [style=small black dot] (266) at (-6.25, -1.25) {};
		\node [style=none] (267) at (0, 0) {$=$};
		\node [style=none] (268) at (8.5, 0.75) {};
		\node [style=none] (269) at (8.5, -1) {};
		\node [style=none] (270) at (9.5, -1) {};
		\node [style=none] (271) at (9.5, 0.75) {};
		\node [style=none] (272) at (9.5, 0) {};
		\node [style=none] (273) at (8.5, 0.5) {};
		\node [style=none] (280) at (6, 0.5) {};
		\node [style=none] (282) at (10.5, 0) {};
		\node [style=none] (296) at (5.25, 0) {};
		\node [style=none] (297) at (5.25, -1.5) {};
		\node [style=none] (298) at (6.25, -1.5) {};
		\node [style=none] (299) at (6.25, 0) {};
		\node [style=none] (300) at (6.25, -1) {};
		\node [style=none] (301) at (5.25, -1.25) {};
		\node [style=small black dot] (304) at (7, -1.5) {};
		\node [style=none] (306) at (6.25, -0.25) {};
		\node [style=none] (307) at (5.75, -0.75) {$d$};
		\node [style=none] (309) at (6.25, -2) {};
		\node [style=small black dot] (311) at (4.25, -1.25) {};
		\node [style=small black dot] (312) at (2.75, -2) {};
		\node [style=small black dot] (313) at (7.75, -0.75) {};
		\node [style=none] (315) at (8.5, -0.75) {};
		\node [style=small black dot] (316) at (-5.75, 0.5) {};
		\node [style=none] (317) at (-6.75, 1) {};
		\node [style=none] (318) at (-8.75, 1) {};
		\node [style=none] (319) at (1.75, 0.25) {};
		\node [style=none] (320) at (1.75, -0.75) {};
		\node [style=none] (321) at (2.75, -0.75) {};
		\node [style=none] (322) at (2.75, 0.25) {};
		\node [style=none] (323) at (1.75, -0.25) {};
		\node [style=none] (324) at (2.75, 0) {};
		\node [style=none] (325) at (0.75, -0.25) {};
		\node [style=none] (326) at (5.25, -0.5) {};
		\node [style=none] (327) at (2.75, -0.5) {};
		\node [style=small black dot] (328) at (3.75, 0.5) {};
		\node [style=none] (329) at (2.75, 1) {};
		\node [style=none] (330) at (0.75, 1) {};
		\node [style=none] (331) at (4, 0.25) {};
		\node [style=none] (332) at (4, -2.25) {};
		\node [style=none] (333) at (7.25, -2.25) {};
		\node [style=none] (334) at (7.25, 0.25) {};
		\node [style=none] (335) at (-5.75, -0.25) {\scriptsize $c_0$};
		\node [style=none] (336) at (-5.75, -1) {\scriptsize $c_0$};
		\node [style=none] (337) at (-3.25, -0.25) {\scriptsize $c_1$};
		\node [style=none] (338) at (-8.5, 1.25) {\scriptsize $c_1$};
		\node [style=none] (339) at (-8.5, 0) {\scriptsize $c_0$};
		\node [style=none] (340) at (-1, 0.25) {\scriptsize $c_0$};
		\node [style=none] (341) at (-3.5, 0.75) {\scriptsize $c_1$};
		\node [style=none] (342) at (4.75, -0.25) {\scriptsize $c_0$};
		\node [style=none] (343) at (4.75, -1) {\scriptsize $c_0$};
		\node [style=none] (344) at (3.25, -1.75) {\scriptsize $c_1$};
		\node [style=none] (345) at (8, -0.25) {\scriptsize $c_1$};
		\node [style=none] (346) at (10.25, 0.25) {\scriptsize $c_0$};
		\node [style=none] (347) at (1, 1.25) {\scriptsize $c_1$};
		\node [style=none] (348) at (1, 0) {\scriptsize $c_0$};
		\node [style=none] (349) at (7.75, 0.75) {\scriptsize $c_1$};
	\end{pgfonlayer}
	\begin{pgfonlayer}{edgelayer}
		\draw (230.center) to (229.center);
		\draw (229.center) to (232.center);
		\draw (232.center) to (231.center);
		\draw (231.center) to (230.center);
		\draw (237.center) to (236.center);
		\draw (236.center) to (239.center);
		\draw (239.center) to (238.center);
		\draw (238.center) to (237.center);
		\draw (245.center) to (241.center);
		\draw (233.center) to (246.center);
		\draw (251.center) to (250.center);
		\draw (250.center) to (253.center);
		\draw (253.center) to (252.center);
		\draw (252.center) to (251.center);
		\draw [in=180, out=0] (257.center) to (256.center);
		\draw [in=120, out=0] (254.center) to (258);
		\draw (258) to (263.center);
		\draw [in=-120, out=0] (264.center) to (258);
		\draw (266) to (255.center);
		\draw (269.center) to (268.center);
		\draw (268.center) to (271.center);
		\draw (271.center) to (270.center);
		\draw (270.center) to (269.center);
		\draw [in=180, out=0] (280.center) to (273.center);
		\draw (272.center) to (282.center);
		\draw (297.center) to (296.center);
		\draw (296.center) to (299.center);
		\draw (299.center) to (298.center);
		\draw (298.center) to (297.center);
		\draw [in=120, out=0] (300.center) to (304);
		\draw [in=-120, out=0] (309.center) to (304);
		\draw (311) to (301.center);
		\draw (312) to (309.center);
		\draw [bend right] (304) to (313);
		\draw [in=120, out=0] (306.center) to (313);
		\draw (313) to (315.center);
		\draw [bend right] (242.center) to (316);
		\draw (316) to (235.center);
		\draw [in=120, out=0] (317.center) to (316);
		\draw (318.center) to (317.center);
		\draw (320.center) to (319.center);
		\draw (319.center) to (322.center);
		\draw (322.center) to (321.center);
		\draw (321.center) to (320.center);
		\draw (325.center) to (323.center);
		\draw [in=180, out=0] (327.center) to (326.center);
		\draw [bend right] (324.center) to (328);
		\draw [in=120, out=0] (329.center) to (328);
		\draw (330.center) to (329.center);
		\draw (328) to (280.center);
		\draw [dashed] (331.center) to (334.center);
		\draw [dashed] (334.center) to (333.center);
		\draw [dashed] (333.center) to (332.center);
		\draw [dashed] (332.center) to (331.center);
	\end{pgfonlayer}
\end{tikzpicture}\]
As cospans of hypergraphs, this corresponds to the convex matching below:
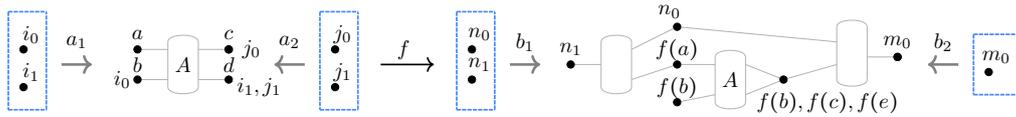
\[
\begin{tikzpicture}
	\begin{pgfonlayer}{nodelayer}
		\node [style=small black dot] (13) at (-4.5, -0.5) {$c_1$};
		\node [style=none] (14) at (-6.5, 0.5) {};
		\node [style=none] (15) at (-6.5, -0.5) {};
		\node [style=none] (16) at (-5.5, -0.5) {};
		\node [style=none] (17) at (-5.5, 0.5) {};
		\node [style=none] (18) at (-5.5, -0.5) {};
		\node [style=none] (38) at (-6.5, 0.5) {};
		\node [style=small black dot] (39) at (-7.5, 0.5) {$c_0$};
		\node [style=none] (96) at (-2.5, 0) {};
		\node [style=none] (97) at (-0.75, 0) {};
		\node [style=none] (103) at (5.5, -0.25) {};
		\node [style=none] (104) at (5.5, -1.25) {};
		\node [style=none] (105) at (6.5, -1.25) {};
		\node [style=none] (106) at (6.5, -0.25) {};
		\node [style=none] (113) at (6.5, -1.25) {};
		\node [style=small black dot] (115) at (4.25, -2.25) {$c_0$};
		\node [style=none] (116) at (5.5, -1.25) {};
		\node [style=none] (117) at (5.5, -0.25) {};
		\node [style=small black dot] (118) at (4.25, -0.25) {$c_0$};
		\node [style=small black dot] (130) at (7.75, -0.5) {$c_1$};
		\node [style=none] (151) at (1.75, 0.5) {};
		\node [style=none] (152) at (1.75, -0.5) {};
		\node [style=none] (153) at (2.75, -0.5) {};
		\node [style=none] (154) at (2.75, 0.5) {};
		\node [style=none] (155) at (2.75, -0.5) {};
		\node [style=small black dot] (156) at (4.25, 1.5) {$c_1$};
		\node [style=none] (157) at (1.75, 0) {};
		\node [style=small black dot] (158) at (0.75, 0) {$c_0$};
		\node [style=small black dot] (162) at (-4.5, 0.5) {};
		\node [style=none] (163) at (-5.5, 0.5) {};
		\node [style=small black dot] (164) at (-4.5, 0.5) {$c_1$};
		\node [style=small black dot] (168) at (11.5, 0.25) {};
		\node [style=none] (169) at (9.5, 0.5) {};
		\node [style=none] (170) at (9.5, -0.5) {};
		\node [style=none] (171) at (10.5, -0.5) {};
		\node [style=none] (172) at (10.5, 0.5) {};
		\node [style=none] (173) at (10.5, 0.25) {};
		\node [style=small black dot] (174) at (11.5, 0.25) {$c_0$};
		\node [style=none] (175) at (9.5, -0.25) {};
		\node [style=none] (181) at (6.5, -0.25) {};
		\node [style=none] (182) at (9.5, 0.5) {};
		\node [style=none] (183) at (2.75, 0.5) {};
		\node [style=none] (186) at (-6, 0) {$d$};
		\node [style=none] (189) at (6, -0.75) {$d$};
		\node [style=none] (191) at (-8.25, 1) {$i_0$};
		\node [style=none] (193) at (-3.75, 1) {$j_0$};
		\node [style=none] (194) at (-3.75, 0) {$j_1$};
		\node [style=none] (196) at (-1.75, 0.5) {$f$};
		\node [style=none] (197) at (4.25, 0.5) {$f(i_0)$};
		\node [style=none] (198) at (4.25, -1.25) {$f(i_1)$};
		\node [style=none] (200) at (9.45, -1.5) {$f(i_1),f(j_0),f(j_1)$};
		\node [style=none] (307) at (-6.5, -0.5) {};
		\node [style=small black dot] (308) at (-7.5, -0.5) {$c_0$};
		\node [style=none] (309) at (-8.25, 0) {$i_1$};
	\end{pgfonlayer}
	\begin{pgfonlayer}{edgelayer}
		\draw [style=gray edge] (15.center) to (14.center);
		\draw [style=gray edge, bend left=90, looseness=0.75] (14.center) to (17.center);
		\draw [style=gray edge] (17.center) to (16.center);
		\draw [style=gray edge, in=270, out=270, looseness=0.75] (16.center) to (15.center);
		\draw [style=gray edge, in=-180, out=0, looseness=0.50] (18.center) to (13);
		\draw [style=gray edge] (39) to (38.center);
		\draw [style=gray edge] (104.center) to (103.center);
		\draw [style=gray edge, bend left=90, looseness=0.75] (103.center) to (106.center);
		\draw [style=gray edge] (106.center) to (105.center);
		\draw [style=gray edge, in=270, out=270, looseness=0.75] (105.center) to (104.center);
		\draw [style=gray edge] (115) to (116.center);
		\draw [style=gray edge] (118) to (117.center);
		\draw [style=gray edge] (152.center) to (151.center);
		\draw [style=gray edge, bend left=90, looseness=0.75] (151.center) to (154.center);
		\draw [style=gray edge] (154.center) to (153.center);
		\draw [style=gray edge, in=270, out=270, looseness=0.75] (153.center) to (152.center);
		\draw [style=gray edge] (158) to (157.center);
		\draw [style=gray edge, in=-180, out=0, looseness=0.50] (163.center) to (162);
		\draw [style=gray edge] (170.center) to (169.center);
		\draw [style=gray edge, bend left=90, looseness=0.75] (169.center) to (172.center);
		\draw [style=gray edge] (172.center) to (171.center);
		\draw [style=gray edge, in=270, out=270, looseness=0.75] (171.center) to (170.center);
		\draw [style=gray edge, in=-180, out=0, looseness=0.50] (173.center) to (168);
		\draw [style=gray edge] (181.center) to (130);
		\draw [style=gray edge] (130) to (175.center);
		\draw [style=gray edge] (156) to (182.center);
		\draw [style=gray edge] (183.center) to (156);
		\draw [style=gray edge] (155.center) to (118);
		\draw [style=gray edge] (113.center) to (130);
		\draw [thick, ->] (96.center) to (97.center);
		\draw [style=gray edge] (308) to (307.center);
	\end{pgfonlayer}
\end{tikzpicture}\]
with the following weak boundary complement:
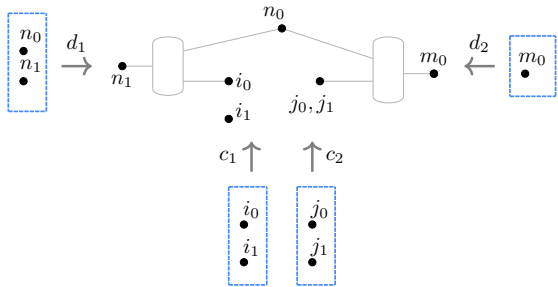
\[
\scalebox{0.9}{\begin{tikzpicture}
	\begin{pgfonlayer}{nodelayer}
		\node [style=none] (201) at (7.5, 2.75) {};
		\node [style=none] (202) at (7.5, 0.75) {};
		\node [style=none] (203) at (9.25, 0.75) {};
		\node [style=none] (204) at (9.25, 2.75) {};
		\node [style=small black dot] (214) at (-1.75, 0) {$c_0$};
		\node [style=small black dot] (217) at (-1.75, 1.25) {$c_0$};
		\node [style=none] (218) at (-7.25, 1.75) {};
		\node [style=none] (219) at (-6.25, 1.75) {};
		\node [style=none] (220) at (-6.25, 1.75) {};
		\node [style=none] (221) at (6, 1.75) {};
		\node [style=none] (222) at (7, 1.75) {};
		\node [style=none] (223) at (-7.25, 1.75) {};
		\node [style=none] (224) at (-6.25, 1.75) {};
		\node [style=small black dot] (225) at (1.25, 1.25) {$c_1$};
		\node [style=none] (226) at (-9.25, 3.5) {};
		\node [style=none] (227) at (-9.25, 0.25) {};
		\node [style=none] (228) at (-7.75, 0.25) {};
		\node [style=none] (229) at (-7.75, 3.5) {};
		\node [style=small black dot] (230) at (-8.75, 2.25) {$c_1$};
		\node [style=small black dot] (231) at (-8.75, 1) {$c_0$};
		\node [style=small black dot] (232) at (8.25, 1.5) {$c_0$};
		\node [style=none] (234) at (-8.25, 1.5) {};
		\node [style=none] (235) at (-8.25, 2.75) {$n_0$};
		\node [style=none] (236) at (-8.25, 1.5) {$n_1$};
		\node [style=none] (237) at (8.5, 2) {$m_0$};
		\node [style=none] (239) at (-7.25, 1.75) {};
		\node [style=none] (240) at (-6.25, 1.75) {};
		\node [style=none] (241) at (-6.25, 1.75) {};
		\node [style=none] (242) at (-4.25, 2.25) {};
		\node [style=none] (243) at (-4.25, 1.25) {};
		\node [style=none] (244) at (-3.25, 1.25) {};
		\node [style=none] (245) at (-3.25, 2.25) {};
		\node [style=none] (246) at (-3.25, 1.25) {};
		\node [style=small black dot] (247) at (0, 3) {$c_1$};
		\node [style=none] (248) at (-4.25, 1.75) {};
		\node [style=small black dot] (249) at (-5.25, 1.75) {$c_0$};
		\node [style=none] (250) at (-0.5, 3.5) {$n_0$};
		\node [style=small black dot] (251) at (5, 1.5) {};
		\node [style=none] (252) at (3, 2.25) {};
		\node [style=none] (253) at (3, 1) {};
		\node [style=none] (254) at (4, 1) {};
		\node [style=none] (255) at (4, 2.25) {};
		\node [style=none] (256) at (4, 1.5) {};
		\node [style=small black dot] (257) at (5, 1.5) {$c_0$};
		\node [style=none] (258) at (3, 1.25) {};
		\node [style=none] (259) at (5, 2.25) {$m_0$};
		\node [style=none] (263) at (3, 2) {};
		\node [style=none] (264) at (-3.25, 2.25) {};
		\node [style=none] (266) at (-5.25, 2.5) {$n_1$};
		\node [style=none] (267) at (-2, -2.25) {};
		\node [style=none] (268) at (-2, -5.5) {};
		\node [style=none] (269) at (-0.5, -5.5) {};
		\node [style=none] (270) at (-0.5, -2.25) {};
		\node [style=small black dot] (271) at (-1.5, -3.5) {$c_0$};
		\node [style=small black dot] (272) at (-1.5, -4.75) {$c_0$};
		\node [style=none] (273) at (-1, -4.25) {};
		\node [style=none] (274) at (-1, -3) {$i_0$};
		\node [style=none] (275) at (-1, -4.25) {$i_1$};
		\node [style=none] (289) at (-1, -1.75) {};
		\node [style=none] (290) at (-1, -0.75) {};
		\node [style=none] (291) at (1, -1.75) {};
		\node [style=none] (292) at (1, -0.75) {};
		\node [style=none] (297) at (-6.75, 2.5) {$d_1$};
		\node [style=none] (298) at (6.5, 2.5) {$d_2$};
		\node [style=none] (299) at (-1.75, -1.25) {};
		\node [style=none] (300) at (-1.75, -1.25) {$c_1$};
		\node [style=none] (301) at (1.75, -1.25) {$c_2$};
		\node [style=none] (302) at (-1, 0.25) {$i_1$};
		\node [style=none] (303) at (-1, 1.5) {$i_0$};
		\node [style=none] (304) at (1, 0.5) {$j_0,j_1$};
		\node [style=none] (305) at (0.5, -2.25) {};
		\node [style=none] (306) at (0.5, -5.5) {};
		\node [style=none] (307) at (2, -5.5) {};
		\node [style=none] (308) at (2, -2.25) {};
		\node [style=small black dot] (309) at (1, -3.5) {$c_1$};
		\node [style=small black dot] (310) at (1, -4.75) {$c_1$};
		\node [style=none] (311) at (1.5, -4.25) {};
		\node [style=none] (312) at (1.5, -3) {$j_0$};
		\node [style=none] (313) at (1.5, -4.25) {$j_1$};
	\end{pgfonlayer}
	\begin{pgfonlayer}{edgelayer}
		\draw [style=hadamard edge] (201.center) to (204.center);
		\draw [style=hadamard edge] (203.center) to (204.center);
		\draw [style=hadamard edge] (201.center) to (202.center);
		\draw [style=hadamard edge] (202.center) to (203.center);
		\draw [style=pointer edge] (223.center) to (224.center);
		\draw [style=pointer edge] (222.center) to (221.center);
		\draw [style=hadamard edge] (226.center) to (229.center);
		\draw [style=hadamard edge] (228.center) to (229.center);
		\draw [style=hadamard edge] (226.center) to (227.center);
		\draw [style=hadamard edge] (227.center) to (228.center);
		\draw [style=gray edge] (243.center) to (242.center);
		\draw [style=gray edge, bend left=90, looseness=0.75] (242.center) to (245.center);
		\draw [style=gray edge] (245.center) to (244.center);
		\draw [style=gray edge, in=270, out=270, looseness=0.75] (244.center) to (243.center);
		\draw [style=gray edge] (249) to (248.center);
		\draw [style=gray edge] (253.center) to (252.center);
		\draw [style=gray edge, bend left=90, looseness=0.75] (252.center) to (255.center);
		\draw [style=gray edge] (255.center) to (254.center);
		\draw [style=gray edge, in=270, out=270, looseness=0.75] (254.center) to (253.center);
		\draw [style=gray edge, in=-180, out=0, looseness=0.50] (256.center) to (251);
		\draw [style=gray edge] (225) to (258.center);
		\draw [style=gray edge] (247) to (263.center);
		\draw [style=gray edge] (264.center) to (247);
		\draw [style=gray edge] (246.center) to (217);
		\draw [style=hadamard edge] (267.center) to (270.center);
		\draw [style=hadamard edge] (269.center) to (270.center);
		\draw [style=hadamard edge] (267.center) to (268.center);
		\draw [style=hadamard edge] (268.center) to (269.center);
		\draw [style=pointer edge] (289.center) to (290.center);
		\draw [style=pointer edge] (291.center) to (292.center);
		\draw [style=hadamard edge] (305.center) to (308.center);
		\draw [style=hadamard edge] (307.center) to (308.center);
		\draw [style=hadamard edge] (305.center) to (306.center);
		\draw [style=hadamard edge] (306.center) to (307.center);
	\end{pgfonlayer}
\end{tikzpicture}}\]
\end{exa}
Note that, contrary to boundary complements in the symmetric monoidal case\linebreak[4]\cite{pt2}, weak boundary complements are not necessarily unique if they exist.
\begin{defi}[Weakly Convex DPO Rewriting] We call a DPO rewriting step as in Definition~\ref{def:DPO} \emph{weakly convex} if $f : L \rightarrow D$ is a convex 
matching and $v'w' \rightarrow C$ is a weak boundary complement in the leftmost pushout square.
\end{defi}


We can now conclude the soundness and completeness of weakly convex DPO rewriting for string diagrams with commutative monoid structure. 
\begin{thm}\label{thm:rew}
Let $\R = \langle l, r \rangle$ be a rewrite rule on $\cprd$. Then,
\begin{equation*}
	\label{eq:th2}
	d \Rightarrow_\R e \textrm{ iff } \iso{d} \rew_{\iso{\R}} \iso{e}
\end{equation*}
\end{thm}
\begin{proof}
For the direction from left to right we proceed as follows. From the definition of rewriting, and given the assumption $d \Rightarrow_\R e$, we have equalities as in~\eqref{eq:stringdiadrew}. We now interpret the string diagrams involved, obtaining right-monogamous cospans:
\begin{equation*}\label{eq:th2cospans}
\begin{aligned}
\left(\cspobj{v}{q_1}{D}{q_2}{w}\right):=\iso{d} \qquad
\left(\cspobj{v}{p_1}{E}{p_2}{w}\right):=\iso{e} \\
\left(\cspobj{v'}{a_1}{L}{a_2}{w'}\right):=\iso{l} \qquad
\left(\cspobj{v'}{b_1}{R}{b_2}{w'}\right):=\iso{r} \\
\left(\cspobj{v}{x_1}{C_1}{x_2}{uv'}\right):=\iso{c_1} \qquad
\left(\cspobj{uw'}{y_1}{C_2}{y_2}{w}\right):=\iso{c_2}
\end{aligned}
\end{equation*}
From the last two cospans above, by simply rearranging nodes on the interface from the left to the right and viceversa, we obtain:
\begin{align*}
\cspobj{v'u}{\tilde{x_1}}{\tilde{C_1}}{\tilde{x_2}}{v} \qquad\qquad
\cspobj{uw'}{\tilde{y_1}}{\tilde{C_2}}{\tilde{y_2}}{w}
\end{align*}
We now define a cospan $\cspobj{v'w'}{}{C}{}{vw}$ as:
\begin{align*}\label{eq:weakboundth2}
\left(\cspobj{v'w'}{z_1}{i+k+j}{z_2}{v'u u w'}\right) \, ; \, \left(\begin{matrix}
   & \cspobj{v'u}{\tilde{x_1}}{\tilde{C_1}}{\tilde{x_2}}{v} &\\
   & \bigoplus & \\
   & \cspobj{u w'}{\tilde{y_1}}{\tilde{C_2}}{\tilde{y_2}}{w}  &
   \end{matrix}\right)
\end{align*}
where $z_1 \colon (v'w') \to (iv' u w')$ is the inclusion map, $z_2 \colon (v' u u w') \to (v' u w')$ is defined as $id_{v'} + \mu_u + id_{w'}$, with $\mu_u \colon u u \to u$ mapping both copies of node $x$ in the word $u u$ to the single corresponding $x$ in $u$. Intuitively, $\cspobj{v'w'}{}{C}{}{vw}$ represents the string diagram where we have rearranged nodes in $v$ and $v'$ on the opposite side. One may verify that:
\begin{equation*}\label{eq:dpoconstr}
    \begin{aligned}
\left(\cspobj{\emptyword}{}{D}{[q_1,q_2]}{v w}\right)=\left(\cspobj{\emptyword}{}{L}{[a_1,a_2]}{v'w'}\right);(\cspobj{v'w'}{}{C}{}{v w})
\\
\left(\cspobj{\emptyword}{}{E}{[p_1,p_2]}{v w}\right)=\left(\cspobj{\emptyword}{}{R}{[b_1,b_2]}{v'w'}\right);(\cspobj{v'w'}{}{C}{}{v w})
\end{aligned}
\end{equation*}
Recall that composition of cospans is obtained via pushouts, hence the two equalities of~\eqref{eq:dpoconstr} yield a DPO rewriting step $\iso{d} \rew_{\iso{\R}} \iso{e}$ as in Definition~\ref{def:DPO}.
Since $l$ is simply a sub-string diagram of $d$, the mapping from $L$ to $D$ is a convex 
match.
Furthermore, note that no two nodes from $v'u$ can be identified with each other, hence $v' \rightarrow C$ is mono, and no node from $v'$ can be identified with any node in $w'$ or $w$. As $w \rightarrow C$ is trivially mono, we have that $C$ is indeed a weak boundary complement.

Now we deal with the converse implication. Assume $\iso{d} \rew_{\iso{\R}} \iso{e}$, where $\iso{d}$, $\iso{e}$, $\iso{l}$, and $\iso{r}$ are defined as the cospans in~\eqref{eq:th2cospans}. 
By assumption, and since composition of cospans is performed via pushouts, there exists a weak boundary complement $\cspobj{v'w'}{[c_1,c_2]}{L^\bot}{[d_1,d_2]}{v w}$ such that
\begin{align*}
\iso{d}&=(\cspobj{\emptyword}{}{v'}{\mu_i}{v'v'});(id_{v'}\oplus \iso{l});(\cspobj{v'w'}{[c_1,c_2]}{L^\bot}{[d_1,d_2]}{v w})\\
\iso{e}&=(\cspobj{\emptyword}{}{v'}{\mu_{v'}}{v'v'});(id_{v'}\oplus \iso{r});(\cspobj{v'w'}{[c_1,c_2]}{L^\bot}{[d_1,d_2]}{v w})
\end{align*}
We now would like to decompose the cospan $\cspobj{v w'}{[c_2,d_1]}{L^\bot}{[c_1,d_2]}{w v'}$ into
\[(\cspobj{vw'}{}{C_1}{}{uv'w'})\,;\,(\cspobj{uv'w'}{id_u\oplus \sigma_{v'}^{w'}}{uw'v'}{}{uw'v'})\,;\,(\cspobj{uw'v'}{}{C_2}{}{wv'})\]
for some $u\in\Sigma^*$, right-monogamous cospans $\cspobj{v}{}{C_1}{}{v'u}$ and $\cspobj{w'u}{}{C_2}{}{w}$, and where $\sigma_{v'}^{w'}:v'w'\to w'v'$ the map that swaps the two components $v'$ and $w'$.  

For this purpose, let $C_1$ be the hypergraph whose set of hyperedges and nodes are all those that are on some path preceding a node in the image of $v'$ but not on a path starting from a node in the image of $w'$; let $C_2$ be a hypergraph whose hyperedges and nodes are those that are on a path starting from any of the nodes in $C_1$ (in $L^\bot$ that is) or in the image of $w'$. Note that these conditions are not exhaustive: the remaining nodes and hyperedges can be either in $C_1$ or $C_2$ with no consequence (we don't require $C_1$ and $C_2$ to be unique in any way). For both of these, let the source and target maps be the appropriate restrictions of the source and target maps of $L^\bot$. For the cospans, the left leg of $\cspobj{v}{}{C_1}{}{v'u}$ is $c_2$; its right leg is the map which restricts to $d_2$ on $v'$ and is the identity on the $u$-nodes shared between $C_1$ and $C_2$ (for some fixed ordering of them to identify them with the word $u$). The right leg of $\cspobj{w'u}{}{C_2}{}{w}$ is $c_1$; its left leg is the identity on the $u$ shared nodes between $C_1$ and $C_2$ as before, and restricts to $d_1$ on the image of $w'$. We have thus built a decomposition of $\cspobj{vw'}{[c_2,d_1]}{L^\bot}{[c_1,d_2]}{wv'}$ as
\[(\cspobj{vw'}{}{C_1}{}{u v'w'})\,;\,(\cspobj{u v'w'}{id_k\oplus \sigma_{v'}^{w'}}{uw'v'}{}{uw'v'})\,;\,(\cspobj{uw'v'}{}{C_2}{}{wv'})\]

By fullness of $\iso{\cdot}$ we have $c_1$, $c_2$ such that $\iso{c_1} =\cspobj{v}{}{C_1}{}{v'u}$ and $\iso{c_2}=\cspobj{w'u}{}{C_2}{}{w}$; moreover we have, by construction:
\begin{align*}
\iso{d} =  \big(id_n\oplus(\cspobj{\emptyword}{}{v'}{\mu_{v'}}{v'v'})\big);\iso{\begin{tikzpicture}
	\begin{pgfonlayer}{nodelayer}
		\node [style=none] (72) at (-0.5, 0.75) {};
		\node [style=none] (73) at (0.5, 0) {};
		\node [style=none] (74) at (-0.5, 0) {};
		\node [style=none] (75) at (0.5, 0.75) {};
		\node [style=none] (76) at (0.5, 0.5) {};
		\node [style=none] (77) at (1.5, 0.5) {};
		\node [style=none] (78) at (1.5, 1.5) {};
		\node [style=none] (79) at (-1.5, 1.5) {};
		\node [style=none] (80) at (-1.5, 0.5) {};
		\node [style=none] (81) at (-0.5, 0.5) {};
		\node [style=none] (82) at (-0.5, 1.5) {};
		\node [style=none] (83) at (0.5, 1.5) {};
		\node [style=none] (85) at (1.5, 1) {};
		\node [style=none] (86) at (2.25, 1) {};
		\node [style=none] (87) at (-1.5, 1) {};
		\node [style=none] (88) at (-3.75, 1) {};
		\node [style=none] (89) at (2, 1.25) {\scriptsize $w$};
		\node [style=none] (90) at (-3.5, 1.25) {\scriptsize $v$};
		\node [style=none] (91) at (2, 0.25) {\scriptsize $v'$};
		\node [style=none] (92) at (-1.5, 0.25) {\scriptsize $w'$};
		\node [style=none] (105) at (-0.5, 1.25) {};
		\node [style=none] (106) at (0.5, 1.25) {};
		\node [style=none] (107) at (0, 1.5) {\scriptsize $u$};
		\node [style=none] (108) at (1, 1) {$c_2$};
		\node [style=none] (109) at (-1, 1) {$c_1$};
		\node [style=none] (110) at (-2, 0) {};
		\node [style=none] (111) at (2.25, 0) {};
		\node [style=none] (112) at (-3.75, -0.75) {};
		\node [style=none] (113) at (2.25, -0.75) {};
		\node [style=none] (114) at (2, -0.5) {\scriptsize $v'$};
		\node [style=none] (116) at (-3, 0.5) {};
		\node [style=none] (117) at (-3, -0.5) {};
		\node [style=none] (118) at (-2, -0.5) {};
		\node [style=none] (119) at (-2, 0.5) {};
		\node [style=none] (120) at (-3, 0) {};
		\node [style=none] (122) at (-2.5, 0) {$l$};
		\node [style=none] (123) at (-3.75, -0.5) {\scriptsize $v'$};
		\node [style=none] (124) at (-3.75, 0) {};
		\node [style=none] (125) at (-3.75, 0.25) {\scriptsize $v'$};
	\end{pgfonlayer}
	\begin{pgfonlayer}{edgelayer}
		\draw (80.center) to (81.center);
		\draw (81.center) to (82.center);
		\draw (83.center) to (76.center);
		\draw (76.center) to (77.center);
		\draw (77.center) to (78.center);
		\draw (79.center) to (80.center);
		\draw [in=180, out=0] (72.center) to (73.center);
		\draw [in=180, out=0] (74.center) to (75.center);
		\draw (88.center) to (87.center);
		\draw (85.center) to (86.center);
		\draw (79.center) to (82.center);
		\draw (78.center) to (83.center);
		\draw (105.center) to (106.center);
		\draw (110.center) to (74.center);
		\draw (73.center) to (111.center);
		\draw (112.center) to (113.center);
		\draw (117.center) to (118.center);
		\draw (118.center) to (119.center);
		\draw (116.center) to (117.center);
		\draw (116.center) to (119.center);
		\draw (124.center) to (120.center);
	\end{pgfonlayer}
\end{tikzpicture}};\big(id_w\oplus\cspobj{v'v'}{\mu_{v'}}{v'}{}{\emptyword})\big)\\
\iso{e} =  \big(id_v\oplus(\cspobj{\emptyword}{}{v'}{\mu_{v'}}{v'v'})\big);\iso{\begin{tikzpicture}
	\begin{pgfonlayer}{nodelayer}
		\node [style=none] (72) at (-0.5, 0.75) {};
		\node [style=none] (73) at (0.5, 0) {};
		\node [style=none] (74) at (-0.5, 0) {};
		\node [style=none] (75) at (0.5, 0.75) {};
		\node [style=none] (76) at (0.5, 0.5) {};
		\node [style=none] (77) at (1.5, 0.5) {};
		\node [style=none] (78) at (1.5, 1.5) {};
		\node [style=none] (79) at (-1.5, 1.5) {};
		\node [style=none] (80) at (-1.5, 0.5) {};
		\node [style=none] (81) at (-0.5, 0.5) {};
		\node [style=none] (82) at (-0.5, 1.5) {};
		\node [style=none] (83) at (0.5, 1.5) {};
		\node [style=none] (85) at (1.5, 1) {};
		\node [style=none] (86) at (2.25, 1) {};
		\node [style=none] (87) at (-1.5, 1) {};
		\node [style=none] (88) at (-3.75, 1) {};
		\node [style=none] (89) at (2, 1.25) {\scriptsize $w$};
		\node [style=none] (90) at (-3.5, 1.25) {\scriptsize $v$};
		\node [style=none] (91) at (2, 0.25) {\scriptsize $v'$};
		\node [style=none] (92) at (-1.5, 0.25) {\scriptsize $w'$};
		\node [style=none] (105) at (-0.5, 1.25) {};
		\node [style=none] (106) at (0.5, 1.25) {};
		\node [style=none] (107) at (0, 1.5) {\scriptsize $u$};
		\node [style=none] (108) at (1, 1) {$c_2$};
		\node [style=none] (109) at (-1, 1) {$c_1$};
		\node [style=none] (110) at (-2, 0) {};
		\node [style=none] (111) at (2.25, 0) {};
		\node [style=none] (112) at (-3.75, -0.75) {};
		\node [style=none] (113) at (2.25, -0.75) {};
		\node [style=none] (114) at (2, -0.5) {\scriptsize $v'$};
		\node [style=none] (116) at (-3, 0.5) {};
		\node [style=none] (117) at (-3, -0.5) {};
		\node [style=none] (118) at (-2, -0.5) {};
		\node [style=none] (119) at (-2, 0.5) {};
		\node [style=none] (120) at (-3, 0) {};
		\node [style=none] (122) at (-2.5, 0) {$r$};
		\node [style=none] (123) at (-3.75, -0.5) {\scriptsize $v'$};
		\node [style=none] (124) at (-3.75, 0) {};
		\node [style=none] (125) at (-3.75, 0.25) {\scriptsize $v'$};
	\end{pgfonlayer}
	\begin{pgfonlayer}{edgelayer}
		\draw (80.center) to (81.center);
		\draw (81.center) to (82.center);
		\draw (83.center) to (76.center);
		\draw (76.center) to (77.center);
		\draw (77.center) to (78.center);
		\draw (79.center) to (80.center);
		\draw [in=180, out=0] (72.center) to (73.center);
		\draw [in=180, out=0] (74.center) to (75.center);
		\draw (88.center) to (87.center);
		\draw (85.center) to (86.center);
		\draw (79.center) to (82.center);
		\draw (78.center) to (83.center);
		\draw (105.center) to (106.center);
		\draw (110.center) to (74.center);
		\draw (73.center) to (111.center);
		\draw (112.center) to (113.center);
		\draw (117.center) to (118.center);
		\draw (118.center) to (119.center);
		\draw (116.center) to (117.center);
		\draw (116.center) to (119.center);
		\draw (124.center) to (120.center);
	\end{pgfonlayer}
\end{tikzpicture}};\big(id_w\oplus(\cspobj{v'v'}{\mu_{v'}}{v'}{}{\emptyword})\big)
\end{align*}
Computing these cospans, we obtain 
\[\iso{d}=\iso{c_1};(\iso{id_u} \oplus \iso{l}); \iso{c_2}\quad \text{ and }\quad \iso{e}=\iso{c_1};(\iso{id_u} \oplus \iso{r}); \iso{c_2}\]
By monoidal functoriality of $\frob{\cdot}$, we have $\iso{d}\!=\!\iso{c_1;(id_u \oplus l); c_2}$ and $\iso{e}\!=\!\iso{c_1;(id_u \oplus r); c_2}$. Finally, since $\iso{\cdot}$ is faithful, we can conclude that $d = c_1;(id_u \oplus l); c_2$ and $e = c_1;(id_u \oplus r); c_2$. This is precisely what it means to  apply the rule $\langle l, r \rangle$ to $d$, so that $d \Rightarrow_\R e$ as we wanted to prove.
\end{proof}




\section{Conclusions and Future Work}\label{sec:conclusions}
The main contribution of this work is twofold. First, with Theorem~\ref{thm:iso}, we identified a combinatorial representation of string diagrams modulo commutative monoid structure. This correspondence relies on introducing a notion of right monogamous cospans, which is intermediate between the `vanilla' cospans characterising string diagrams modulo Frobenius structure, and the monogamous cospans characterising string diagrams modulo symmetric monoidal structure. The characterisation result relies on a factorisation result for right monogamous cospans, and requires some sophistication: compared with similar theorems in~\cite{pt1,pt2} the increased complexity is due to the fact that there is additional structure to consider both on the side of string diagrams (in contrast with~\cite[Theorem 25]{pt2}, which only accounts for symmetric monoidal structure) and on the side of hypergraphs (in contrast with~\cite[Theorem 4.1]{pt1}, which accounts for generic hypergraph without monogamy conditions). 

Note that the work of Fritz and Liang~\cite{Fritz}, which appeared at the same time as a preprint of our work~\cite{arXivversion}, provides a result dual to Theorem~\ref{thm:iso}: instead of monoids, they consider props with a chosen commutative \emph{comonoid} structure---called `CD-categories' or `gs-categories'. On the side of hypergraphs, instead of restricting  monogamy to right-monogamy, they consider \emph{left-monogamy}, which is essentially the dual notion. Moreover, their result is more general than our first characterisation theorem, since it proves the 2-categorical universal property of the free gs-monoidal category over a chosen signature, and does not require the target gs-monoidal categories in the associated equivalence to be strict~\cite[Theorem 4.1]{Fritz}. The combinatorial heart of their proof relies on stripping `pieces' (discard, copy, or hyperedge) of a given left-monogamous cospan of hypergraph from the right and showing that any two equivalent cospans can be stripped of the same pieces in the same order. This is not dissimilar to our factorisation into levels, though Fritz and Liang induct over a certain measure of complexity of diagrams that is different from our approach. Along the same lines, Corradini and Gadducci had constructed the free (single-coloured) gs-monoidal signature over some signature~\cite[Theorem 23]{corradini1999algebraic}, but their proof contains a subtle gap, as pointed out by Fritz and Liang~\cite[Section 4]{Fritz}. Since our proof uses a different approach and does not rely on the same problematic assumption, it is not susceptible to the same issue. 

The second contribution of our paper, Theorem~\ref{thm:rew}, showed a correspondence between string diagrams rewriting modulo commutative monoid structure and a certain variant of DPO hypergraph rewriting. In order to ensure soundness and completeness, we introduced a suitable restriction of DPO rewriting, called \emph{weakly convex} to echo the convex rewriting characterising string diagrams in a symmetric monoidal category~\cite{pt2}. A subtlety of this result was identifying a suitably weak notion of boundary complement. Even though weak boundary complements are not unique `on-the-nose' as the boundary complements considered in the more restrictive setting of convex rewriting, they are sufficiently well-behaved for the purpose of establishing the correspondence with string diagram rewriting.

Going forward, other interesting directions to pursue are the study of confluence as in~\cite{pt3}, now in the presence of commutative (co)monoids. We are also interested in characterising notions of rewriting modulo structures intermediate between commutative monoid and Frobenius algebra---comparison with the very recent work on rewriting in traced comonoid structure~\cite{GhicaKaye23} seems particularly promising in this regard. In terms of case studies, as mentioned in the introduction, our work paves the way for the study of rewriting for theories which do not host a Frobenius structure, yet include commutative (co)monoids. Crucially, the equations of commutative (co)monoids immediately lead to non-terminating rewrite systems if taken naively as rewrite rules. Categories of matrix-like structures, based on bialgebra or Hopf algebras, well-known to be incompatible with the axioms of Frobenius algebras (see \emph{eg.}~\cite{Zanasi15} for an overview), would seem to be a particularly fitting candidate for further investigation.


\bibliographystyle{alphaurl}
\bibliography{sample}

\newpage

\appendix

\section{String diagrams for SMCs}\label{app}

As explained in the main body of the paper, Joyal and Street have shown that the morphisms of the free SMC over a given signature can be described as certain graphs, thereby justifying drawing them as string diagrams~\cite[Theorem 2.3]{JOYAL199155}. We recall here the more precise statement of their result and show how their graphs correspond precisely to monogamous acyclic cospans of hypergraphs. This will allow us to show Proposition~\ref{prop:monogamous-interpret-faithful}.

\begin{defi}\label{def:JS-graph}
A \emph{JS-graph} is a directed acyclic graph\footnote{Joyal and Street use the terms \emph{oriented} instead of directed, and \emph{progressive} instead of acyclic. We renamed terminology to match the one used in the main text of our paper. Their graphs also allow for multiple edges and self-loops.} with an anchored boundary and a valuation.

A \emph{boundary} is a distinguished set of nodes of degree one. The boundary nodes with out-degree one are called the \emph{inputs} of $G$, and those of in-degree one are its \emph{outputs}. Non-boundary nodes are called \emph{inner} nodes. 

The \emph{domain} (resp. \emph{codomain}) of a JS-graph is the set of edges connected to an input (resp. output) node. Such a graph is \emph{anchored} when its domain and codomain are each linearly ordered\footnote{The authors also require the graph to be \emph{polarised}, that is, the input and output nodes are each linearly ordered, which is implied by the anchored condition.}.  

Finally, a \emph{valuation} is an assignment of colours of $\mathcal{C}$  to each edge of the graph and an operation of the chosen signature $\Sigma$ to each inner node of the graph.\footnote{The authors call signatures \emph{tensor schemes}.}
\end{defi}

Like cospan of hypergraphs, the graphs of Joyal and Street are defined up to isomorphism. Their notion of isomorphism is the usual graph isomorphism with the added requirement that the valuation and the anchoring of boundary edges be preserved. This is equivalent to the fact that we only consider cospans of hypergraphs up to isomorphism.

The composition $G; H$ of two JS-graphs $G$ and $H$ such that the codomain of $G$ matches that of $H$ is described as follows by Joyal and Street: the graph is obtained from the disjoint union of $G$ and $H$ by identifying the output nodes of $G$ and the input nodes of $H$; the inner nodes and edges are those of $G$ and $H$, except for the edges of $G$ which have target a boundary node, and the edges of $H$ which have sources a boundary node; these edges pair up via the corresponding boundary node, each pair corresponding an edge to $G; H$; as paired edges have equal values, we obtain a valuation on $G; H$.

JS-graphs can also be composed in parallel. Joyal and Street define the monoidal product $G_1\otimes G_2$ as the disjoint union of the two graphs, with boundary nodes those of $G_1$ and $G_2$, whose valuation restricts to $G_1$ and $G_2$ to give their valuations. 

JS-graphs with the operations of composition and monoidal product define a symmetric monoidal category whose objects are words over the generating colours $\mathcal{C}$ and morphisms are JS-graphs with domain and codomain given by the valuation at their domain and codomain~\cite[Section 2]{JOYAL199155}. In fact, JS-graphs whose nodes are valued in the chosen signature $\Sigma$ and whose edges are valued in the chosen set of colours $\mathcal{C}$ form a $\mathcal{C}$-coloured prop, which we call $\JSGraph_{\Sigma,\mathcal{C}}$.

Joyal and Street have shown that JS-graphs are string diagrams for the free SMC on a given set of colours and signature. The following is a simple reformulation in our language of~\cite[Theorem 2.3]{JOYAL199155}.
\begin{thm}\label{thm:JS-graph-diagrams}
There is an isomorphism $G$ of coloured props between $\mathbf{S}_{\Sigma,\mathcal{C}}$ and $\JSGraph_{\Sigma,\mathcal{C}}$.
\end{thm}

It is not difficult to establish a one-to-one correspondence between JS-graphs and monogamous acyclic cospans of hypergraphs. 
\begin{defi}
Given a JS-graph $G$, let $F(G) = (\cspobj{u}{f}{H}{g}{v})$ be the cospan of hypergraphs such that
\begin{itemize}
\item the $(k,l)$-hyperedges of $H$ are the \emph{inner} nodes of $G$ with in-degreee $k$ and outdegree $l$, with label in $\Sigma$ given by the corresponding valuation of $G$;
\item the hypernodes of $H$ are the edges of $G$ with colour label given by the corresponding valuation of $G$;
\item a hypernode $x$ of $H$ is a source of some hyperedge $h$, if the corresponding edge $x$ in $G$ has target $h$ in $G$; similarly, a hypernode $x$ of $H$ is a target of some hyperedge $h$, if the corresponding edge $x$ in $G$ has source $h$ in $G$;
\item the map $f$ sends the $n$-th colour in $u$ to the $n$-th input node of $G$; similarly, $g$ maps the $n$-th colour in $v$ to the $n$-th output node of $G$.
\end{itemize}
\end{defi}
\begin{prop}\label{prop:jsgraph-rmacs}
$F\from \JSGraph_{\Sigma,\mathcal{C}}\to \rmacsphyp$ is a faithful morphism of $\mathcal{C}$-coloured props.
\end{prop}
\begin{proof}
Given two JS-graphs $G$ and $H$, by definition of $G;H$ we have that $F(G;H)$ is isomorphic to the cospan of hypergraphs obtained by composing (via pushout) the two cospans $F(G)$ and $F(H)$: indeed, it is the disjoint sum of the two hypergraphs, with the boundary nodes identified. Similarly, $F(G_1)\otimes F(G_2)$ is clearly isomorphic to $F(G_1\otimes G_2)$.

Let $G$ and $G'$ be two JS-graphs with the same boundary, such that $F(G)$ is isomorphic to $F(G')$ as cospans of hypergraphs---call $f$ this isomorphism. We write $f_\star$ for its action on nodes and $f_{k,l}$ for its action on hyperedges. Let us define an isomorphism $g = (g_V,g_E)$ between $G$ and $G'$ from this data. First, $g_V$ sends an inner node $x$ of $G$ with in-degree $k$ and out-degree $l$ to $f_{k,l}(x)$ (since $x$ is a hyperedge of $F(G)$ by definition), which is a hyperedge of $F(G')$ and therefore a node of $G'$ by definition of $F(G')$. On boundary nodes $g_V$ is the identity. Then, $g_E$ maps an edge $e$ of $G$ to $f_\star(e)$ (since $e$ is a hypernode of $F(G)$ by definition), which is a hypernode of $F(G')$ and therefore an edge of $G'$ by definition of $F(G')$. This does define a homomorphism: if $e$ is an edge of $G$ with source $x$ and target $x'$, then $g_E(e) = f_\star(e)$ has source $f_{k,l}(x) = g_V(x)$ and target $f_{k',l'}(x')= g_V(x')$ as required, because $f$ is a hypergraph-homomorphism. Moreover, $g$ trivially maps boundary nodes to boundary nodes, and preserves valuations because $f$ preserves labels.

Finally, $g_E$ is one-to-one since $f_\star$ is, and $g_V$ is also one-to-one since it is the disjoint sum of all $f_{k,l}$ for all pairs $(k,l)$ of in-degrees and-out-degrees of nodes of $G$, and each $f_{k,l}$ is assumed one-to-one too. Thus $F$ is faithful. 
\end{proof}

As a result, using Joyal and Street's characterisation of string diagrams as JS-graphs (Theorem~\ref{thm:JS-graph-diagrams}), we can show that $\sfun: \mathbf{S}_{\Sigma,\mathcal{C}} \rightarrow \rmacsphyp$ is a faithful $\mathcal{C}$-coloured prop morphism whose image consists precisely of the monogamous acyclic cospans.
\begin{proof}[Proof of Proposition~\ref{prop:monogamous-interpret-faithful}]
Recall the isomorphism $G$ of Theorem~\ref{thm:JS-graph-diagrams}. Because $\mathbf{S}_{\Sigma,\mathcal{C}}$ is free, $G$ is fully characterised by its image on elements of $\Sigma$: each generator $o\in\Sigma_{m,n}$ is mapped to a JS-graph with a single inner node with value $o$, $m$ input nodes and $n$ output nodes. Clearly, $\sfun[o] = FG(o)$, and therefore more generally we have $\sfun = FG$. Since $\sfun$ is the composite of two faithful morphisms, it is also faithful. Finally, the image of $\sfun$ are precisely the monogamous acyclic cospans of hypergraphs, as can be checked on generators once again. 
\end{proof}
\end{document}